\documentclass[%
reprint,
superscriptaddress,
amsmath,amssymb,
aps,longbibliography,
]{revtex4-1}
\usepackage[breaklinks=true]{hyperref}

\usepackage{microtype}
\usepackage{multirow}
\usepackage{graphicx}%
\usepackage{verbatim}
\usepackage[version=4]{mhchem}
\usepackage{xcolor}
\usepackage{physics}
\usepackage[normalem]{ulem}
\usepackage{pdfpages}
\usepackage{pgffor}
\usepackage{booktabs}
\AtBeginDocument{
    \heavyrulewidth=.08em
    \lightrulewidth=.05em
    \cmidrulewidth=.03em
    \belowrulesep=.65ex
    \belowbottomsep=0pt
    \aboverulesep=.4ex
    \abovetopsep=0pt
    \cmidrulesep=\doublerulesep
    \cmidrulekern=.5em
    \defaultaddspace=.5em
}

\definecolor{verde}{rgb}{0.,0.6,0}
\definecolor{orange}{rgb}{1.0,0.5,0.0}

\newif\ifhighlight

\newcommand{\highlight}{\highlighttrue}
\highlight

\newcommand{\editor}[2]{%
  \expandafter\newcommand\csname #1note\endcsname[1]{%
    \textcolor{#2}{(\textbf{#1note:} \textsc{##1})}}%
  \expandafter\newcommand\csname #1\endcsname[1]{%
    \ifhighlight\textcolor{#2}{##1} \else ##1\fi}%
  \expandafter\newcommand\csname #1cancel\endcsname[1]{%
    \ifhighlight\textcolor{#2}{\sout{##1}}\fi}%
  \expandafter\newcommand\csname #1change\endcsname[2]{%
    \ifhighlight\textcolor{#2}{\sout{##1} ##2}\else ##2\fi}%
  \newenvironment{#1text}{\ifhighlight\color{#2}\fi}{\color{black}}
}

\editor{MK}{orange}
\editor{MS}{verde}

\newcommand*\dif{\mathop{}\!\mathrm{d}}

\makeatletter
\AtBeginDocument{\let\LS@rot\@undefined}
\makeatother

\begin{document}

\title{How to Train a Shallow Ensemble}

\author{Moritz Sch\"afer}
\affiliation{Laboratory of Computational Science and Modeling, Institut des Mat\'eriaux, \'Ecole Polytechnique F\'ed\'erale de Lausanne, 1015 Lausanne, Switzerland}
\affiliation{Institute for Theoretical Chemistry, University of Stuttgart, Pfaffenwaldring 55, 70569 Stuttgart, Germany}

\author{Matthias Kellner}
\affiliation{Laboratory of Computational Science and Modeling, Institut des Mat\'eriaux, \'Ecole Polytechnique F\'ed\'erale de Lausanne, 1015 Lausanne, Switzerland}

\author{Johannes K\"astner}
\affiliation{Institute for Theoretical Chemistry, University of Stuttgart, Pfaffenwaldring 55, 70569 Stuttgart, Germany}

\author{Michele Ceriotti}
\email{michele.ceriotti@epfl.ch}
\affiliation{Laboratory of Computational Science and Modeling, Institut des Mat\'eriaux, \'Ecole Polytechnique F\'ed\'erale de Lausanne, 1015 Lausanne, Switzerland}

\newcommand{\CSM}[1]{{\color{red}#1}}

\date{\today}%

\begin{abstract}

\emph{Shallow ensembles} provide a convenient strategy for uncertainty quantification in machine learning interatomic potentials, that is computationally efficient because the different ensemble members share a large part of the model weights. 
In this work, we systematically investigate training strategies for shallow ensembles to balance calibration performance with computational cost.
We first demonstrate that explicit optimization of a negative log-likelihood (NLL) loss improves calibration with respect to approaches based on ensembles of randomly initialized models, or on a last-layer Laplace approximation. 
However, models trained solely on energy objectives yield miscalibrated force estimates. We show that explicitly modeling force uncertainties via an NLL objective is essential for reliable calibration, though it typically incurs a significant computational overhead. To address this, we validate an efficient protocol: full-model fine-tuning of a shallow ensemble originally trained with a probabilistic energy loss, or one sampled from the Laplace posterior. This approach results in negligible reduction in calibration quality compared to training from scratch, while reducing training time by up to 96~\%. 
We evaluate this protocol across a diverse range of materials, including amorphous carbon, ionic liquids (BMIM), liquid water (\ce{H2O}), barium titanate (BaTiO$_3$), and a model tetrapeptide (Ac-Ala3-NHMe), establishing practical guidelines for reliable uncertainty quantification in atomistic machine learning.

\end{abstract}

\maketitle

\newcommand{\nens}{n_\text{ens}}

\section{Introduction}

Introducing machine learning (ML) surrogate models and machine learning interatomic potentials (MLIPs) into first-principles atomistic modeling workflows should always be approached with care.
Most ML models originate in statistical learning frameworks and therefore introduce additional uncertainty in their predictions, either from limited knowledge (epistemic uncertainty)\cite{kendallWhatUncertaintiesWe2017}, irreducible noise in the training data (aleatoric uncertainty)\cite{aldossarySilicoChemicalExperiments2024,hullermeierAleatoricEpistemicUncertainty2021b} and the inability of the chosen ML architecture to capture complex physical interactions (model misspecification)\cite{perezUncertaintyQuantificationMisspecified2025}. 
These uncertainties add to the pre-existing discrepancies of observables computed with the electronic-structure method targeted for ML acceleration and experimental observations\cite{herbstPosterioriErrorEstimation2020,pernotLongRoadCalibrated2022a}.

In practice, these additional sources of uncertainty can negate any advantages gained from using faster ML models.
The successful deployment of ML models and trustworthy interpretation of ML-accelerated simulation, therefore, relies not only on accurate point estimates but also on the ability to quantify the reliability of predictions using well-calibrated uncertainty estimates \cite{lakshminarayananSimpleScalablePredictive2017} and, ultimately, practical ways to propagate predicted model uncertainties to derived quantities\cite{carleoMachineLearningPhysical2019,imba+21jcp}.
Beyond data generation via active learning procedures \cite{smithLessMoreSampling2018, kulichenkoUncertaintydrivenDynamicsActive2023, zillsMachineLearningdrivenInvestigation2024a, schaferEnhancedRepresentationBasedSampling2026}, calibrated uncertainties are essential for production simulations, where they can flag unreliable results or be propagated through downstream workflows, for example, quantifying the error on average thermodynamic quantities\cite{imba+21jcp}.

Neural-network-based (NN) ML models gained popularity because they offer favourable asymptotic scaling for both training and inference, and they are supported by highly optimized implementations in general machine-learning frameworks.
In contrast to models such as Gaussian Process Regressors (GPR), which provide built-in posterior inference \cite{rasmussenGaussianProcessesMachine2008}, traditional NNs that are comprised of one set of model weights from maximum likelihood, or MAP training,  are point estimators and must be equipped with approximate uncertainty quantification (UQ) schemes \cite{gawlikowskiSurveyUncertaintyDeep2023,grasselliUncertaintyEraMachine2025}.
Approximate NN UQ schemes typically either approximate the model posterior or try to estimate quantities that can serve as proxies to the confidence of model predictions.

Common strategies to approximate the weight posterior include Bayesian Neural Networks \cite{mackayPracticalBayesianFramework1992} and Monte Carlo dropout \cite{galDropoutBayesianApproximation} and Deep Ensembles \cite{lakshminarayananSimpleScalablePredictive2017}.
Alternative approaches seek to reduce computational cost via direct Mean-Variance Estimation (MVE) \cite{nixEstimatingMeanVariance1994,aminiDeepEvidentialRegression2020a,tanSinglemodelUncertaintyQuantification2023} or by utilizing distance-based metrics in latent space, such as conformal prediction \cite{hoFlexibleUncertaintyCalibration2025,zhuFastUncertaintyEstimates2023,garainUncertaintyCalibrationMolecular2025}.
However, for MLIPs, full ensembles consisting of independently trained models are often favored for their robustness and simplicity\cite{smithLessMoreSampling2018}.

We recently introduced the ``direct propagation of shallow ensembles'' (DPOSE) scheme \cite{kellnerUncertaintyQuantificationDirect2024a}, an ensemble-based ML model UQ scheme, striking a good balance between accuracy and evaluation cost, which can be applied to any architecture and has low implementation complexity.
DPOSE reduces the overhead of traditional full ensembles by sharing the model backbone and ensembling only the last layer, with a joint training procedure of all ensemble members simultaneously, using a probabilistic Gaussian negative log-likelihood (NLL) loss function.
By integrating uncertainty awareness into the training procedure, DPOSE ensures well-calibrated uncertainty estimates with negligible additional training and evaluation cost. DPOSE has been successfully applied in materials modelling applications such as propagating the uncertainties of the universal machine-learning interatomic potential PET-MAD to melting-point calculations \cite{mazitovPETMADLightweightUniversal2025a}, performing active learning in surface catalysis \cite{kempenBreakingScalingRelations2025}, modelling general organic reactions \cite{chemrxiv-2025-f1hgn-v4}, and detecting out-of-domain samples \cite{vinchurkarUncertaintyQuantificationGraph2025}.

An alternative approach to generate shallow ensembles is the Last Layer Prediction Rigidity (LLPR)\cite{chon+23jctc,bigiPredictionRigidityFormalism2024a,chon+25fd} formalism.
It is based on a Laplace approximation of the last layer posterior, and unlike DPOSE, it is applied post-training to a single, MSE-trained MLIP.
In its original formulation, the LLPR method was only constructed for direct predictions, like energies, but not derivatives like forces.
However, this extension was performed for the MACE-MP0 foundation model\cite{batatiaFoundationModelAtomistic2025}.
While the DPOSE method was not initially designed with explicit training for force uncertainty, it still showed potential for providing reliable force estimates. However, the actual impact of adding force uncertainty-aware training to the DPOSE model remained untested until now.
The last-layer-based nature of the two approaches naturally raises the question of their similarity and efficacy.
Specifically, it is unclear whether DPOSE is merely a Monte Carlo approximation of the same posterior targeted by LLPR, or whether the calibration of uncertainties arises from the learned features of the whole model.
Furthermore, including a probabilistic force loss poses some challenges during training as computing the associated uncertainties requires evaluating the Jacobian of ensemble member energies with respect to coordinates, which, unlike the probabilistic energy training, incurs a high cost.
It is therefore crucial to explore efficient ways for force uncertainty-aware training.

In this work, we explore the design space of DPOSE training strategies for neural network (NN) based MLIPs.
We compare the quality of shallow ensemble and LLPR uncertainties, discuss the necessary steps required for well-calibrated force uncertainty estimates, and conclude by providing practical recommendations for training efficient, uncertainty-aware potentials.

\section{Methods}

\subsection{Machine learning interatomic potentials}

Accurate interatomic potentials are central to atomistic modelling, enabling the connection between microscopic structure and macroscopic observables via statistical physics.
MLIPs trained on quantum chemical reference data describe potential energy surfaces with high accuracy while significantly reducing computational cost.
Unlike the underlying electronic structure methods (e.g., DFT), MLIPs typically achieve linear scaling with system size by decomposing the total potential energy $E(A, \boldsymbol{\theta})$ into atomic contributions $\varepsilon( A_i, \boldsymbol{\theta})$\cite{behlerGeneralizedNeuralNetworkRepresentation2007a}:

\begin{align}\label{eq:energy_sum}
    E(A, \boldsymbol{\theta}) = \sum_i^{N_{\text{atoms}}} \varepsilon( A_i, \boldsymbol{\theta})
\end{align}

These atomic contributions depend on the local atomic environment $A_i$ and the learnable model parameters $\boldsymbol{\theta}$.
The atomic forces $\boldsymbol{F}_i$ required for molecular dynamics are obtained analytically as the negative gradient of this potential energy surface with respect to atomic positions $\boldsymbol{r}_i$:

\begin{align}\label{eq:forces}
    \boldsymbol{F}_i(A, \boldsymbol{\theta}) = -\nabla_{\boldsymbol{r}_i} E(A, \boldsymbol{\theta})
\end{align}

While architectures differ in how they encode $A_i$, the models used in this work share some common architectural motifs.
First, interatomic distances are expanded within a spherical cutoff using radial basis functions.
The radial two-body information may be augmented by angular and many-body descriptors and (non-)linear transformations in order to enhance the representational fidelity needed to represent complex atomistic environments\cite{schuttEquivariantMessagePassing2021b,pozdnyakovIncompletenessAtomicStructure2020b}.

Crucially for the uncertainty quantification methods discussed later, the final atomic energy is predicted by mapping this learned feature representation $\boldsymbol{h}_i$ through a linear readout layer:

\begin{align}\label{eq:last-layer}
    \varepsilon( A_i, \boldsymbol{\theta}) = \boldsymbol{w}^\top \boldsymbol{h}_i + b
\end{align}

In this work, we use Gaussian Moment Neural Networks (GMNN) \cite{zaverkinGaussianMomentsPhysically2020,zaverkinFastSampleEfficientInteratomic2021} through their efficient implementation in the \texttt{apax} package~\cite{schaferApaxFlexiblePerformant2025}.
We perform some additional experiments using So3krates \cite{frankSo3kratesEquivariantAttention2022} and a NequIP-style\cite{batznerE3equivariantGraphNeural2022} potential, referred to as EquivMP, on selected datasets to test the transferability of our findings to different model architectures.
A more detailed description of the three architectures can be found in SI Section~\ref{sec:architectures}.
All trainings and model evaluations are performed with \texttt{IPSuite}\cite{zillsCollaborationMachineLearnedPotentials2024}.

\subsection{Uncertainty estimation}

To quantify the uncertainty of a model's predictions, we move from single-point estimates to a probabilistic framework.
In a rigorous Bayesian treatment, the model parameters are random variables, and the predictive distribution for a target $y$ is obtained by marginalizing over these parameters.

\begin{equation} \label{eq:bayes_integral}
    p(y|A, \mathcal{D}) = \int p(y|A, \boldsymbol{\theta}) p(\boldsymbol{\theta}|\mathcal{D}) \dif\boldsymbol{\theta}
\end{equation}

As this integral is intractable for deep neural networks, practical methods must find suitable approximations, such as stochastic gradient Markov chain Monte Carlo\cite{welling2011bayesian,thalerScalableBayesianUncertainty2023a}.
Ensembling independently trained models is one of the most common strategies for MLIPs. 
Due to the non-convexity of the neural network parameter optimization problem, different random weight initializations are sufficient to make models converge to distinct local minima of the loss landscape\cite{fortLargeScaleStructure2019,fortDeepEnsemblesLoss2020}.
Consequently, each model provides different predictions $y^{(k)}(A)$ for property $y$ of the inputs $A$.
Mean, and error can then be estimated as:

\begin{equation}
    \bar{y}(A) = \frac{1}{\nens}\sum^{\nens}_{k=1}y^{(k)}(A)
\end{equation}
\begin{equation}
    \sigma^2_{\nens}(A) = \frac{1}{\nens - 1} \sum^{\nens}_{k=1}{[y^{(k)}(A) - \bar{y}(A)]^2}.
\end{equation}

While conceptually simple, this strategy incurs an $\nens$-fold increase in asymptotic training and inference time.
In most atomistic modeling tasks, such a slowdown is not acceptable, as other sources of uncertainties at a fixed computational budget may increase, such as sampling uncertainties when only $\nens$ times shorter simulations may be run.

\subsection{Efficient Last-Layer Approximations}

To circumvent the cost of a full Bayesian treatment and full ensembles, we utilize last-layer approximations.
The cost of ensembling can be reduced by committees that partially or fully share weights up to the readout layer via the DPOSE approach\cite{kellnerUncertaintyQuantificationDirect2024a}.
In DPOSE, a shallow ensemble consisting of multiple output units of the readout layer in Equation~\ref{eq:last-layer} is trained in an end-to-end manner by minimizing the Gaussian NLL loss:

\begin{equation}\label{eq:nll_loss}
   \text{NLL}(\Delta y,\sigma) = \frac{1}{2}\Big[\frac{\Delta y^2}{\sigma^2} + \ln(2\pi\sigma^2)\Big] 
\end{equation}

\noindent
which is summed over the training set and depends on the empirical error $\Delta y$ and the predicted ensemble variance $\sigma^2$ of each sample.
In this formulation, all parameters are jointly optimized, allowing uncertainty information to propagate into the deeper layers during training.
For forward properties like energy, the computational overhead is simply that of a larger last-layer readout.

As an analytical alternative, the LLPR framework applies a Laplace approximation to the posterior of the last-layer based on a single model trained with an MSE loss.

\begin{equation}\label{eq:mse_loss}
    \text{MSE}(\Delta y) = \frac{1}{N} \sum_{i=1}^{N} (\Delta y_i)^2
\end{equation}

The parameters at the loss minimum correspond to the maximum a posteriori (MAP) estimate $\theta_{\text{MAP}}$.
Assuming that $\theta_{\text{MAP}}$ is a local minimum of the loss function, a second-order Taylor expansion of the posterior distribution $p(\theta|\mathcal{D})$ around $\theta_{\text{MAP}}$ yields:

\begin{equation}
   p(\theta|\mathcal{D}) \approx \mathcal{N}(\theta|\theta_{\text{MAP}};\Sigma_{\text{LL}}) 
\end{equation}

The covariance $\Sigma_{\text{LL}}$ is the inverse Hessian of the loss with respect to the parameters at the MAP solution.
For the last-layer weights, the covariance $\Sigma_{\text{LL}}$ is approximated as

\begin{equation}
    \Sigma_{\text{LL}} \approx \alpha^2(\mathbf{H} + \eta^2\mathbf{I})^{-1}
\end{equation}

\noindent
where $\mathbf{H}$ is the generalized Gauss-Newton approximation of the Hessian for the training loss, $\alpha$ is a calibration scalar, and $\eta^2\mathbf{I}$ is a regularization term.
This approximation is used to avoid the high computational cost of exact second derivatives.
The calculation of the approximated Hessian is discussed in more detail in Section~\ref{sec:llpr_detail} as the concrete contributions are dependent on the loss function.
An analytic uncertainty estimate for a query structure $A$ with aggregated features $\boldsymbol{h}(A)  = \sum_{i \in S} \boldsymbol{h}_i$ is then given by:

\begin{equation} \label{eq:analytic_energy_uq}
    \sigma_{E}^2(A) = \boldsymbol{h}(A)^\top \Sigma_{\text{LL}} \boldsymbol{h}(A)
\end{equation}

It is also possible to draw samples from the last layer posterior distribution. By sampling an ensemble of last-layer weights from the LLPR posterior, we can construct models that are architecturally identical to a Shallow Ensemble.

\begin{equation}
    \boldsymbol{\theta}_{\text{LL,ens}} \sim \mathcal{N}(\boldsymbol{\theta}_{\text{LL,MAP}}, \boldsymbol{\Sigma}_{\text{LL}})
    \label{eq:sample_llpr_posterior}
\end{equation}

In practice, the mean of the sampled $\theta_{\text{LL,ens}}$ is constrained to exactly the MAP solution $\theta_{\text{MAP,LL}}$ as discussed in Section~\ref{sec:llpr_detail}.
The explicit committee predictions, just like for a DPOSE ensemble, can then be practically used to propagate the committee predictions through arbitrarily complex workflows.

\subsection{Estimating Force Uncertainty}

Applications, such as active learning, profit strongly from calibrated estimators of the force uncertainty\cite{zhangActiveLearningUniformly2019,podryabinkinActiveLearningLinearly2017a,buiCalibratedUncertaintySampling2025,heidSpatiallyResolvedUncertainties2024a}.
For Shallow Ensembles, we quantify this uncertainty as the standard deviation of the ensemble force predictions.
The force prediction for the $k$-th ensemble member is given by the gradient of its energy with respect to atomic coordinates:

\begin{equation}
    \mathbf{F}_i^{(k)} = -\nabla_{\mathbf{r}_i} E^{(k)}(A)
\end{equation}

When training ensembles end-to-end with force uncertainties, we treat each force component as an independent random variable.
Consequently, the NLL loss is computed component-wise using Equation~\ref{eq:nll_loss}.
It should be noted that Willow et al.\cite{willow2025bayesian} use a block diagonal approximation for force uncertainty informed NLL losses, which considers the covariance of force components on each atom in the context of a mean-variance estimation model.
While they report improvements in force uncertainty calibration compared to only using an energy NLL loss, it remains an open question whether the added complexity of modelling intra-atomic covariance is strictly necessary for calibration compared to a simpler component-wise independent approach.

In the LLPR framework, force uncertainties can be obtained from variance propagation.
Since forces are the negative gradient of the energy, the uncertainty depends on the Jacobian of the learned features $\boldsymbol{h}(A)$ with respect to the atomic coordinates $\mathbf{r}$.
The analytic force variance is given by:

\begin{equation}
    \boldsymbol{\sigma}_{F}^2(A) = \text{diag}\Big(\mathbf{J}(A) \boldsymbol{\Sigma}_{\text{LL}} \mathbf{J}(A)^\top\Big)
\end{equation}

where $\mathbf{J}(A)$ is the feature-coordinate Jacobian.
While this analytic approach has been applied to foundation models like MACE-MP0~\cite{batatiaFoundationModelAtomistic2025}, it relies on a covariance $\boldsymbol{\Sigma}_{\text{LL}}$ constructed solely from the energy Hessian.
In this work, we include the force loss contributions in $\boldsymbol{\Sigma}_{\text{LL}}$ and we refer to Section~\ref{sec:llpr_detail} for more information.

While energy UQ is cheap for both Shallow Ensembles and LLPR, force uncertainty estimation incurs a significant computational cost.
For shallow ensembles, the evaluation cost of derivative quantities scales linearly with the ensemble size $\nens$ (modulo efficient vectorization), as gradients must be computed for each ensemble member, regardless of the degree of weight sharing.
Computing analytic LLPR force uncertainty estimates requires computing the full Jacobian matrix $\mathbf{J} \in \mathbb{R}^{3N_{\text{atoms}} \times  D }$.
For models with large feature vectors, this is prohibitively expensive compared to standard backpropagation.
Table~\ref{tab:evaluation_cost} compares the training and evaluation costs of these strategies.

\begin{table}[ht]
\centering
\caption{Computational scaling of training and evaluation for different uncertainty strategies. $\nens$: Ensemble size, $D$: Last-layer feature dimension.}
\label{tab:evaluation_cost}
\begin{tabular}{lcccc}
    \toprule
    & \multicolumn{2}{c}{\textbf{Training Cost}} & \multicolumn{2}{c}{\textbf{Evaluation Cost}} \\
    \cmidrule(lr){2-3} \cmidrule(lr){4-5}
    Strategy & Energy & Forces & Energy & Forces \\
    \midrule
    \textbf{Full Ensemble}      & $\nens$ & $\nens$ & $\nens$ & $\nens$ \\
    \textbf{Shallow Ensemble}   & $\sim1$ & $\nens$ & $\sim1$ & $\nens$ \\
    \textbf{LLPR}               & $1$     & $1$     & $\sim1$ & $D$     \\
    \bottomrule
\end{tabular}
\end{table}

To avoid the high cost of analytic LLPR evaluation, we sample a committee from the posterior (Eq. \ref{eq:sample_llpr_posterior}), replacing the force-feature Jacobian computation ($D$ backprops) with gradients of a small ensemble ($\nens \ll D$).
For inference, \texttt{apax}\cite{schaferApaxFlexiblePerformant2025} further optimizes performance by computing force uncertainties only for logged configurations, otherwise evaluating the mean energy gradient at single-model cost.
However, this optimization does not apply to training with a probabilistic force loss, which requires evaluating ensemble variance at every step.
Beyond efficiency, sampled ensembles serve as initializations for fine-tuning. Ultimately, all strategies yield an identical structure: a shared backbone acting as a feature map, followed by a committee of last-layer weights.
The key difference between all approaches is to what extent the uncertainty information is permitted to flow into the hidden layers during the learning process, and which uncertainties are explicitly considered in the loss function.
To distinguish the various ways we construct uncertainty estimators, we summarize the abbreviations used in this work in Table~\ref{tab:abbrev}.

\begin{table}[htbp]
    \centering
    \caption{Summary of uncertainty estimation strategies, their training objectives, and Hessian contributions.}
    \label{tab:abbrev}
    \begin{tabular}{l l l}
        \toprule
        \textbf{Method} & \textbf{Training Objective} & \textbf{Hessian contrib.} \\
        \midrule
        SE$_{E}$ & $\text{NLL}_E + \text{MSE}_F$ & Implicit \\
        SE$_{E,F}$ & $\text{NLL}_E + \text{NLL}_F$ & Implicit \\
        LLPR$_{E}$ & $\text{MSE}_E + \text{MSE}_F$ & Energy Hessian \\
        LLPR$_{E,F}$ & $\text{MSE}_E + \text{MSE}_F$ & Energy + Force Hessian  \\
        \bottomrule
    \end{tabular}
\end{table}

\subsection{Post-hoc Calibration and Evaluation of Uncertainty Estimates}

Even with rigorous training objectives and posterior descriptions, uncertainty estimates from neural networks may be miscalibrated \cite{pernotLongRoadCalibrated2022a}.
To address this, we apply a post-hoc calibration step, in a similar spirit to temperature scaling approaches in classification\cite{guoCalibrationModernNeural2017}. 
In the regression setting, an analogous correction can be achieved by scaling the predicted variances $\sigma(A)^{2}$ by a scalar factor $\alpha^{2}$.
Minimizing the NLL with respect to $\alpha$ on a held-out validation set yields a closed-form analytical solution for the calibration factor\cite{musi+19jctc}.

\begin{equation}
    \alpha^{2} = \frac{1}{n_{\textrm{val}}}\sum_{A\in\text{val}}\frac{\Delta y(A)^{2}}{\sigma(A)^{2}}
    \label{eq:calibration}
\end{equation}

\noindent
where $\Delta y(A)$ is the prediction error on the validation sample $A$.
Technically, adding a post-hoc calibration factor to DPOSE uncertainty estimates is not strictly necessary, since DPOSE is trained by directly minimizing an NLL objective; nevertheless, for best comparability between LLPR uncertainty and other uncertainty estimators that are fine-tuned or trained with an NLL loss, we also apply the same post-hoc calibration procedure to these NLL-optimized estimators, and, as expected, find that the resulting calibration factors consistently lie close to 1.
For energy and force predictions, a shared calibration factor can be motivated by interpreting calibration as a linear rescaling of the ensemble members around the mean,
\begin{equation}
    E^{(k)}_{\text{cal}}(A) = \bar{E}(A) + \alpha [ E^{(k)}(A) - \bar{E}(A) ]
\end{equation}
Since the force is the negative gradient of the energy, the scaling of the ensemble energies with a scalar calibration factor propagates directly to the ensemble of forces and hence their calibration.
Consequently, if miscalibration arises purely from a global scaling error in the energy landscape, a single scalar $\alpha$ is sufficient to calibrate both targets.
Alternatively, the calibration may be decoupled by optimizing separate factors $\alpha_E$ and $\alpha_F$ for energies and forces, respectively.
This decoupling accounts for the noise scales of the different targets, even though it makes the energies and forces of the calibrated ensemble members inconsistent.
While more advanced post-hoc calibration techniques exist\cite{hoFlexibleUncertaintyCalibration2025}, we focus here on scalar rescaling to isolate the effects of the underlying feature representation. 

To evaluate the quality of uncertainty estimates, the standard metric is the NLL of the ground truth $y_{\text{ref}}$ given the model's predictive distribution.
However, the absolute magnitude of the NLL is dominated by the model's predictive accuracy (RMSE), making it difficult to compare uncertainty strategies across datasets with different scales.
To overcome this limitation, we utilize a normalized metric, the Relative Log-Likelihood (RLL) introduced by \citeauthor{kellnerUncertaintyQuantificationDirect2024a}\cite{kellnerUncertaintyQuantificationDirect2024a}.

\begin{equation}\label{eq:rll}
    \text{RLL} = \frac{
    \text{NLL}_{\text{model}} - \text{NLL}_{\text{base}}
    }{
    \text{NLL}_{\text{oracle}} - \text{NLL}_{\text{base}}
    } \cdot 100\%.
\end{equation}

The RLL rescales the NLL score between a worst-case baseline and a best-case oracle.
The baseline term $\text{NLL}_{\text{base}}$ corresponds to a constant uncertainty estimate equal to the test set RMSE, $\sigma(A) = \text{RMSE}$.
The oracle term $\text{NLL}_{\text{oracle}}$ represents the ideal uncertainty estimate where $\sigma(A) = |\Delta y(A)|$.
Consequently, an RLL of 0\% indicates no information gain over a constant error assumption, while 100\% indicates perfect uncertainty quantification.
A variety of other metrics have been developed to analyse the quality of uncertainty estimation strategies\cite{tanSinglemodelUncertaintyQuantification2023, hoFlexibleUncertaintyCalibration2025, beckMultiheadCommitteesEnable2025}.
We refer to references~\cite{rasmussenUncertainUncertaintiesComparison2023, pernotLongRoadCalibrated2022a} for a more detailed discussion on comparing uncertainty estimators and calibration in atomistic modelling.
Throughout this work, we will use the RLL as the metric to compare uncertainty estimators and additional plots of predicted uncertainties $\sigma(A)$ against prediction errors $|\Delta y(A)|$ as an additional tool to visually inspect the quality of uncertainty estimators.

\subsection{Description of Datasets}

We evaluate our methods on five diverse benchmark datasets covering molecular liquids, solids, and flexible biomolecules.
First, the \textbf{[Bmim]BF$_4$ (BMIM)} ionic liquid dataset~\cite{zillsMachineLearningdrivenInvestigation2024a} contains 1320 training and 100 validation structures of 10 ion pairs.
Originally generated to compute diffusion coefficients, the training set was constructed via active learning across diverse temperatures and densities, while the test set was sampled from a classical force field trajectory at constant volume (500~K).
Reference labels were computed using the B97-3C functional~\cite{brandenburgB973cRevisedLowcost2018,beckeDensityfunctionalThermochemistrySystematic1997} with D3 dispersion corrections~\cite{grimmeConsistentAccurateInitio2010}.

For solid-state systems, we utilize the \textbf{Barium Titanate (\ce{BaTiO3})} dataset~\cite{gigl+22npjcm}, which consists of 1458 structures sampled from the four para- and ferroelectric phases.
These structures were originally generated to study phase transitions, with reference data computed at the PBEsol level of theory.
Additionally, we consider a \textbf{liquid water (\ce{H2O})} dataset~\cite{chengInitioThermodynamicsLiquid2019} containing 1593 structures generated to study the phase diagram and quantum nuclear effects.
Configurations were sampled from high-temperature quenches of classical molecular dynamics simulations and ambient-condition path integral simulations, with labels computed at the revPBE0-D3 level of theory.

To assess performance on flexible biological molecules, we use the \textbf{Ac-Ala3-NHMe (Ala4)} tetrapeptide dataset, a subset of the MD22 benchmark~\cite{chmielaAccurateGlobalMachine2023a} computed at the PBE+MBD level of theory.
Finally, as a challenging test for disordered environments with high energetic variability, we select a subset of the \textbf{Carbon} GAP-2017 dataset~\cite{deringerMachineLearningBased2017a}, consisting of all structures containing 64 atoms originally generated to study amorphous carbon.

\section{Results}

\subsection{Comparing Shallow Ensemble and LLPR Energy Uncertainty Estimates}

We begin by evaluating the capacity of last-layer methods to estimate energy uncertainty across the five benchmark datasets. 
For this purpose, we compare Shallow Ensemble trained with an NLL loss for energy and MSE loss for forces (SE$_{E}$), and the LLPR approach using only energy loss contributions to the Hessian (LLPR$_{E}$) with an underlying base model that was trained with MSE losses for both targets.
In both cases, the models are post-hoc calibrated using Equation~\ref{eq:calibration}.
To ensure statistical robustness, all training runs are repeated with three random seeds, and reported metrics represent the mean of these results.
In the case of BMIM, we note a distributional shift between the validation and test datasets.
The post-hoc calibration based on the validation dataset results in systematically underconfident uncertainty estimates for both uncertainty estimation approaches.
By reshuffling the dataset splits, the distributional shift can be avoided.
We trained SE$_E$ models on both splits and found the one trained on the reshuffled split to achieve significantly better RLLs.
The comparison of the two models is shown in SI Figure~\ref{fig:bmim_reshuffled}.
For all experiments in the main text, we use the reshuffled version of the dataset.
In the case of Ala4, we computed a learning curve of the energy RLL with respect to the training set size, which can be found in Section~\ref{sec:ala4_learning}.
Among the datasets considered here, it is the most flexible system, where accurate uncertainty quantification requires sampling a diverse range of torsional states that are less repetitive than the local environments found in bulk liquid or solid-state systems.
The learning curve is shown in Figure~\ref{fig:ala4_learning_curve}.
The resulting prediction accuracy and calibrated Energy RLLs are summarized in Table~\ref{tab:llpr_vs_se_energy}.

\begin{table}[t]
\centering
\caption{
    Comparison of Energy MAE [meV/atom], Force MAE [meV/Å], and Energy RLL [\%] for LLPR$_{E}$ and SE$_{E}$.
    }
\label{tab:llpr_vs_se_energy}
\begin{tabular}{lrrrrrr}
    \toprule
     & \multicolumn{2}{c}{E MAE} & \multicolumn{2}{c}{F MAE} & \multicolumn{2}{c}{E RLL} \\
    \cmidrule(lr){2-3} \cmidrule(lr){4-5} \cmidrule(lr){6-7}
     & LLPR$_{E}$ & SE$_{E}$ & LLPR$_{E}$ & SE$_{E}$ & LLPR$_{E}$ & SE$_{E}$ \\
    \midrule
    Ala$_4$       & 1.0 & 1.1 & 42.2 & 67.7 & -1.5 & 4.7 \\
    \ce{H2O}      & 1.2 & 1.1 & 44.9 & 46.0 & 22.5 & 49.8 \\
    BaTiO$_3$     & 0.2 & 0.2 & 6.6 & 13.9 & 41.5 & 73.8 \\
    BMIM          & 0.7 & 0.7 & 55.6 & 55.6 & -77.5 & 64.8 \\
    Carbon        & 19.6 & 23.8 & 435.5 & 422.6 & -16.4 & 31.7 \\
    \bottomrule
\end{tabular}
\end{table}

\begin{figure}
 \centering
 \includegraphics[width=\columnwidth]{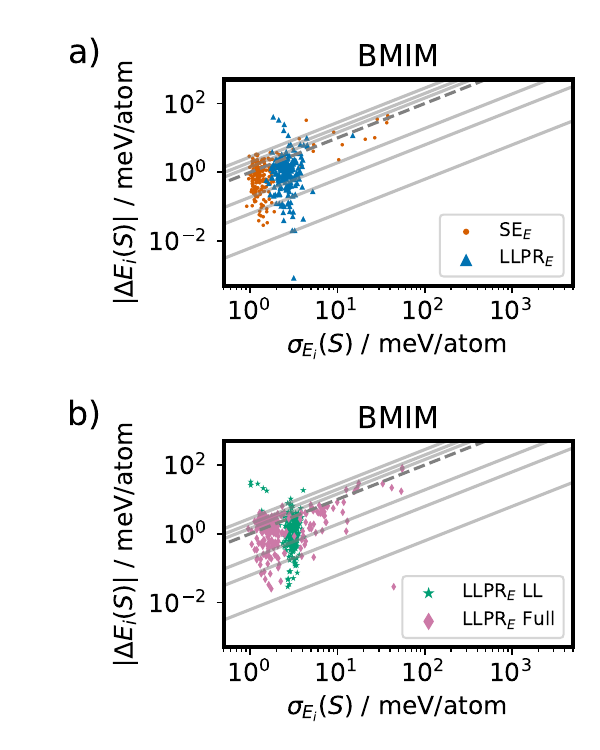} 
 \caption{
    Predicted-empirical error plots of various energy uncertainty estimation approaches for the BMIM dataset. Panel a) compares SE$_{E}$ and LLPR$_{E}$, Panel b) compares last-layer and full model fine tuning for ensembles sampled from the LLPR$_{E}$ posterior.
 }
 \label{fig:energy_cal}
\end{figure}

We find that the inclusion of the energy NLL leaves the quality of energy predictions largely unchanged compared to the MSE baseline (LLPR$_{E}$).
However, we observe a trade-off in force accuracy for Ala$_4$ and \ce{BaTiO3}.
Despite this trade-off, the calibration benefits of the Shallow Ensemble are substantial.
The Shallow Ensemble achieves consistently positive energy RLLs across all five benchmark datasets, indicating that it consistently provides uncertainty estimates that are more informative than a constant baseline.
In contrast, the LLPR$_{E}$ approach exhibits significant failure modes, yielding negative RLLs for Ala$_4$ , Amorphous Carbon, and most notably BMIM.
To understand the mechanisms behind these scores, we visualize the relationship between predicted uncertainty and empirical error in Figure~\ref{fig:energy_cal}a) for the BMIM dataset.

While the predicted uncertainties of SE$_E$ align well with the diagonal, LLPR$_{E}$ exhibits a vertical ``tail'' of points where the predicted uncertainty remains low even as the error spikes to $>$50~meV/atom.
These overconfident outliers correspond to structures in the very low-density regime that were originally constructed during active learning (see Figure~\ref{fig:bmim_reshuffled}b)).
The fact that LLPR$_{E}$ assigns these structures the same low uncertainty suggests a feature rigidity issue where the pre-trained MSE features have mapped these distinct outliers to the same region of latent space as the easier bulk structures, making them indistinguishable to the last-layer posterior.

We investigate whether these failures stem from the Laplace approximation itself or the underlying feature representation by comparing last-layer fine-tuning (LL FT), which updates only the readout weights while keeping the backbone frozen, and full-model fine-tuning (Full FT), which allows the internal feature representation to adapt by relaxing all model parameters.
Both fine-tuning strategies use the NLL loss.
Table~\ref{tab:e_ft} summarizes the energy RLLs obtained via these two fine-tuning protocols, and Figure~\ref{fig:energy_cal}b) portrays the empirical-predicted error plots.

\begin{table}[t]
\centering
\caption{Relative energy log likelihoods RLL [\%] of fine-tuned LLPR$_{E}$ models.}
\label{tab:e_ft}
\begin{tabular}{lrr}
    \toprule
    LLPR$_{E}$ &  LL FT & Full FT \\
    \midrule
    Ala$_4$        & -1.0   & 1.1  \\
    \ce{H2O}       & 28.6   & 34.9 \\
    BaTiO$_3$      & 14.8   & 5.4  \\
    BMIM           & -137.6 & 65.6 \\
    Carbon         & 3.0    & 9.0  \\
    \bottomrule
\end{tabular}
\end{table}

We observe that fine-tuning the LLPR$_{E}$ initialization with an energy NLL loss yields consistent improvements in calibration across all datasets except \ce{BaTiO3}.
For Carbon, a simple last-layer fine-tuning is sufficient to turn the negative RLL positive, suggesting that for some systems, the features are adequate and only the last-layer posterior needs adjustment.
However, the limitations of the fixed backbone are starkly visible in the BMIM dataset.
Here, Last-Layer Fine-Tuning fails to solve the issue and further degrades the RLL to -137.6.
It implies that the failure is not due to the Gaussian approximation of the posterior, but rather due to feature collapse in the backbone.
The linear head cannot find a set of weights that separates the outliers from the bulk because their feature representations are effectively identical.
Consequently, to achieve reliable calibration across all datasets, it is necessary to update the backbone via Full-Model Fine-Tuning.
As shown in Figure~\ref{fig:energy_cal}b), Full FT successfully eliminates the overconfident tail observed in the frozen model as it allows the backbone to move these outliers to a distinct region of the latent space.
By allowing the NLL loss to update the representation, the model learns to disentangle high-error configurations from the bulk, assigning them the higher variance they warrant. This restores the diagonal correlation and recovers an RLL of 65.6, effectively matching the quality of the more expensive scratch-trained Shallow Ensemble (64.8).

\subsection{Calibrated Force Uncertainties}

While energy uncertainty is critical for global stability, many downstream applications, such as active learning, rely on accurate force uncertainty estimates to detect local instability\cite{schranCommitteeNeuralNetwork2020}.
We therefore investigate whether models optimized or calibrated solely for energy uncertainty can provide reliable force uncertainty estimates.

We first evaluate the force calibration of SE$_E$ and LLPR$_E$.
As shown in the left column of Table~\ref{tab:independent_alpha}, the direct application of these energy-calibrated models to force uncertainty estimation yields poor results, with negative RLLs across almost all datasets.

\begin{table}[t]
\centering
\caption{Comparison of force RLLs [\%] for LLPR$_{E}$ and SE$_{E}$ across five benchmark datasets using energy-only and independent energy/ force post-hoc calibration.}
\label{tab:independent_alpha}
\begin{tabular}{lrrrr}
    \toprule
     & \multicolumn{2}{c}{Shared $\alpha_E$} & \multicolumn{2}{c}{Independent $\alpha_{E/F}$} \\
    \cmidrule(lr){2-3} \cmidrule(lr){4-5}
     & LLPR$_{E}$ & SE$_{E}$ & LLPR$_{E}$ & SE$_{E}$ \\
    \midrule
    Ala$_4$         & -11.9 & 9.7    & 10.0  & 11.2 \\
    \ce{H2O}        & -23.9 & 8.6    & 23.3  & 15.9 \\
    \ce{BaTiO3}     & -45.7 & -640.4 & 23.1  & 45.1 \\
    BMIM            & -45.8 & -597.6 & -40.8 & -41.6 \\
    Carbon          & -41.0 & -33.6  & -2.9  & -9.4 \\
    \bottomrule
\end{tabular}
\end{table}

To remedy this, we explore using independent calibration factors for energy and force uncertainties as a low-cost post-hoc fix.
By optimizing separate scalars $\alpha_E$ and $\alpha_F$ on the validation set, we account for the differing units and error scales.
As seen in the right column of Table~\ref{tab:independent_alpha}, this simple decoupling restores positive force RLLs for three of the datasets for both approaches.
Notably, for Water, this recovers the result from Ref.~\cite{kellnerUncertaintyQuantificationDirect2024a}, where SE$_{E}$ yielded calibrated forces.
However, for the challenging Carbon and BMIM datasets, independent calibration fails to produce reliable estimates, with RLLs remaining negative.

\begin{figure}
 \centering
 \includegraphics[width=\linewidth]{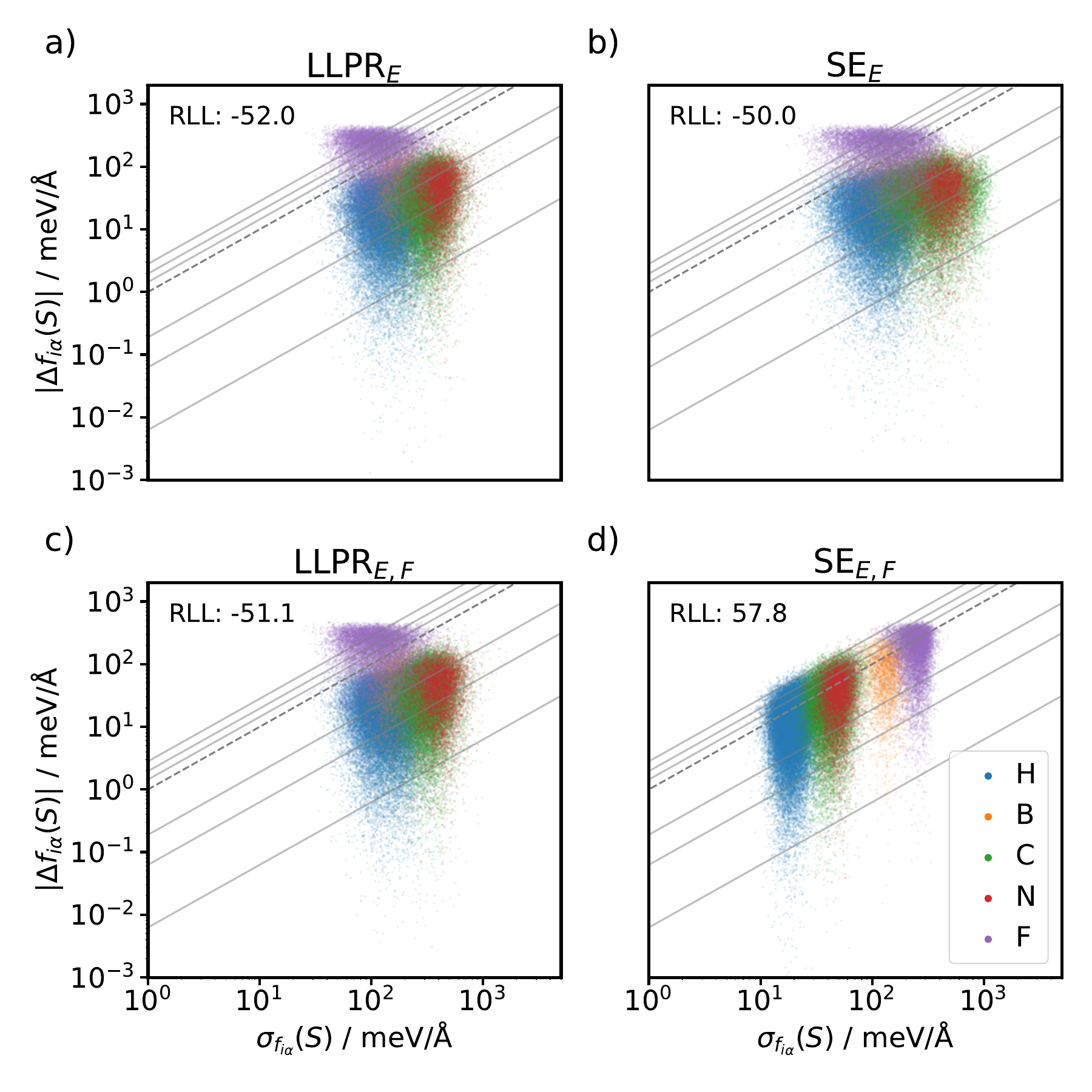} 
 \caption{
    Predicted-empirical error parity plots for force components in the reshuffled BMIM test set. 
    Points are colored by element type to visualize species-dependent calibration.
    Panels (a) and (b) show energy-only calibrated models (LLPR$_{E}$ and SE$_{E}$), which exhibit systematic miscalibration for Boron and Fluorine atoms (anion).
    Panels (c) and (d) show the corresponding force-informed models (LLPR$_{E,F}$ and SE$_{E,F}$).
 }
 \label{fig:bmim_failure}
\end{figure}

Closer inspection of the worst case, BMIM in Figure~\ref{fig:bmim_failure}a) and b) reveals that far from being an issue of global miscalibration, an element-dependent clustering of force uncertainty estimates arises.
The models are systematically overconfident for Boron and Fluorine atoms (the anion) while describing the organic cation's uncertainty well.
These findings align with recent work by Ho et al.~\cite{hoFlexibleUncertaintyCalibration2025}, who identified element-wise variations in the reliability of LLPR$_{E}$ force uncertainties, although the effect is even more pronounced here.
Hence, a single scalar cannot simultaneously correct these disjoint distributions.
To ensure that the results are not specific to a particular choice of model architecture, we evaluate both approaches using two other model architectures, So3krates and a NequIP-like equivariant message-passing model.
As shown in Figure~\ref{fig:bmim_arch}, the element-specific miscalibration of LLPR$_{E}$ is reproduced with two other architectures.

To address this issue, we replace the MSE loss function for the forces with an NLL loss for the shallow ensemble and add the force loss contribution to the Hessian for LLPR. We refer to these models as SE$_{E,F}$ and LLPR$_{E,F}$, respectively.
While SE$_{E}$ is practically unrestricted in the size of the ensemble, as the cost is just that of an additional linear layer, calculating all gradients of ensemble energies with respect to atomic positions \textit{via} backpropagation incurs a significant increase in training times.
Consequently, we performed a convergence test of the calibration metrics with respect to ensemble size and found an ensemble with 32 members to be sufficient for all datasets.
The associated empirical-predicted error plots are displayed in Section~\ref{fig:nens_convergence}.

As shown in Figure~\ref{fig:bmim_failure}c) and d), the improvements of LLPR$_{E,F}$ over LLPR$_{E}$ are minute, while the Shallow Ensemble trained with Force NLL successfully removes the element-specific miscalibration.
The energy and force RLLs of the two force-informed approaches are listed in Table~\ref{tab:alphas} as well as the calibration factors.
The corresponding prediction metrics are summarized in Table~\ref{tab:force_nll_metrics}.
Here, too, we find the inclusion of the NLL loss does not significantly affect the prediction metrics.
For the \ce{BaTiO3} dataset, we found LLPR$_{E,F}$ to be highly sensitive to initialization.
The method produced a valid uncertainty estimate in only one of three seeds; the other two seeds yielded degenerate solutions with RLLs of -37.9 and -104.7.
We report the successful seed to illustrate the potential performance on this dataset, but emphasize that robust application to \ce{BaTiO3} likely requires further stabilization techniques.

\begin{table}[htbp]
    \centering
    \small
    \setlength{\tabcolsep}{4pt}
    \caption{Vertical comparison of Energy and Force RLLs [\%] and calibration factors ($\alpha$) for LLPR$_{E,F}$ and SE$_{E,F}$.}
    \label{tab:alphas}
    \begin{tabular}{llrrrr}
        \toprule
        Dataset & Method & E RLL & F RLL & $\alpha_E$ & $\alpha_F$ \\
        \midrule
        \multirow{2}{*}{Ala$_4$} 
          & LLPR$_{E,F}$ & -1.8 & 10.8 & 1.6 & 1.1 \\
          & SE$_{E,F}$   &  2.0 & 36.8 & 0.9 & 1.0 \\ 
        \addlinespace
        \multirow{2}{*}{\ce{H2O}} 
          & LLPR$_{E,F}$ & 21.4 & 23.6 & 2.6 & 1.0 \\
          & SE$_{E,F}$   & 51.1 & 43.8 & 0.9 & 1.1 \\ 
        \addlinespace
        \multirow{2}{*}{\ce{BaTiO3}} 
          & LLPR$_{E,F}$ & 38.2 & 28.0 & 0.4 & 0.5 \\ %
          & SE$_{E,F}$   & 21.4 &   69.3 & 0.9 & 1.2 \\ 
        \addlinespace
        \multirow{2}{*}{BMIM} 
          & LLPR$_{E,F}$ & -73.9 & -39.8 & 0.9 & 1.0 \\
          & SE$_{E,F}$   &  74.1 &  59.6 & 0.6 & 1.0 \\ 
        \addlinespace
        \multirow{2}{*}{Carbon} 
          & LLPR$_{E,F}$ & -15.8 & -3.2 & 2.0 & 1.0 \\
          & SE$_{E,F}$   &  21.0 &  5.4 & 1.0 & 1.1 \\ 
        \bottomrule
    \end{tabular}
\end{table}

SE$_{E,F}$ achieves massive RLL improvements across the board and achieves positive RLLs for BMIM and Carbon, demonstrating the benefits of force-uncertainty aware training.
Further, the two calibration factors are much closer to 1.0 for SE$_{E,F}$, indicating that the models are well-calibrated even without further post-hoc scaling.
We attribute the failure modes of LLPR$_{E,F}$ to feature rigidity.
Because LLPR$_{E,F}$ is applied post hoc, it relies entirely on the quality of the pretrained features.
In contrast, SE$_{E,F}$ explicitly incorporates uncertainty during training, which encourages the backbone to learn representations that are more suitable for calibration.

\begin{figure}
 \centering
 \includegraphics[width=\columnwidth]{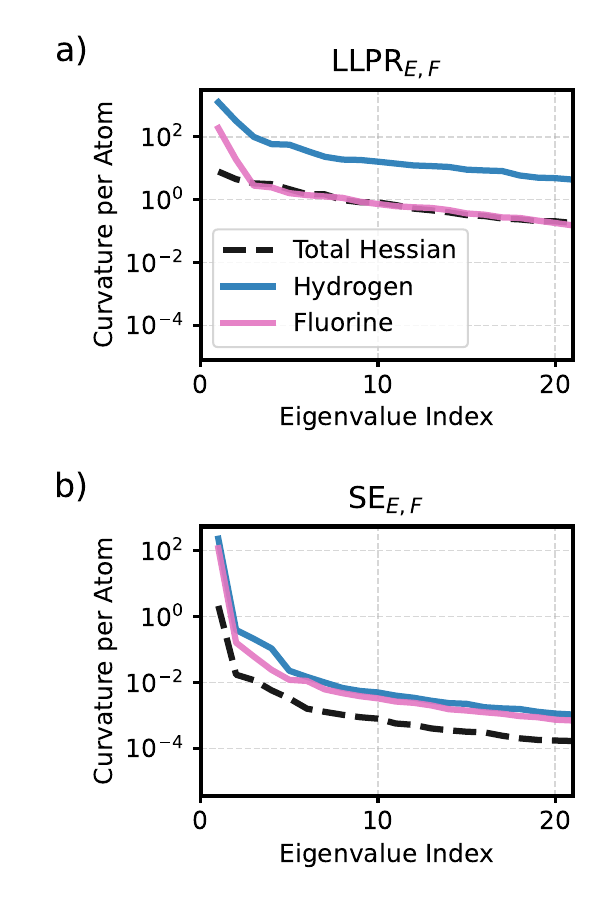} 
 \caption{
    Normalized eigenvalue spectra of the per-element force loss Hessian with respect to the last-layer weights for a) the MSE-trained model used for LLPR$_{E,F}$ and b) the NLL-trained SE$_{E,F}$.
 }
 \label{fig:hessian_mismatch}
\end{figure}

To investigate the origin of the element-specific miscalibration observed for LLPR$_{E,F}$ on the BMIM dataset, we analyzed the curvature of the loss landscape with respect to the last-layer parameters. Specifically, we computed the Hessian of the force loss in the Gauss-Newton approximation.
Figure~\ref{fig:hessian_mismatch} displays the eigenvalue spectra of the total Hessian (computed over all atoms) alongside the spectra of the partial Hessians computed exclusively for Carbon (cation) and Fluorine (anion) atoms.

For the MSE-trained model used in LLPR$_{E,F}$, we observe a significant spectral mismatch.
The eigenvalue spectrum of the Fluorine partial Hessians deviates considerably from the Full Hessian and the Carbon Hessian.
This indicates that the sensitivity of the model parameters with respect to fluorine forces is fundamentally different from the global average.
Since the Laplace approximation approximates the posterior using the inverse of the full Hessian, this mismatch results in a covariance matrix that is ill-suited for fluorine, leading to the systematic miscalibration observed in the empirical-predicted error plots.
We also assess the suitability of SE$_{E,F}$ features by applying LLPR$_{E,F}$ to an SE$_{E,F}$ model.
While this outperforms the MSE-trained baseline, it also shows the element-specific miscalibration.
Further, we consider the impact of omitting the energy contribution to the LLPR Hessian and confirm that the miscalibration still persists.
The resulting empirical-predicted error plot is shown in Figure~\ref{fig:bmim_cross}.
Consequently, the superior performance of end-to-end trained shallow ensembles stems from the joint flexibility of the backbone features and the last layer.

\section{Efficient Generation of Last-Layer Ensembles}

\begin{figure*}
  \centering
 \includegraphics[width=2.0\columnwidth]{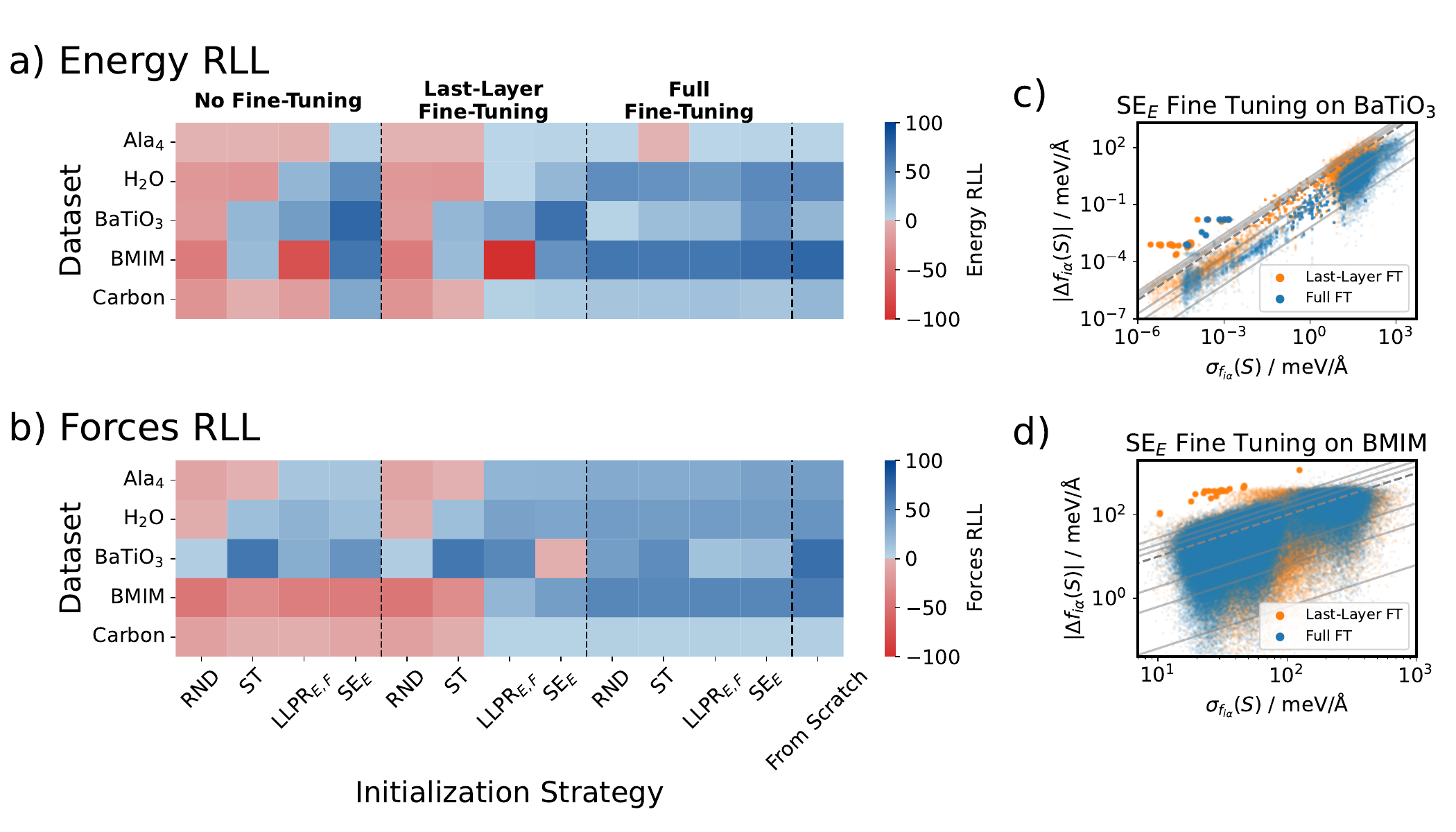} 
 \caption{
    Relative log likelihoods of different last-layer ensemble initialization and fine-tuning strategies compared to training shallow ensembles from scratch.
    a) Relative Energy log-likelihoods across 5 datasets.
    b) Relative Force log-likelihoods. Both are averaged over 3 random seeds. 
    Negative RLLs are only displayed down to -100 for easier interpretability.
    Panels c) and d) show the SE$_{E}$ fine-tuning results for  BMIM and \ce{BaTiO3} with highlighted outliers.
 }
 \label{fig:ensemble_ft}
\end{figure*}

While training shallow ensembles from scratch with force NLL loss yields well-calibrated uncertainty estimates, the computational cost is significant.
Since the training time scales linearly with the ensemble size $n_{\text{ens}}$ (provided the GPU is saturated), training such ensembles from scratch can be prohibitively expensive for large models.
We measured training times for single models and ensembles of sizes $n_\text{ens} \in \left( 4,8,10,16, 32, 64, 96\right)$ for the BMIM dataset.
The time per epoch is plotted against ensemble size in Fig.~\ref{fig:ens_scaling} for three model architectures.
We find that the timings increase by factors of around 3, 7, and 17 for GMNN, So3krates, and EquivMP, respectively, at $n_\text{ens} = 32$ compared to single models.
To address this, we investigate strategies to construct well-calibrated ensembles by initializing them on top of a pre-trained base model, followed by a low-cost fine-tuning phase.
All fine-tuning experiments in the remainder of this work are using the joint energy-forces NLL loss.

We consider the following ensemble construction strategies:
\textbf{Random isotropic initialization (RND)}  - Last-layer weights are sampled from an isotropic Gaussian centered at the MAP parameters.
In combination with post-hoc calibration, this method effectively calculates a prediction's uncertainty by the feature magnitudes and represents a natural baseline.
\textbf{Subsampling trained heads (ST)}  - Independent output heads are trained on overlapping subsets of the original training data while sharing the pretrained backbone.
This increases diversity while keeping training cheap.
This strategy has been proposed by Beck and coworkers, training independent multilayer perceptron heads of a universal MACE forcefield as a practical means for uncertainty quantification\cite{beckMultiheadCommitteesEnable2025}.
Unlike \citeauthor{beckMultiheadCommitteesEnable2025}, we only train the last-layer with subsampled training sets, not entire readout heads.
\textbf{LLPR posterior sampling (LLPR$_{E,F}$)} - Ensemble members are sampled from the Laplace posterior of a pretrained MSE model.
As shown in Section C, LLPR$_{E,F}$ alone can yield systematic miscalibration (e.g., BMIM), but fine-tuning the resulting ensemble significantly improves calibration at minimal cost.
\textbf{Hybrid energy–force objective (SE$_{E}$)} - Ensembles are trained with an NLL loss on energies and an MSE loss on forces.
Since force variances are not explicitly computed, this approach is almost identical in cost to training a single MSE model, yet the features are already suitable for energy uncertainty estimation.

Following initialization, we apply two fine-tuning protocols, last-layer fine-tuning and full model fine-tuning.
Figure~\ref{fig:ensemble_ft} presents the energy (a) and force (b) RLLs for these strategies across the five benchmark datasets.
The rightmost column (``From Scratch'') serves as the target scores that the fine-tuning approaches aim to achieve at a reduced cost.
The results are grouped by the fine-tuning method used, including the RLLs on initialisation for completeness.

Without fine-tuning, the baseline RND initialization performs poorly, yielding negative RLLs for both energies and forces across all datasets.
The ST initialization achieves positive RLLs for some datasets, but lags behind LLPR$_{E,F}$ and SE$_{E}$.
This poor performance compared to the findings by \citeauthor{beckMultiheadCommitteesEnable2025}\cite{beckMultiheadCommitteesEnable2025} likely stems from our restriction to the last-layer or their use of MACE, which may offer a better internal representation.
Last-layer fine-tuning based on NLL  leaves the results of these two approaches essentially unchanged.
The LLPR$_{E,F}$ results generally improve, but the energy RLLs remain negative for two datasets.

While SE$_{E,F}$ achieves the overall best results for last-layer fine-tuning, we observe a surprising quality degradation for the \ce{BaTiO3} force RLL.
For one random seed, the Force RLL drops to -70, despite the other seeds performing well at around $\sim 30$.
A closer inspection reveals that the degradation is caused by a few outliers in the low error regime $< 0.1$ meV/Å, which dominate the RLL calculation as shown in Figure~\ref{fig:ensemble_ft}c).
However, this instability is largely irrelevant in active learning applications, which focus on identifying high-error configurations. Furthermore, last-layer fine-tuning is already sufficient to remove the element-specific miscalibration observed for BMIM in the case of LLPR$_{E,F}$ and SE$_{E}$, as shown in Figure~\ref{fig:ensemble_ft}d).

Relaxing the backbone in full model fine-tuning serves as an equalizer for all initialization strategies.
By relaxing the backbone parameters, the models overcome inadequate representations present in the initialization strategies.
With the exception of ST energy RLLs for Ala$_4$, all approaches achieve positive energy and force RLLs across all datasets.
Most notably, the SE$_{E}$ variant effectively matches the performance of the expensive ``From Scratch`` SE$_{E,F}$ training, only deviating for \ce{BaTiO3}, where it achieves a slightly better energy and slightly worse force RLL.
Here, too, the cause is a few outliers in the low force error regime.

\begin{figure}
 \centering
 \includegraphics[width=\columnwidth]{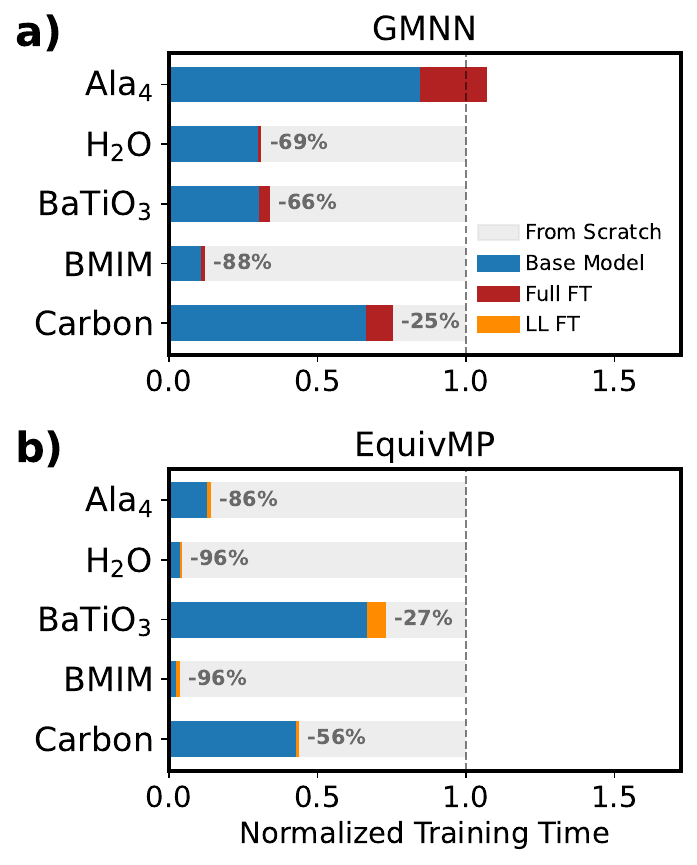} 
 \caption{
    Training time savings in percent for SE$_{E}$ training followed by fine-tuning compared to training with a force NLL loss from scratch across the 5 benchmark datasets.
    Panel a) shows the savings for GMNN with full-model fine tuning (red) and panel b) shows the savings for EquivMP with last-layer fine tuning (orange).
    The base model training time fractions are indicated in blue.
 }
 \label{fig:speedup}
\end{figure}

Compared to learning from scratch using a force NLL, all the fine-tuning strategies afford considerably reduced computational effort.
Since the pre-trained models are already near the MSE optimum, the fine-tuning phases converge rapidly.
Figure~\ref{fig:speedup} compares the total training times of the fine-tuning approaches for GMNN and EquivMP to shallow ensembles trained from scratch. For both model architectures, there is a rough dependence of the time saved on the size of the structures in the training dataset, although the trend is not monotonic due to different training convergence behaviors.
In the case of the slim GMNN model and the dataset with the smallest structures, Ala$_4$ (42 atoms), no time is to be saved with the fine-tuning approach.
However, for the dataset with the largest structures, BMIM (300 atoms), the training time savings are huge at almost 90~\%.
As the EquivMP model is significantly more computationally intensive, we observe a training time savings across the board at over 80~\% for three of the datasets.
In both cases, a fine-tuning approach offers an affordable and robust pathway to well-calibrated uncertainty estimates.

Based on these results, we make two recommendations for practical purposes.
In cases where an existing MSE-trained model is available (e.g., a foundation model), we recommend initializing ensembles via LLPR$_{E,F}$ followed by full model fine-tuning.
If no prior model is available, training a shallow ensemble with the hybrid loss (energy NLL + force MSE) followed by full model fine-tuning provides an even more accurate alternative.

\section{Conclusions and Outlook}

In this work, we have systematically investigated last-layer approaches to uncertainty quantification in MLIPs, focusing on shallow ensembles and LLPR-based posterior approximations on a wide range of datasets.

We first compared the two approaches in terms of their energy uncertainty estimation.
There, we found the shallow ensemble to yield reliable uncertainty estimates across all experiments, while LLPR$_{E}$ estimates exhibit some failure modes.
The rigidity of the features used in LLPR$_{E}$ resulted in some highly miscalibrated outliers for the BMIM dataset.
We demonstrated that this failure results from a combination of the last-layer Laplace approximation itself and the features used therein.
While computing LLPR on a more robust SE$_{E,F}$ backbone improves results, and last-layer fine-tuning helps on some datasets, neither approach fully resolves the issues on its own.

Secondly, we established that a probabilistic treatment of forces during training is necessary for reliable uncertainty quantification.
We found that the energy-centric approaches yield miscalibration for most of the datasets.
While independently post-hoc calibrating energy and force uncertainty estimates improves the results somewhat, pathological cases where the shape of the uncertainty distribution needs to adapt are not resolved.
Spectral analysis of the loss Hessian revealed the cause to be a mismatch between the overall Hessian and the element-specific loss curvature.
Explicitly incorporating the force NLL loss was the only variant of the approaches considered here that ensured robust calibration across chemical species.

Finally, we identified efficient strategies for generating last-layer ensembles with calibrated force uncertainty estimates without the cost of training from scratch.
We explored four initialisation schemes, including an isotropic random last-layer initialization, subsampling trained output heads, LLPR$_{E,F}$ posterior sampling, and an energy-only shallow ensemble, followed by fine-tuning with the joint energy-force NLL objective.
Our results show that random isotropic initialization and subsampling trained output heads do not result in well-calibrated uncertainty estimates out of the box or after last-layer fine-tuning.
LLPR$_{E,F}$ and energy-only shallow ensembles are greatly improved by last-layer fine-tuning, but there are some cases where they are not reliable, either.
We found full model fine-tuning to give the highest quality approximation to the quality of uncertainty estimates of a shallow ensemble trained from scratch, with the energy-only shallow ensemble achieving overall the best-calibrated results.
These initialization strategies can reduce the training times of force uncertainty-aware shallow ensembles by up to 96~\% on the datasets considered here, depending on the size of the structures in the dataset.
These results could be extended by the use of advanced post-hoc calibration techniques, such as the one introduced by \citeauthor{hoFlexibleUncertaintyCalibration2025}~\cite{hoFlexibleUncertaintyCalibration2025}, to further improve the reliability of uncertainty estimates for atomistic machine learning.
This direction will be explored in future work.

Taken together, these findings establish practical guidelines for training last-layer ensembles for MLIPs from a methodological side.
We have provided easily accessible implementations for the \texttt{apax}/\texttt{IPSuite} software stack and provide a Jax-based LLPR implementation with \texttt{Laplax}.
When training resources are abundant, shallow ensembles with probabilistic losses remain the gold standard.
However, when training time is a concern, we recommend using an energy-only shallow ensemble for training new models, followed by full model fine-tuning as the most reliable last-layer uncertainty estimation strategy.
If expensive pre-trained models exist, using LLPR initialized ensembles and full-model fine-tuning is a viable alternative.
Overall, our analysis confirms the practical viability of last-layer methods to achieve scalable uncertainty quantification in atomistic machine learning.

\section{Data Availability}

The workflow script, as well as all training input files and analysis scripts needed to reproduce the work presented here, can be found at \url{https://github.com/M-R-Schaefer/last_layer_ensemble_data}.
All data generated during the iterative training and production simulations are stored on an S3-object storage.
It can be obtained by cloning the repository and executing \texttt{dvc pull} in the repository folder.
\texttt{apax} is an open source package and can be installed \textit{via} \texttt{pip install apax}.
The implementation of LLPR is available on GitHub under \url{https://github.com/M-R-Schaefer/laplax}.

\begin{acknowledgments}
We thank Marcel Langer for the provision of a development version of So3krates. The authors would like to thank Filippo Bigi and Sanggyu Chong for insightful discussions and early tests of LLPR uncertainty estimates in the BMIM dataset.  M.K. and M.C. acknowledge financial support by the Swiss National Science Foundation (Project 200020\_214879). J.K., and M.S. acknowledge support by the Deutsche Forschungsgemeinschaft (DFG, German Research Foundation) in the framework of the priority program SPP 2363, “Utilization and Development of Machine Learning for Molecular Applications - Molecular Machine Learning” Project No. 497249646.
Further funding was provided by Deutsche Forschungsgemeinschaft (DFG, German Research Foundation) under Germany's Excellence Strategy - EXC 2075 – 390740016. We acknowledge the support by the Stuttgart Center for Simulation Science (SimTech).
\end{acknowledgments}

\clearpage

\onecolumngrid
\part*{Supplementary Information}

\setcounter{secnumdepth}{3}           %
\renewcommand{\thesection}{S\arabic{section}}
\setcounter{section}{0}

\renewcommand{\thefigure}{S\arabic{figure}}
\setcounter{figure}{0}

\renewcommand{\thetable}{S\arabic{table}}
\setcounter{table}{0}

\section{Model Architectures}\label{sec:architectures}
In this work, we use three different model architectures in total, each differing in the way local atomic environments are represented and implemented in or interfaced with \texttt{apax}.
Consequently, certain aspects of the implementations are shared between the three architectures.
Here, we discuss the common aspects of these models before explaining the differences between their internal representations.

In each case, the pairwise distances between each atom and its local neighbours are expanded in a radial basis Psi, using Gaussian or Bessel functions.
From this neighborhood expansion, the three architectures differ in how they proceed to further build up the representation of the local atomic environment $G$.
Once obtained, the representation passes through a readout neural network.
$E_i = \mathrm{NN}(\boldsymbol{G}_i)$ and adjusted by element-specific scaling and shifting parameters, $\sigma_{Z_i}$ and $\mu_{Z_i}$.
\begin{align}\label{eq:scale_shift}
    E_i = \sigma_{Z_i} \cdot \text{NN}(\boldsymbol{G}_i) + \mu_{Z_i}
\end{align}
Finally, the atomic energies $E_i$ are summed up as in Equation~\ref{eq:energy_sum}.

In the GMNN architecture, element-pair-specific parameters $\beta$ are used to form linear combinations of the radial basis functions, and angular information is captured by Cartesian moments, i.e., polynomials of the unit distance vectors up to some rotation order $L$.

\begin{align}\label{eq:basis_fn}
    \Psi_{i, L, n} = \sum_{j \neq i} R_{Z_i, Z_j, n'}(r_{ij}, \beta_{Z_i, Z_j, n', n}) \boldsymbol{\hat{r}}_{ij}^{\otimes L}
\end{align}

The invariant descriptor $\boldsymbol{G}$ is obtained from fully contracting the equivariant features $\Psi_{i, L, n}$ according to
\begin{align}\label{eq:contraction}
    G_{i, n_1, n_2} &= (\Psi_{i, 1, n_1})_a (\Psi_{i, 1, n_2})_a  \nonumber \\ 
    &\;\; \vdots \\
    G_{i, n_1, n_2, n_3} &= (\Psi_{i, 1, n_1})_a (\Psi_{i, 3, n_2})_{a,b,c} (\Psi_{i, 2, n_3})_{b,c}. \nonumber
\end{align}

The EquivMP architecture is an equivariant message passing architecture similar to NequIP.
 Based on per-atom features $h_i$, equivariant messages are constructed using Clebsch-Gordan products with spherical harmonics.

\begin{align}
m_{ij}^{(l_1)} = \sum_{l_2, l_3} C^{l_1}_{l_2 l_3} h_j^{(l_2)} \otimes Y^{l_3}(\hat{r}_{ij})
\end{align}

Atom $i$ aggregates the messages from all its neighbors and uses the result to update its own feature vector.
Updates for the feature vector of atom $i$ are computed as 

\begin{align}
h_i' = h_i + \text{UpdateFunc}\left( \sum_{j \neq i} \phi(r_{ij}) \cdot m_{ij} \right)
\end{align}

The So3krates architecture builds on previous equivariant messages but avoids the computational complexity of full Clebsch-Gordan products.
By separating messages into invariant and equivariant parts, the update function for the equivariant features is greatly simplified.
Instead, the learning of complex interatomic interactions is performed by the Euclidean attention mechanism, whereby the $l=0$ output of CG products is used to compute weights for the messages.

\section{Details of the LLPR Implementation}\label{sec:llpr_detail}

The posterior covariance $\Sigma_{\text{LL}}$ is computed from the inverse Hessian of the loss.
The overall Hessian $\mathbf{H}$ is constructed either from only the Hessian of the energy loss, $\mathbf{H}_E$, or, if the force loss is also considered, additively from the contributions of the Hessians of the energy and force loss:

\begin{equation}
    \mathbf{H} = \mathbf{H}_E + \mathbf{H}_F
\end{equation}

The individual contributions are calculated from a generalized Gauss-Newton approximation using the last layer features.
The energy contribution is derived from the network sensitivity with respect to the energy, $g_E$, which for the last-layer corresponds to the aggregated feature vector $h(A)$:

\begin{equation}
    \mathbf{H}_E = \sum_{A \in \mathcal{D}} \mathbf{g}_E(A) \mathbf{g}_E(A)^\top ,\quad \mathbf{g}_E(A)= \boldsymbol{h}(A) 
\end{equation}

The force contribution $\mathbf{H}_F$ consists of the Jacobian of the forces with respect to the weights.
Since forces are gradients of the energy, these sensitivities are the feature-coordinate Jacobians:

\begin{equation}
    \mathbf{H}_F =
    \sum_{A \in \mathcal{D}} \sum_{i=1}^{N_A} \sum_{\alpha \in {x,y,z}}
    \mathbf{g}_{F_{i,\alpha}}(A) \mathbf{g}_{F_{i,\alpha}}(A)^\top ,\quad
    \mathbf{g}_{F_{i,\alpha}}(A) = \frac{\partial \boldsymbol{F}_{i,\alpha}}{\partial \boldsymbol{w_{\text{LL}}}}
\end{equation}

A small regularization term $\eta^2\mathbf{I}$ is added to the thus approximated Hessian to ensure the matrix is positive definite.

\begin{equation}
    \Sigma_{\text{LL}} = \alpha^2(\mathbf{H} + \eta^2\mathbf{I})^{-1}
\end{equation}

Where $\eta$ is iteratively increased until positive definiteness is achieved.
In cases where the assumptions of the GGN approximation fail, for example, if early stopping terminates training when the parameters are not at a minimum of the training loss surface, $H$ may be singular.
In these cases, $\eta$ may increase to such an extent that the Hessian and thus the covariance become isotropic.
After post-hoc calibration, LLPR will give identical uncertainty estimates as RND.

A last-layer ensemble can be initialized from the LLPR posterior by sampling weights from the multivariate Gaussian distribution over the last-layer parameters.

\begin{equation}
    \boldsymbol{\tilde{w}}^{(k)} \sim \mathcal{N}(\boldsymbol{w}_{\text{MAP}}, \boldsymbol{\Sigma}_{LL})    
\end{equation}

The mean of this distribution, the maximum a posteriori parameters theta, corresponds to the last-layer parameters from a model trained with an MSE loss $\boldsymbol{w}_{\text{MAP}}$.
Due to the finite number of samples, the empirical mean of the raw samples will generally not equal $\boldsymbol{w}_{\text{MAP}}$.
To correct for this finite sampling error, we explicitly subtract the empirical mean from each sample and shift by $\mathbf{\theta}_{\text{MAP}}$.

\begin{equation}
    \boldsymbol{w}^{(k)} = \boldsymbol{\tilde{w}}^{(k)} - \frac{1}{n_{\text{ens}}} \sum_{k^\prime} \boldsymbol{\tilde{w}}^{(k^{\prime})} + \boldsymbol{w}_{\text{MAP}}
\end{equation}

\section{Training Details}\label{si:hypers}

GMNN models used a radial neighborhood cutoff of 5 Å.
The neighborhood density was expanded in 7 Bessel functions and contracted to 5 radial functions via Equation~\ref{eq:basis_fn}.
A readout neural network with two hidden layers using 128 and 64 units was used.
The Adam optimizer was used with learning rates of 0.0075 for the neural network layers and 0.002 for all other parameters.
During training, the learning rate was adjusted with a cyclic cosine learning rate scheduler using a period of 50 epochs and a decay factor of 0.93.

The hyperparameters for the So3krates and EquivMP models were obtained from a hyperparameter search.
For the So3krates model, we use a radial distance cutoff of 5.0 Å and 8 Bessel functions for the neighborhood expansion.
For the internal feature representation, 96 units are used, and the maximum angular degree is set to 3.
After 2 message passing steps, predictions are made by a readout neural network using 128 and 32 hidden units.
For the EquivMP model, 24 Gaussian basis functions were used with a cutoff of 5.0 Å.
The internal representation used 24 features and $l_{\text{max}}=2$.
After 2 message passing layers, a readout neural network consisting of hidden layers with 64 and 64 units. 
All models were trained for a maximum of 4000 epochs with an early stopping patience of 250 epochs.
The only exception is GMNN models trained on the Ala$_4$ dataset, where a maximum of 1000 epochs was used in combination with a linear learning rate schedule.
For the fine-tuning runs, the maximum number of epochs was reduced to 200.
A batch size of 1 was used for all experiments.

The loss functions used throughout this work are the mean squared error (MSE) and the negative log-likelihood (NLL).
Here, we state the full version of the combined energy-force losses.
The hybrid energy-force loss function consists of the energy term of the NLL loss and the force term of the MSE loss.

\begin{equation}
    \text{MSE}_{\text{combined}} = \sum_{A \in \mathcal{D}} \left( E_A - \hat{E}_A \right)^2 + \sum_{A \in \mathcal{D}} \sum_{i=1}^{N_A} \sum_{\alpha \in \{x,y,z\}} \left( F_{i,\alpha} - \hat{F}_{i,\alpha} \right)^2
\end{equation}

\begin{equation}
    \text{NLL}_{\text{combined}} = \sum_{A \in \mathcal{D}} \left[ 
    \frac{1}{2} \left( \frac{(E_A - \hat{E}_A)^2}{\sigma_{E,A}^2} + \ln(2\pi\sigma_{E,A}^2) \right) 
    + 
    \frac{1}{2} \sum_{i=1}^{N_A} \sum_{\alpha \in \{x,y,z\}} \left( \frac{(F_{i,\alpha} - \hat{F}_{i,\alpha})^2}{\sigma_{F,i,\alpha}^2} + \ln(2\pi\sigma_{F,i,\alpha}^2) \right) 
    \right]
\end{equation}

\newpage
\pagebreak

\section{Alanine Tetrapeptide Learning Curve}\label{sec:ala4_learning}

\begin{figure}[bht]
 \centering
 \includegraphics[width=8cm]{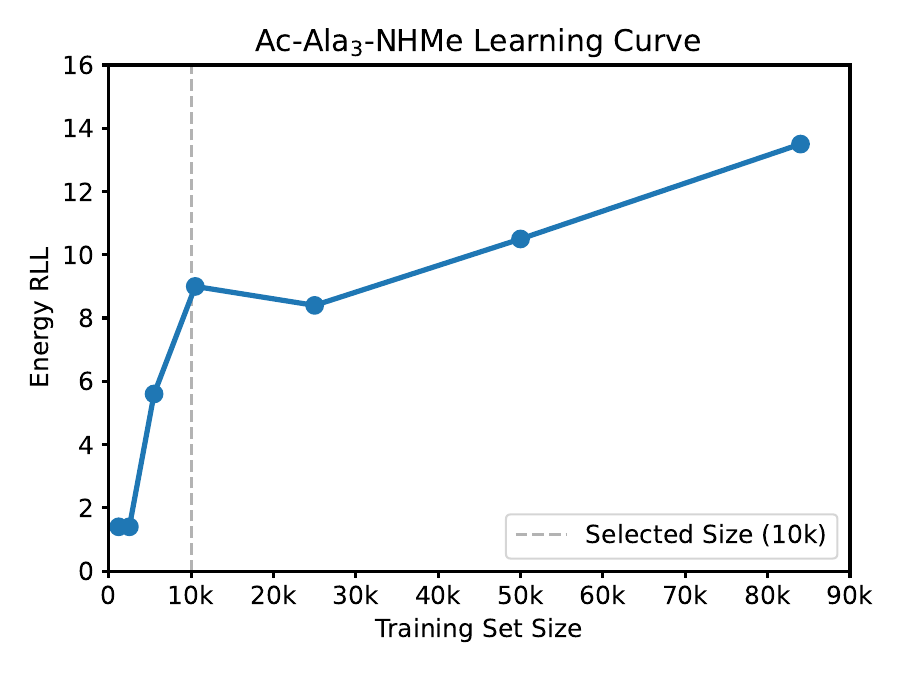} 
 \caption{
    Learning curve of the Energy RLL for the Ala$_4$ dataset using an SE$_E$ model.
    The vertical dashed line marks the training set size of 10,000 structures chosen for the experiments in the main text, representing the point where calibration performance stabilizes.
 }
 \label{fig:ala4_learning_curve}
\end{figure}

Unlike the condensed-phase systems that we investigated, Ala$_4$ features a diverse conformational space requiring sufficient sampling for calibration.
We computed the Energy RLL learning curve in Figure~\ref{fig:ala4_learning_curve} to determine the necessary dataset size.
While RLL continues to improve marginally up to the full 84,000 structures, we observe a stabilization of performance around 10,000 samples.
Consequently, we selected a subset of 10,000 structures for all Ala$_4$ experiments in the main text, striking a good trade-off between reliable uncertainty quantification and computational efficiency.

\section{Further Analysis of [Bmim]BF$_4$ Energy Uncertainties}

Using the original dataset splits for the BMIM dataset resulted in difficulties for the uncertainty quantification methods considered here to achieve positive energy RLLs.
We identified the cause of this to be a distribution shift between the original training and test datasets.
The empirical-predicted uncertainty plot for the original and reshuffled BMIM test sets using an SE$_{E}$ model.

\begin{figure}[h!]
 \centering
 \includegraphics[width=16cm]{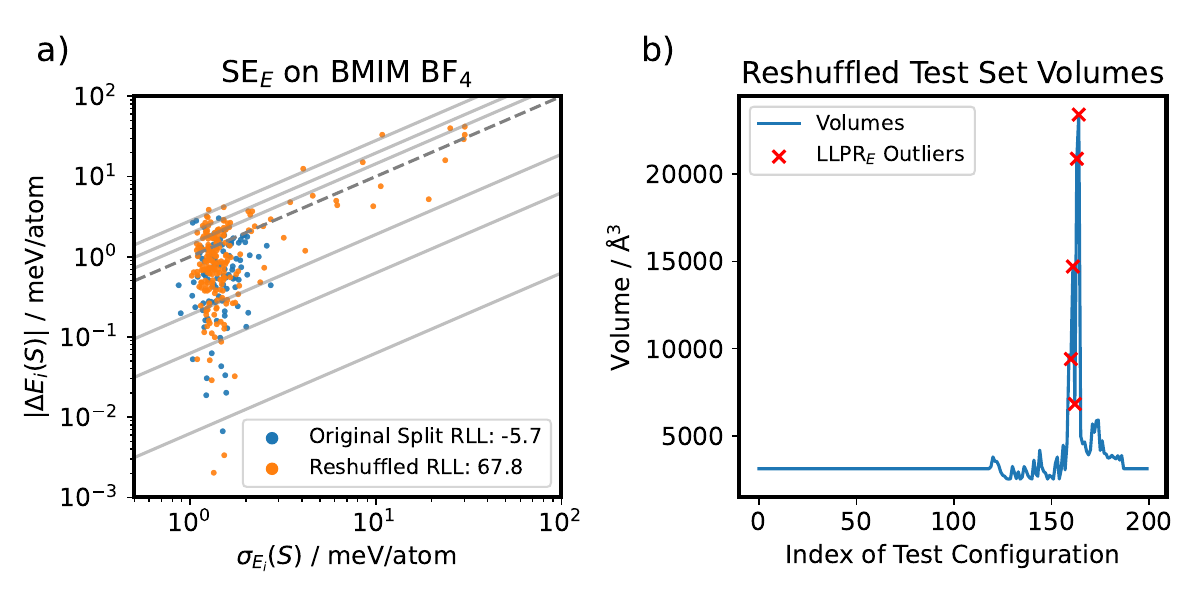} 
 \caption{
    a) Test set empirical-predicted energy error plot of an SE$_{E}$ GMNN model for the original train/validation/test split and the reshuffled versions of the BMIM dataset.
    b) Cell volume of the structures in the reshuffled BMIM test set, with those structures yielding the worst-calibrated energy uncertainty estimates by LLPR$_{E}$ highlighted in red.
 }
 \label{fig:bmim_reshuffled}
\end{figure}

In the main text, we observe that LLPR$_{E}$ struggled to calibrate energy uncertainties when the data points are not well represented in the training data.
Figure~\ref{fig:bmim_reshuffled}b) illustrates that the origin of this failure consists entirely of structures with anomalously large simulation cell volumes, corresponding to dissociated ion pairs or gas-phase-like regions.

\section{Convergence tests}
\begin{figure}[h!]
 \centering
 \includegraphics[width=0.9\linewidth]{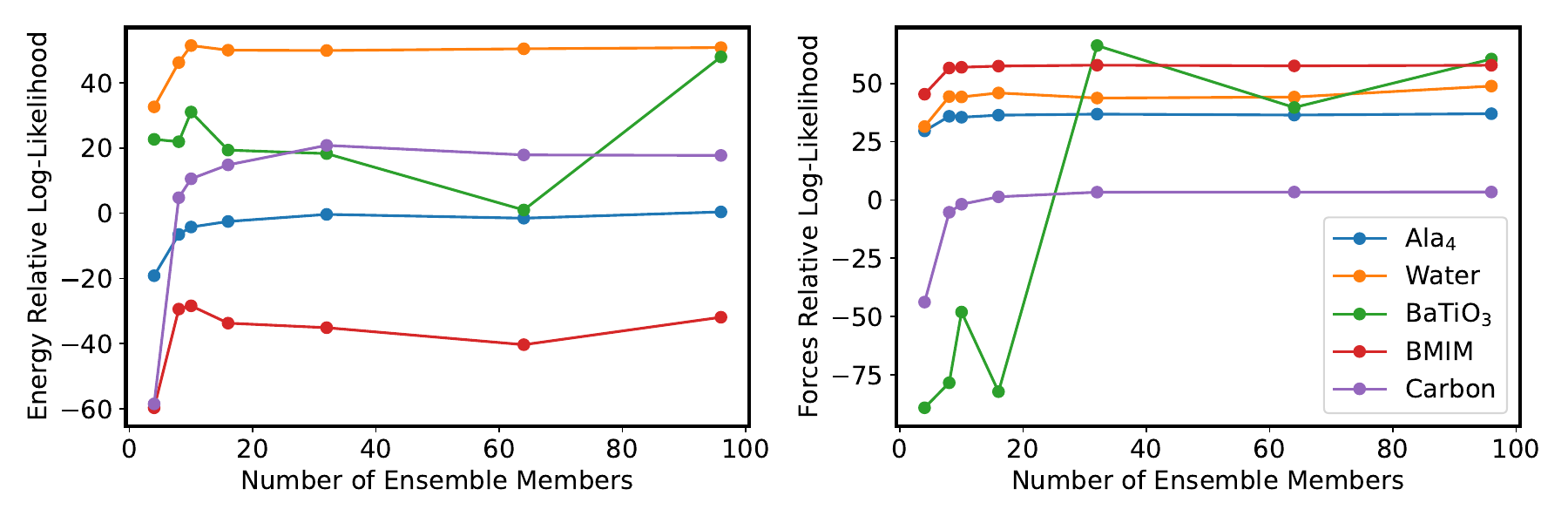} 
 \caption{
    Test set negative log-likelihood and relative log-likelihood for GMNN SE$_{E,F}$ models trained with an increasing number of shallow ensemble members for all five datasets. Note that the convergence curves were recorded without post-hoc calibration and reshuffling of the BMIM dataset and serve purely as an indicator of the convergence of SE model uncertainty estimates with respect to the number of ensemble members.
 }
 \label{fig:nens_convergence}
\end{figure}

We assess the convergence of shallow ensemble uncertainty estimates with respect to the number of ensemble members.
For energy NLL training, the number of ensemble members is practically unrestricted.
It is thus possible to choose a large number and be reasonably confident that the resulting uncertainty metrics are converged with respect to ensemble size.
Including the force NLL term restricts the ensemble size somewhat and considerably slows down training.
As a result, one should choose the smallest ensemble that converges metrics like NLL or RLL.
We train GMNN models with $N_{ens} \in \left( 4,8,10,16, 32, 64\right)$ on the water and BMIM datasets.
The resulting NLLs and RLLs averaged over three random seeds corresponding to each ensemble size are displayed in Figure~\ref{fig:nens_convergence}.

We find that for all datasets, $N_{ens}=32$ is sufficient for obtaining converged uncertainty quantification metrics when averaged over three random seeds.
However, different datasets show different variance in the training results.
BaTiO3 in particular can yield miscalibrated results occasionally, even at 64 ensemble members.

\section{Extended Analysis of the NLL Force Loss}

\begin{figure}[bt]
 \centering
 \includegraphics[width=\linewidth]{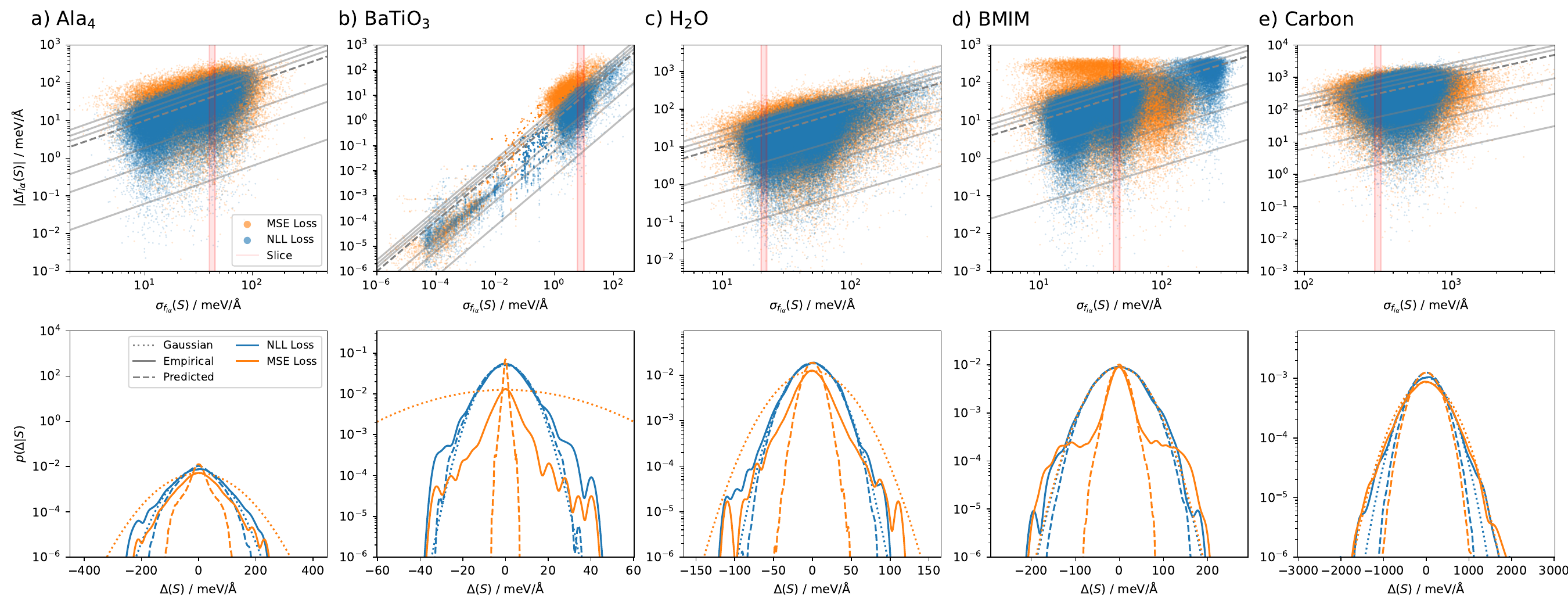} 
 \caption{
    Predicted-empirical error plots and distribution along selected uncertainty slices for SE$_E$ (orange scatter plots) and SE$_{E,F}$ (blue scatter plots) models trained on the a) Ala$_4$, b) BaTiO$_3$,  c) H$_2$O, d) BMIM, and e) Carbon datasets.
 }
 \label{fig:predicted_empirical}
\end{figure}

In the main text, we analysed the effect of combining an NLL energy loss with an NLL force loss instead of an MSE force loss and showed the empirical-predicted error plots for the case of the BMIM dataset.
To provide a more complete picture of the behavior of the NLL force loss, we provide the prediction metrics of the MSE and NLL trained models in Table~\ref{tab:force_nll_metrics} and the corresponding empirical-predicted error plots for the remaining four datasets in Figure~\ref{fig:predicted_empirical} (left panels) and the distribution of empirical and predicted errors for selected slices of uncertainties (right panels).

\begin{table}[bt]
\centering
\caption{Test set energy [meV/atom] and force MAEs [meV/Å] for LLPR$_{E,F}$ and SE$_{E,F}$ on the five benchmark datasets.}
\label{tab:force_nll_metrics}
\begin{tabular}{lrrrr}
    \toprule
     & \multicolumn{2}{c}{E MAE} & \multicolumn{2}{c}{F MAE} \\
    \cmidrule(lr){2-3} \cmidrule(lr){4-5}
     & LLPR$_{E,F}$ & SE$_{E,F}$ & LLPR$_{E,F}$ & SE$_{E,F}$ \\
    \midrule
    Ala$_4$     & 1.0  & 0.9  & 42.2  & 40.2  \\
    \ce{H2O}    & 1.2  & 2.0  & 44.9  & 41.1  \\
    \ce{BaTiO3} & 0.2  & 0.5  & 6.6   & 7.4   \\
    BMIM        & 0.7  & 1.1  & 55.6  & 50.0  \\
    Carbon      & 19.6 & 25.9 & 435.5 & 421.9 \\
    \bottomrule
\end{tabular}
\end{table}

We find that the results are consistent with those for BMIM shown in the main text.
Across all datasets, models trained with an NLL force loss exhibit a sharper distribution of predicted uncertainty and overall better calibration.
The only difference with regard to the case of water is that the distribution along the uncertainty slices does not necessarily conform to a Gaussian.

\section{Extended Analysis of the [Bmim]BF$_4$ Miscalibration}\label{sec:llpr_bmim}

\begin{figure}[b]
 \centering
 \includegraphics[width=0.5\columnwidth]{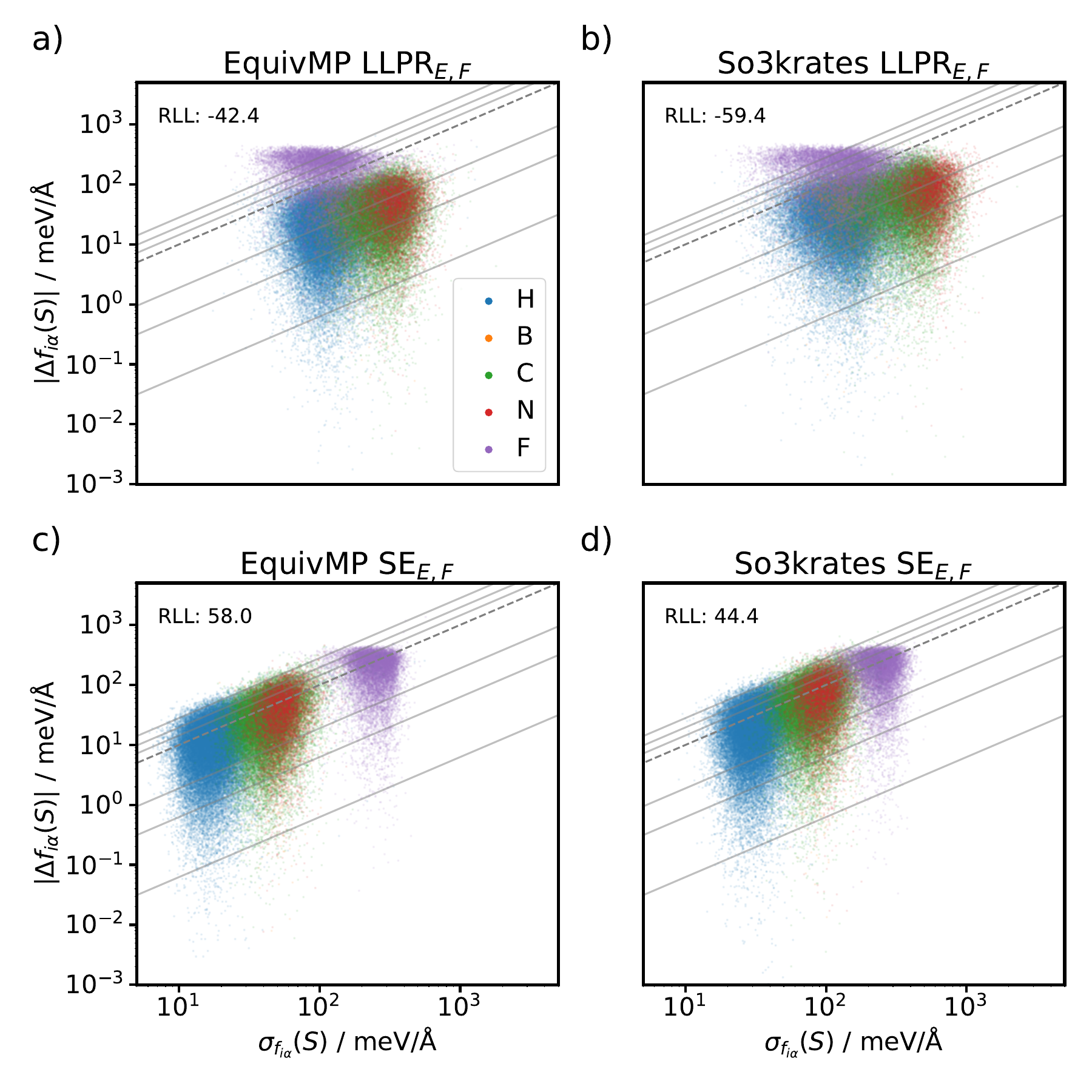} 
 \caption{
    Comparison of force uncertainty calibration across different model architectures on the BMIM dataset.
    The predicted-empirical plots are shown for EquivMP (left panels) and So3krates (right panels) with LLPR results shown on the left and shallow ensemble results on the right.
    The force components are colored by the type of element they belong to.
 }
 \label{fig:bmim_arch}
\end{figure}

To confirm that the systematic miscalibration of fluorine and boron forces in [Bmim]BF$_4$ is not specific to the GMNN architecture, we extended our analysis to the EquivMP and So3krates models.
Figure~\ref{fig:bmim_arch} shows the predicted-empirical force error plots for the BMIM dataset for LLPR$_{E,F}$ and SE$_{E,F}$ using the two additional architectures. We find that the LLPR$_{E,F}$ results for BMIM are consistent across all three architectures.
This consistency across different architectures demonstrates that the observed failure mode is not a limitation of model expressivity.
Rather, it confirms that the limitation is fundamental to the application of using the averaged covariance matrix for elements with significantly different individual force loss Hessians.
Training an EquivMP model as a shallow ensemble reproduces the well-calibrated results of GMNN in the main text.

\begin{figure}[b]
 \centering
 \includegraphics[width=0.7\columnwidth]{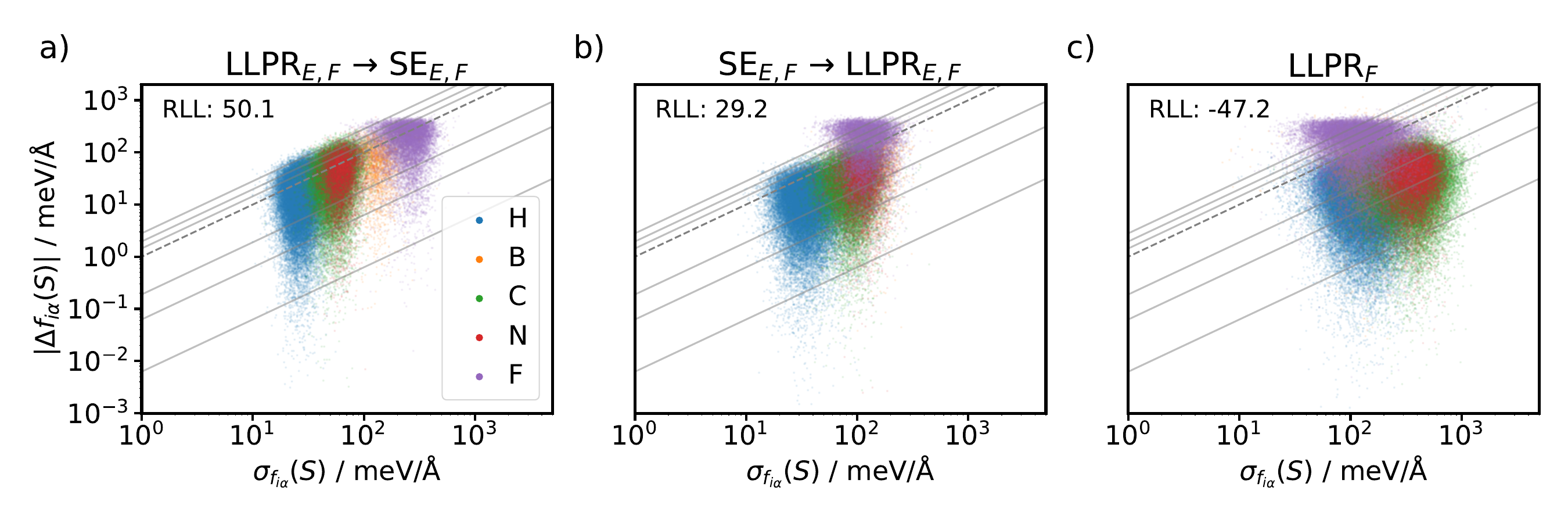} 
 \caption{
    LLPR BMIM force uncertainty ablation tests with a) an LLPR$_{E,F}$ model fine tuned with a combined NLL loss, b) LLPR$_{E,F}$ applied to an SE$_{E,F}$, and c) LLPR$_F$.
    The force components are colored by the type of element they belong to.
 }
 \label{fig:bmim_cross}
\end{figure}

To further isolate the role of feature representation, we performed a cross-validation experiment.
Fine-tuning an LLPR-initialized ensemble (LLPR$_{E,F}$ $\xrightarrow{}$ SE$_{E,F}$) eliminates the anion miscalibration (restoring RLL 50.1).
In contrast, applying the LLPR approximation to an SE-trained backbone (SE$_{E,F}$ $\xrightarrow{}$ LLPR$_{E,F}$) retains a positive RLL of 29.2, demonstrating that SE features are robust to the Gaussian approximation; the element-specific miscalibration is reintroduced.
This confirms that the failure of LLPR$_{E,F}$ arises from feature rigidity, which can be resolved by allowing the representation to adapt via probabilistic training.
Finally, we confirm that element-wise BMIM miscalibration persists even when using only the force loss contribution to the Hessian by constructing an LLPR$_F$ model.
The resulting empirical-predicted error plots are displayed in Figure~\ref{fig:bmim_cross}.

\clearpage

\section{Computational Scaling of Shallow Ensembles}

\begin{figure}[h]
 \centering
 \includegraphics[width=0.75\linewidth]{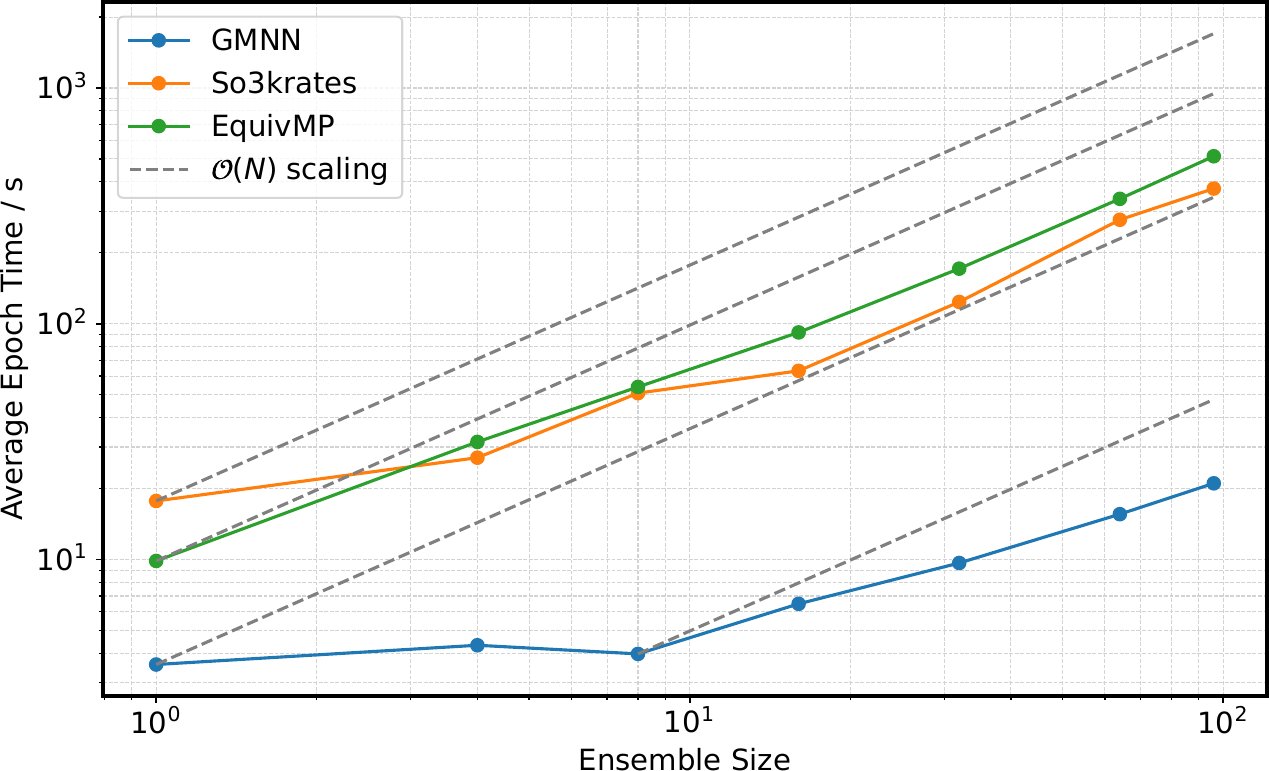} 
 \caption{
    Average epoch times of the GMNN, So3krates, and EquivMP architectures in relation to the size of the ensemble when trained as SE$_{E,F}$ models.
    An ensemble size of 1 represents a single model trained with MSE losses.
    The grey dashed lines represent linear scaling.
 }
 \label{fig:ens_scaling}
\end{figure}

Training shallow ensembles from scratch with force NLL loss slows down training considerably.
We illustrate this for three architectures, GMNN, So3krates, and EquivMP.
For each architecture, we train ensembles of size $N_{ens} \in \left( 4,8,10,16, 32, 64, 96\right)$ for the BMIM dataset.
All training runs were conducted on an A100 GPU.
The average epoch times compared to a single model are visualized in Fig.~\ref{fig:ens_scaling}. 
As the GMNN model is the slimmest of the three, it does not saturate the GPU for small ensemble sizes.
Epoch times grow linearly for $N_{ens} \geq 8$ as smaller ensembles do not fully saturate the GPU, resulting in sub-linear scaling.
For the chosen ensemble size of 32 members, the GMNN, So3krates, and EquivMP models are 2.7, 6.9, and 17 times slower than a single model.

\section{Analysing Last-Layer Covariances}

To understand why last-layer fine-tuning improves uncertainty estimates in some cases but not in others is challenging when both the backbone features and the last-layer covariance change simultaneously.
We therefore restrict the analysis to last-layer fine-tuning only.
Since the RND, ST, and LLPR$_{E,F}$ models share the same frozen backbone, we focus on a comparison between RND, which does not benefit from last-layer fine-tuning, and LLPR$_{E,F}$, which does.

\begin{figure}
 \centering
 \includegraphics[width=\columnwidth]{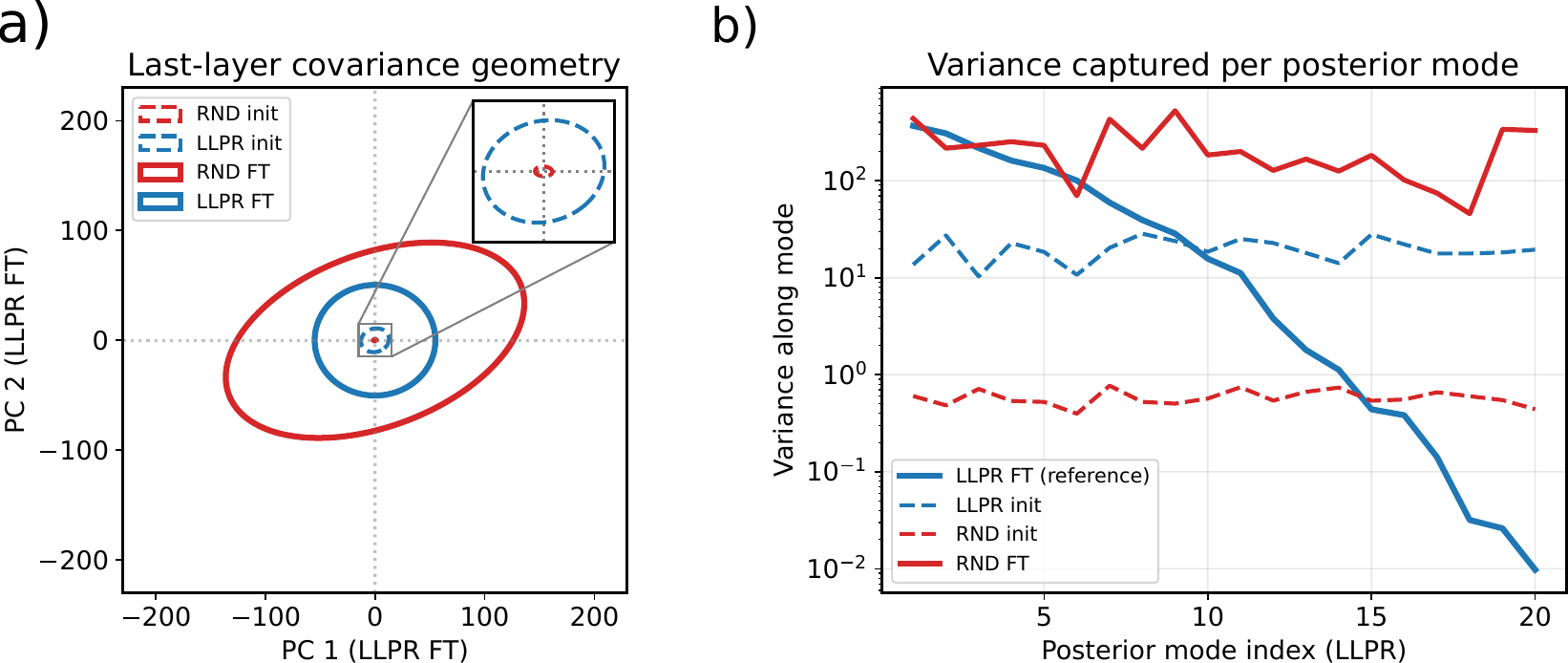} 
 \caption{
    Evolution of ensemble weight distributions in principal component of LLPR$_{E,F}$ space before and after NLL fine-tuning.
    The ellipses represent the $2\sigma$ confidence intervals.
    Left Column: Comparison of initialization strategies based on the same MSE-trained backbone.
    Right Column: Comparison of the SE$_{E}$ model against the ``From Scratch'' gold standard.
    The PCA basis is defined by the ``From Scratch'' weights.
 }
 \label{fig:ll_pca}
\end{figure}

Figure~\ref{fig:ll_pca}(a) shows projections of last-layer weights onto the first two principal components of the fine-tuned LLPR$_{E,F}$ weights.
The LLPR$_{E,F}$ covariance is already aligned with this basis at initialization, and fine-tuning primarily changes the scale of the uncertainty while preserving its orientation.
In contrast, the RND covariance is initially both misaligned and its scale is too small.
Although fine-tuning substantially increases its overall variance, the covariance remains misaligned with respect to the posterior principal components.

To analyse this effect further, Figure~\ref{fig:ll_pca}(b) shows the variance captured along each principal component of the fine-tuned LLPR$_{E,F}$ covariance.
Specifically, we computed the principal components of fine-tuned centered LLPR$_{E,F}$ weight samples and projected other last-layer weight ensembles onto this fixed basis.
For each method, we then measure the variance along each mode.
By construction, RND assigns nearly uniform variance across all modes at initialization, and last-layer fine-tuning does not introduce meaningful structure, suggesting that the optimization primarily increases variance magnitude without aligning it with the directions relevant to the data.
In contrast, the LLPR$_{E,F}$ covariance can adapt both in scale and shape during the NLL fine-tuning as the method is based on the Hessian.

These results demonstrate that uncertainty quality depends primarily on alignment with posterior geometry rather than on overall variance magnitude.
While fine-tuning increases the parameter-space variance of the RND model, this variance is allocated to directions that are not relevant for the data likelihood and therefore does not translate into improved predictive uncertainty.
In contrast, the curvature-aware Laplace approximation concentrates uncertainty along meaningful posterior directions, explaining its substantially better uncertainty quantification performance.

\begin{thebibliography}{63}%
\makeatletter
\providecommand \@ifxundefined [1]{%
 \@ifx{#1\undefined}
}%
\providecommand \@ifnum [1]{%
 \ifnum #1\expandafter \@firstoftwo
 \else \expandafter \@secondoftwo
 \fi
}%
\providecommand \@ifx [1]{%
 \ifx #1\expandafter \@firstoftwo
 \else \expandafter \@secondoftwo
 \fi
}%
\providecommand \natexlab [1]{#1}%
\providecommand \enquote  [1]{``#1''}%
\providecommand \bibnamefont  [1]{#1}%
\providecommand \bibfnamefont [1]{#1}%
\providecommand \citenamefont [1]{#1}%
\providecommand \href@noop [0]{\@secondoftwo}%
\providecommand \href [0]{\begingroup \@sanitize@url \@href}%
\providecommand \@href[1]{\@@startlink{#1}\@@href}%
\providecommand \@@href[1]{\endgroup#1\@@endlink}%
\providecommand \@sanitize@url [0]{\catcode `\\12\catcode `\$12\catcode
  `\&12\catcode `\#12\catcode `\^12\catcode `\_12\catcode `\%12\relax}%
\providecommand \@@startlink[1]{}%
\providecommand \@@endlink[0]{}%
\providecommand \url  [0]{\begingroup\@sanitize@url \@url }%
\providecommand \@url [1]{\endgroup\@href {#1}{\urlprefix }}%
\providecommand \urlprefix  [0]{URL }%
\providecommand \Eprint [0]{\href }%
\providecommand \doibase [0]{http://dx.doi.org/}%
\providecommand \selectlanguage [0]{\@gobble}%
\providecommand \bibinfo  [0]{\@secondoftwo}%
\providecommand \bibfield  [0]{\@secondoftwo}%
\providecommand \translation [1]{[#1]}%
\providecommand \BibitemOpen [0]{}%
\providecommand \bibitemStop [0]{}%
\providecommand \bibitemNoStop [0]{.\EOS\space}%
\providecommand \EOS [0]{\spacefactor3000\relax}%
\providecommand \BibitemShut  [1]{\csname bibitem#1\endcsname}%
\let\auto@bib@innerbib\@empty
\bibitem [{\citenamefont {Kendall}\ and\ \citenamefont
  {Gal}(2017)}]{kendallWhatUncertaintiesWe2017}%
  \BibitemOpen
  \bibfield  {author} {\bibinfo {author} {\bibfnamefont {Alex}\ \bibnamefont
  {Kendall}}\ and\ \bibinfo {author} {\bibfnamefont {Yarin}\ \bibnamefont
  {Gal}},\ }\bibfield  {title} {\enquote {\bibinfo {title} {What uncertainties
  do we need in {{Bayesian}} deep learning for computer vision?}}\ }in\
  \href@noop {} {\emph {\bibinfo {booktitle} {Proceedings of the 31st
  {{International Conference}} on {{Neural Information Processing
  Systems}}}}},\ \bibinfo {series and number} {{{NIPS}}'17}\ (\bibinfo
  {publisher} {Curran Associates Inc.},\ \bibinfo {year} {2017})\ pp.\ \bibinfo
  {pages} {5580--5590}\BibitemShut {NoStop}%
\bibitem [{\citenamefont {Aldossary}\ \emph {et~al.}(2024)\citenamefont
  {Aldossary}, \citenamefont {Campos-Gonzalez-Angulo}, \citenamefont
  {Pablo-García}, \citenamefont {Leong}, \citenamefont {Rajaonson},
  \citenamefont {Thiede}, \citenamefont {Tom}, \citenamefont {Wang},
  \citenamefont {Avagliano},\ and\ \citenamefont
  {Aspuru-Guzik}}]{aldossarySilicoChemicalExperiments2024}%
  \BibitemOpen
  \bibfield  {author} {\bibinfo {author} {\bibfnamefont {Abdulrahman}\
  \bibnamefont {Aldossary}}, \bibinfo {author} {\bibfnamefont {Jorge~Arturo}\
  \bibnamefont {Campos-Gonzalez-Angulo}}, \bibinfo {author} {\bibfnamefont
  {Sergio}\ \bibnamefont {Pablo-García}}, \bibinfo {author} {\bibfnamefont
  {Shi~Xuan}\ \bibnamefont {Leong}}, \bibinfo {author} {\bibfnamefont
  {Ella~Miray}\ \bibnamefont {Rajaonson}}, \bibinfo {author} {\bibfnamefont
  {Luca}\ \bibnamefont {Thiede}}, \bibinfo {author} {\bibfnamefont {Gary}\
  \bibnamefont {Tom}}, \bibinfo {author} {\bibfnamefont {Andrew}\ \bibnamefont
  {Wang}}, \bibinfo {author} {\bibfnamefont {Davide}\ \bibnamefont
  {Avagliano}}, \ and\ \bibinfo {author} {\bibfnamefont {Alán}\ \bibnamefont
  {Aspuru-Guzik}},\ }\bibfield  {title} {\enquote {\bibinfo {title} {In
  {{Silico Chemical Experiments}} in the {{Age}} of {{AI}}: {{From Quantum
  Chemistry}} to {{Machine Learning}} and {{Back}}},}\ }\href {\doibase
  10.1002/adma.202402369} {\bibfield  {journal} {\bibinfo  {journal} {Advanced
  Materials}\ }\textbf {\bibinfo {volume} {36}},\ \bibinfo {pages} {2402369}
  (\bibinfo {year} {2024})}\BibitemShut {NoStop}%
\bibitem [{\citenamefont {H{\"u}llermeier}\ and\ \citenamefont
  {Waegeman}(2021)}]{hullermeierAleatoricEpistemicUncertainty2021b}%
  \BibitemOpen
  \bibfield  {author} {\bibinfo {author} {\bibfnamefont {Eyke}\ \bibnamefont
  {H{\"u}llermeier}}\ and\ \bibinfo {author} {\bibfnamefont {Willem}\
  \bibnamefont {Waegeman}},\ }\bibfield  {title} {\enquote {\bibinfo {title}
  {Aleatoric and epistemic uncertainty in machine learning: An introduction to
  concepts and methods},}\ }\href {\doibase 10.1007/s10994-021-05946-3}
  {\bibfield  {journal} {\bibinfo  {journal} {Machine Learning}\ }\textbf
  {\bibinfo {volume} {110}},\ \bibinfo {pages} {457--506} (\bibinfo {year}
  {2021})}\BibitemShut {NoStop}%
\bibitem [{\citenamefont {Perez}\ \emph {et~al.}(2025)\citenamefont {Perez},
  \citenamefont {Subramanyam}, \citenamefont {Maliyov},\ and\ \citenamefont
  {Swinburne}}]{perezUncertaintyQuantificationMisspecified2025}%
  \BibitemOpen
  \bibfield  {author} {\bibinfo {author} {\bibfnamefont {Danny}\ \bibnamefont
  {Perez}}, \bibinfo {author} {\bibfnamefont {Aparna P.~A.}\ \bibnamefont
  {Subramanyam}}, \bibinfo {author} {\bibfnamefont {Ivan}\ \bibnamefont
  {Maliyov}}, \ and\ \bibinfo {author} {\bibfnamefont {Thomas~D.}\ \bibnamefont
  {Swinburne}},\ }\bibfield  {title} {\enquote {\bibinfo {title} {Uncertainty
  quantification for misspecified machine learned interatomic potentials},}\
  }\href {\doibase 10.1038/s41524-025-01758-4} {\bibfield  {journal} {\bibinfo
  {journal} {npj Computational Materials}\ }\textbf {\bibinfo {volume} {11}},\
  \bibinfo {pages} {263} (\bibinfo {year} {2025})}\BibitemShut {NoStop}%
\bibitem [{\citenamefont {Herbst}\ \emph {et~al.}(2020)\citenamefont {Herbst},
  \citenamefont {Levitt},\ and\ \citenamefont
  {Cancès}}]{herbstPosterioriErrorEstimation2020}%
  \BibitemOpen
  \bibfield  {author} {\bibinfo {author} {\bibfnamefont {Michael~F.}\
  \bibnamefont {Herbst}}, \bibinfo {author} {\bibfnamefont {Antoine}\
  \bibnamefont {Levitt}}, \ and\ \bibinfo {author} {\bibfnamefont {Eric}\
  \bibnamefont {Cancès}},\ }\bibfield  {title} {\enquote {\bibinfo {title} {A
  posteriori error estimation for the non-self-consistent {{Kohn}}–{{Sham}}
  equations},}\ }\href {\doibase 10.1039/D0FD00048E} {\bibfield  {journal}
  {\bibinfo  {journal} {Faraday Discussions}\ }\textbf {\bibinfo {volume}
  {224}},\ \bibinfo {pages} {227--246} (\bibinfo {year} {2020})}\BibitemShut
  {NoStop}%
\bibitem [{\citenamefont {Pernot}(2022)}]{pernotLongRoadCalibrated2022a}%
  \BibitemOpen
  \bibfield  {author} {\bibinfo {author} {\bibfnamefont {Pascal}\ \bibnamefont
  {Pernot}},\ }\bibfield  {title} {\enquote {\bibinfo {title} {The long road to
  calibrated prediction uncertainty in computational chemistry},}\ }\href
  {\doibase 10.1063/5.0084302} {\bibfield  {journal} {\bibinfo  {journal} {The
  Journal of Chemical Physics}\ }\textbf {\bibinfo {volume} {156}},\ \bibinfo
  {pages} {114109} (\bibinfo {year} {2022})}\BibitemShut {NoStop}%
\bibitem [{\citenamefont {Lakshminarayanan}\ \emph {et~al.}(2017)\citenamefont
  {Lakshminarayanan}, \citenamefont {Pritzel},\ and\ \citenamefont
  {Blundell}}]{lakshminarayananSimpleScalablePredictive2017}%
  \BibitemOpen
  \bibfield  {author} {\bibinfo {author} {\bibfnamefont {Balaji}\ \bibnamefont
  {Lakshminarayanan}}, \bibinfo {author} {\bibfnamefont {Alexander}\
  \bibnamefont {Pritzel}}, \ and\ \bibinfo {author} {\bibfnamefont {Charles}\
  \bibnamefont {Blundell}},\ }\bibfield  {title} {\enquote {\bibinfo {title}
  {Simple and scalable predictive uncertainty estimation using deep
  ensembles},}\ }in\ \href@noop {} {\emph {\bibinfo {booktitle} {Proceedings of
  the 31st {{International Conference}} on {{Neural Information Processing
  Systems}}}}},\ \bibinfo {series and number} {{{NIPS}}'17}\ (\bibinfo
  {publisher} {Curran Associates Inc.},\ \bibinfo {year} {2017})\ pp.\ \bibinfo
  {pages} {6405--6416}\BibitemShut {NoStop}%
\bibitem [{\citenamefont {Carleo}\ \emph {et~al.}(2019)\citenamefont {Carleo},
  \citenamefont {Cirac}, \citenamefont {Cranmer}, \citenamefont {Daudet},
  \citenamefont {Schuld}, \citenamefont {Tishby}, \citenamefont
  {{Vogt-Maranto}},\ and\ \citenamefont
  {Zdeborov{\'a}}}]{carleoMachineLearningPhysical2019}%
  \BibitemOpen
  \bibfield  {author} {\bibinfo {author} {\bibfnamefont {Giuseppe}\
  \bibnamefont {Carleo}}, \bibinfo {author} {\bibfnamefont {Ignacio}\
  \bibnamefont {Cirac}}, \bibinfo {author} {\bibfnamefont {Kyle}\ \bibnamefont
  {Cranmer}}, \bibinfo {author} {\bibfnamefont {Laurent}\ \bibnamefont
  {Daudet}}, \bibinfo {author} {\bibfnamefont {Maria}\ \bibnamefont {Schuld}},
  \bibinfo {author} {\bibfnamefont {Naftali}\ \bibnamefont {Tishby}}, \bibinfo
  {author} {\bibfnamefont {Leslie}\ \bibnamefont {{Vogt-Maranto}}}, \ and\
  \bibinfo {author} {\bibfnamefont {Lenka}\ \bibnamefont {Zdeborov{\'a}}},\
  }\bibfield  {title} {\enquote {\bibinfo {title} {Machine learning and the
  physical sciences},}\ }\href {\doibase 10.1103/RevModPhys.91.045002}
  {\bibfield  {journal} {\bibinfo  {journal} {Reviews of Modern Physics}\
  }\textbf {\bibinfo {volume} {91}},\ \bibinfo {pages} {045002} (\bibinfo
  {year} {2019})}\BibitemShut {NoStop}%
\bibitem [{\citenamefont {Imbalzano}\ \emph {et~al.}(2021)\citenamefont
  {Imbalzano}, \citenamefont {Zhuang}, \citenamefont {Kapil}, \citenamefont
  {Rossi}, \citenamefont {Engel}, \citenamefont {Grasselli},\ and\
  \citenamefont {Ceriotti}}]{imba+21jcp}%
  \BibitemOpen
  \bibfield  {author} {\bibinfo {author} {\bibfnamefont {Giulio}\ \bibnamefont
  {Imbalzano}}, \bibinfo {author} {\bibfnamefont {Yongbin}\ \bibnamefont
  {Zhuang}}, \bibinfo {author} {\bibfnamefont {Venkat}\ \bibnamefont {Kapil}},
  \bibinfo {author} {\bibfnamefont {Kevin}\ \bibnamefont {Rossi}}, \bibinfo
  {author} {\bibfnamefont {Edgar~A.}\ \bibnamefont {Engel}}, \bibinfo {author}
  {\bibfnamefont {Federico}\ \bibnamefont {Grasselli}}, \ and\ \bibinfo
  {author} {\bibfnamefont {Michele}\ \bibnamefont {Ceriotti}},\ }\bibfield
  {title} {\enquote {\bibinfo {title} {Uncertainty estimation for molecular
  dynamics and sampling},}\ }\href {\doibase 10.1063/5.0036522} {\bibfield
  {journal} {\bibinfo  {journal} {J. Chem. Phys.}\ }\textbf {\bibinfo {volume}
  {154}},\ \bibinfo {pages} {074102} (\bibinfo {year} {2021})}\BibitemShut
  {NoStop}%
\bibitem [{\citenamefont {Smith}\ \emph {et~al.}(2018)\citenamefont {Smith},
  \citenamefont {Nebgen}, \citenamefont {Lubbers}, \citenamefont {Isayev},\
  and\ \citenamefont {Roitberg}}]{smithLessMoreSampling2018}%
  \BibitemOpen
  \bibfield  {author} {\bibinfo {author} {\bibfnamefont {Justin~S.}\
  \bibnamefont {Smith}}, \bibinfo {author} {\bibfnamefont {Ben}\ \bibnamefont
  {Nebgen}}, \bibinfo {author} {\bibfnamefont {Nicholas}\ \bibnamefont
  {Lubbers}}, \bibinfo {author} {\bibfnamefont {Olexandr}\ \bibnamefont
  {Isayev}}, \ and\ \bibinfo {author} {\bibfnamefont {Adrian~E.}\ \bibnamefont
  {Roitberg}},\ }\bibfield  {title} {\enquote {\bibinfo {title} {Less is more:
  {{Sampling}} chemical space with active learning},}\ }\href {\doibase
  10.1063/1.5023802} {\bibfield  {journal} {\bibinfo  {journal} {The Journal of
  Chemical Physics}\ }\textbf {\bibinfo {volume} {148}},\ \bibinfo {pages}
  {241733} (\bibinfo {year} {2018})}\BibitemShut {NoStop}%
\bibitem [{\citenamefont {Kulichenko}\ \emph {et~al.}(2023)\citenamefont
  {Kulichenko}, \citenamefont {Barros}, \citenamefont {Lubbers}, \citenamefont
  {Li}, \citenamefont {Messerly}, \citenamefont {Tretiak}, \citenamefont
  {Smith},\ and\ \citenamefont
  {Nebgen}}]{kulichenkoUncertaintydrivenDynamicsActive2023}%
  \BibitemOpen
  \bibfield  {author} {\bibinfo {author} {\bibfnamefont {Maksim}\ \bibnamefont
  {Kulichenko}}, \bibinfo {author} {\bibfnamefont {Kipton}\ \bibnamefont
  {Barros}}, \bibinfo {author} {\bibfnamefont {Nicholas}\ \bibnamefont
  {Lubbers}}, \bibinfo {author} {\bibfnamefont {Ying~Wai}\ \bibnamefont {Li}},
  \bibinfo {author} {\bibfnamefont {Richard}\ \bibnamefont {Messerly}},
  \bibinfo {author} {\bibfnamefont {Sergei}\ \bibnamefont {Tretiak}}, \bibinfo
  {author} {\bibfnamefont {Justin~S.}\ \bibnamefont {Smith}}, \ and\ \bibinfo
  {author} {\bibfnamefont {Benjamin}\ \bibnamefont {Nebgen}},\ }\bibfield
  {title} {\enquote {\bibinfo {title} {Uncertainty-driven dynamics for active
  learning of interatomic potentials},}\ }\href {\doibase
  10.1038/s43588-023-00406-5} {\bibfield  {journal} {\bibinfo  {journal}
  {Nature Computational Science}\ ,\ \bibinfo {pages} {1--10}} (\bibinfo {year}
  {2023})}\BibitemShut {NoStop}%
\bibitem [{\citenamefont {Zills}\ \emph
  {et~al.}(2024{\natexlab{a}})\citenamefont {Zills}, \citenamefont
  {René~Schäfer}, \citenamefont {Tovey}, \citenamefont {Kästner},\ and\
  \citenamefont {Holm}}]{zillsMachineLearningdrivenInvestigation2024a}%
  \BibitemOpen
  \bibfield  {author} {\bibinfo {author} {\bibfnamefont {Fabian}\ \bibnamefont
  {Zills}}, \bibinfo {author} {\bibfnamefont {Moritz}\ \bibnamefont
  {René~Schäfer}}, \bibinfo {author} {\bibfnamefont {Samuel}\ \bibnamefont
  {Tovey}}, \bibinfo {author} {\bibfnamefont {Johannes}\ \bibnamefont
  {Kästner}}, \ and\ \bibinfo {author} {\bibfnamefont {Christian}\
  \bibnamefont {Holm}},\ }\bibfield  {title} {\enquote {\bibinfo {title}
  {Machine learning-driven investigation of the structure and dynamics of the
  {{BMIM-BF}} 4 room temperature ionic liquid},}\ }\href {\doibase
  10.1039/D4FD00025K} {\bibfield  {journal} {\bibinfo  {journal} {Faraday
  Discussions}\ }\textbf {\bibinfo {volume} {253}},\ \bibinfo {pages}
  {129--145} (\bibinfo {year} {2024}{\natexlab{a}})}\BibitemShut {NoStop}%
\bibitem [{\citenamefont {Schäfer}\ and\ \citenamefont
  {Kästner}(2026)}]{schaferEnhancedRepresentationBasedSampling2026}%
  \BibitemOpen
  \bibfield  {author} {\bibinfo {author} {\bibfnamefont {Moritz~R.}\
  \bibnamefont {Schäfer}}\ and\ \bibinfo {author} {\bibfnamefont {Johannes}\
  \bibnamefont {Kästner}},\ }\bibfield  {title} {\enquote {\bibinfo {title}
  {Enhanced {{Representation-Based Sampling}} for the {{Efficient Generation}}
  of {{Data Sets}} for {{Machine-Learned Interatomic Potentials}}},}\ }\href
  {\doibase 10.1021/acs.jctc.5c01767} {\bibfield  {journal} {\bibinfo
  {journal} {Journal of Chemical Theory and Computation}\ } (\bibinfo {year}
  {2026}),\ 10.1021/acs.jctc.5c01767},\ \bibinfo {note} {in press}\BibitemShut
  {NoStop}%
\bibitem [{\citenamefont {Rasmussen}\ and\ \citenamefont
  {Williams}(2008)}]{rasmussenGaussianProcessesMachine2008}%
  \BibitemOpen
  \bibfield  {author} {\bibinfo {author} {\bibfnamefont {Carl~Edward}\
  \bibnamefont {Rasmussen}}\ and\ \bibinfo {author} {\bibfnamefont {Christopher
  K.~I.}\ \bibnamefont {Williams}},\ }\href@noop {} {\emph {\bibinfo {title}
  {Gaussian Processes for Machine Learning}}},\ \bibinfo {edition} {3rd}\ ed.,\
  Adaptive Computation and Machine Learning\ (\bibinfo  {publisher} {{MIT
  Press}},\ \bibinfo {address} {{Cambridge, Mass.}},\ \bibinfo {year} {2008})\
  pp.\ \bibinfo {pages} {23, 177}\BibitemShut {NoStop}%
\bibitem [{\citenamefont {Gawlikowski}\ \emph {et~al.}(2023)\citenamefont
  {Gawlikowski}, \citenamefont {Tassi}, \citenamefont {Ali}, \citenamefont
  {Lee}, \citenamefont {Humt}, \citenamefont {Feng}, \citenamefont {Kruspe},
  \citenamefont {Triebel}, \citenamefont {Jung}, \citenamefont {Roscher},
  \citenamefont {Shahzad}, \citenamefont {Yang}, \citenamefont {Bamler},\ and\
  \citenamefont {Zhu}}]{gawlikowskiSurveyUncertaintyDeep2023}%
  \BibitemOpen
  \bibfield  {author} {\bibinfo {author} {\bibfnamefont {Jakob}\ \bibnamefont
  {Gawlikowski}}, \bibinfo {author} {\bibfnamefont {Cedrique
  Rovile~Njieutcheu}\ \bibnamefont {Tassi}}, \bibinfo {author} {\bibfnamefont
  {Mohsin}\ \bibnamefont {Ali}}, \bibinfo {author} {\bibfnamefont {Jongseok}\
  \bibnamefont {Lee}}, \bibinfo {author} {\bibfnamefont {Matthias}\
  \bibnamefont {Humt}}, \bibinfo {author} {\bibfnamefont {Jianxiang}\
  \bibnamefont {Feng}}, \bibinfo {author} {\bibfnamefont {Anna}\ \bibnamefont
  {Kruspe}}, \bibinfo {author} {\bibfnamefont {Rudolph}\ \bibnamefont
  {Triebel}}, \bibinfo {author} {\bibfnamefont {Peter}\ \bibnamefont {Jung}},
  \bibinfo {author} {\bibfnamefont {Ribana}\ \bibnamefont {Roscher}}, \bibinfo
  {author} {\bibfnamefont {Muhammad}\ \bibnamefont {Shahzad}}, \bibinfo
  {author} {\bibfnamefont {Wen}\ \bibnamefont {Yang}}, \bibinfo {author}
  {\bibfnamefont {Richard}\ \bibnamefont {Bamler}}, \ and\ \bibinfo {author}
  {\bibfnamefont {Xiao~Xiang}\ \bibnamefont {Zhu}},\ }\bibfield  {title}
  {\enquote {\bibinfo {title} {A survey of uncertainty in deep neural
  networks},}\ }\href {\doibase 10.1007/s10462-023-10562-9} {\bibfield
  {journal} {\bibinfo  {journal} {Artificial Intelligence Review}\ }\textbf
  {\bibinfo {volume} {56}},\ \bibinfo {pages} {1513--1589} (\bibinfo {year}
  {2023})}\BibitemShut {NoStop}%
\bibitem [{\citenamefont {Grasselli}\ \emph {et~al.}(2025)\citenamefont
  {Grasselli}, \citenamefont {Chong}, \citenamefont {Kapil}, \citenamefont
  {Bonfanti},\ and\ \citenamefont
  {Rossi}}]{grasselliUncertaintyEraMachine2025}%
  \BibitemOpen
  \bibfield  {author} {\bibinfo {author} {\bibfnamefont {Federico}\
  \bibnamefont {Grasselli}}, \bibinfo {author} {\bibfnamefont {Sanggyu}\
  \bibnamefont {Chong}}, \bibinfo {author} {\bibfnamefont {Venkat}\
  \bibnamefont {Kapil}}, \bibinfo {author} {\bibfnamefont {Silvia}\
  \bibnamefont {Bonfanti}}, \ and\ \bibinfo {author} {\bibfnamefont {Kevin}\
  \bibnamefont {Rossi}},\ }\href {\doibase 10.48550/arXiv.2503.09196} {\enquote
  {\bibinfo {title} {Uncertainty in the era of machine learning for atomistic
  modeling},}\ } (\bibinfo {year} {2025}),\ \Eprint
  {http://arxiv.org/abs/2503.09196} {2503.09196} \BibitemShut {NoStop}%
\bibitem [{\citenamefont
  {MacKay}(1992)}]{mackayPracticalBayesianFramework1992}%
  \BibitemOpen
  \bibfield  {author} {\bibinfo {author} {\bibfnamefont {David J.~C.}\
  \bibnamefont {MacKay}},\ }\bibfield  {title} {\enquote {\bibinfo {title} {A
  {{Practical Bayesian Framework}} for {{Backpropagation Networks}}},}\ }\href
  {\doibase 10.1162/neco.1992.4.3.448} {\bibfield  {journal} {\bibinfo
  {journal} {Neural Computation}\ }\textbf {\bibinfo {volume} {4}},\ \bibinfo
  {pages} {448--472} (\bibinfo {year} {1992})}\BibitemShut {NoStop}%
\bibitem [{\citenamefont {Gal}\ and\ \citenamefont
  {Ghahramani}(2016)}]{galDropoutBayesianApproximation}%
  \BibitemOpen
  \bibfield  {author} {\bibinfo {author} {\bibfnamefont {Yarin}\ \bibnamefont
  {Gal}}\ and\ \bibinfo {author} {\bibfnamefont {Zoubin}\ \bibnamefont
  {Ghahramani}},\ }\bibfield  {title} {\enquote {\bibinfo {title} {Dropout as a
  bayesian approximation: Representing model uncertainty in deep learning},}\
  }in\ \href {https://proceedings.mlr.press/v48/gal16.html} {\emph {\bibinfo
  {booktitle} {Proceedings of The 33rd International Conference on Machine
  Learning}}},\ \bibinfo {series} {Proceedings of Machine Learning Research},
  Vol.~\bibinfo {volume} {48},\ \bibinfo {editor} {edited by\ \bibinfo {editor}
  {\bibfnamefont {Maria~Florina}\ \bibnamefont {Balcan}}\ and\ \bibinfo
  {editor} {\bibfnamefont {Kilian~Q.}\ \bibnamefont {Weinberger}}}\ (\bibinfo
  {publisher} {PMLR},\ \bibinfo {address} {New York, New York, USA},\ \bibinfo
  {year} {2016})\ pp.\ \bibinfo {pages} {1050--1059}\BibitemShut {NoStop}%
\bibitem [{\citenamefont {Nix}\ and\ \citenamefont
  {Weigend}(1994)}]{nixEstimatingMeanVariance1994}%
  \BibitemOpen
  \bibfield  {author} {\bibinfo {author} {\bibfnamefont {D.A.}\ \bibnamefont
  {Nix}}\ and\ \bibinfo {author} {\bibfnamefont {A.S.}\ \bibnamefont
  {Weigend}},\ }\bibfield  {title} {\enquote {\bibinfo {title} {Estimating the
  mean and variance of the target probability distribution},}\ }in\ \href
  {\doibase 10.1109/ICNN.1994.374138} {\emph {\bibinfo {booktitle} {Proceedings
  of 1994 {{IEEE International Conference}} on {{Neural Networks}}
  ({{ICNN}}'94)}}},\ Vol.~\bibinfo {volume} {1}\ (\bibinfo {year} {1994})\ pp.\
  \bibinfo {pages} {55--60 vol.1}\BibitemShut {NoStop}%
\bibitem [{\citenamefont {Amini}\ \emph {et~al.}(2020)\citenamefont {Amini},
  \citenamefont {Schwarting}, \citenamefont {Soleimany},\ and\ \citenamefont
  {Rus}}]{aminiDeepEvidentialRegression2020a}%
  \BibitemOpen
  \bibfield  {author} {\bibinfo {author} {\bibfnamefont {Alexander}\
  \bibnamefont {Amini}}, \bibinfo {author} {\bibfnamefont {Wilko}\ \bibnamefont
  {Schwarting}}, \bibinfo {author} {\bibfnamefont {Ava}\ \bibnamefont
  {Soleimany}}, \ and\ \bibinfo {author} {\bibfnamefont {Daniela}\ \bibnamefont
  {Rus}},\ }\bibfield  {title} {\enquote {\bibinfo {title} {Deep evidential
  regression},}\ }in\ \href
  {https://proceedings.neurips.cc/paper_files/paper/2020/file/aab085461de182608ee9f607f3f7d18f-Paper.pdf}
  {\emph {\bibinfo {booktitle} {Advances in Neural Information Processing
  Systems}}},\ Vol.~\bibinfo {volume} {33},\ \bibinfo {editor} {edited by\
  \bibinfo {editor} {\bibfnamefont {H.}~\bibnamefont {Larochelle}}, \bibinfo
  {editor} {\bibfnamefont {M.}~\bibnamefont {Ranzato}}, \bibinfo {editor}
  {\bibfnamefont {R.}~\bibnamefont {Hadsell}}, \bibinfo {editor} {\bibfnamefont
  {M.F.}\ \bibnamefont {Balcan}}, \ and\ \bibinfo {editor} {\bibfnamefont
  {H.}~\bibnamefont {Lin}}}\ (\bibinfo  {publisher} {Curran Associates, Inc.},\
  \bibinfo {year} {2020})\ pp.\ \bibinfo {pages} {14927--14937}\BibitemShut
  {NoStop}%
\bibitem [{\citenamefont {Tan}\ \emph {et~al.}(2023)\citenamefont {Tan},
  \citenamefont {Urata}, \citenamefont {Goldman}, \citenamefont {Dietschreit},\
  and\ \citenamefont
  {{G{\'o}mez-Bombarelli}}}]{tanSinglemodelUncertaintyQuantification2023}%
  \BibitemOpen
  \bibfield  {author} {\bibinfo {author} {\bibfnamefont {Aik~Rui}\ \bibnamefont
  {Tan}}, \bibinfo {author} {\bibfnamefont {Shingo}\ \bibnamefont {Urata}},
  \bibinfo {author} {\bibfnamefont {Samuel}\ \bibnamefont {Goldman}}, \bibinfo
  {author} {\bibfnamefont {Johannes C.~B.}\ \bibnamefont {Dietschreit}}, \ and\
  \bibinfo {author} {\bibfnamefont {Rafael}\ \bibnamefont
  {{G{\'o}mez-Bombarelli}}},\ }\bibfield  {title} {\enquote {\bibinfo {title}
  {Single-model uncertainty quantification in neural network potentials does
  not consistently outperform model ensembles},}\ }\href {\doibase
  10.1038/s41524-023-01180-8} {\bibfield  {journal} {\bibinfo  {journal} {npj
  Computational Materials}\ }\textbf {\bibinfo {volume} {9}},\ \bibinfo {pages}
  {225} (\bibinfo {year} {2023})}\BibitemShut {NoStop}%
\bibitem [{\citenamefont {Ho}\ \emph {et~al.}(2025)\citenamefont {Ho},
  \citenamefont {Ortner},\ and\ \citenamefont
  {Wang}}]{hoFlexibleUncertaintyCalibration2025}%
  \BibitemOpen
  \bibfield  {author} {\bibinfo {author} {\bibfnamefont {Cheuk~Hin}\
  \bibnamefont {Ho}}, \bibinfo {author} {\bibfnamefont {Christoph}\
  \bibnamefont {Ortner}}, \ and\ \bibinfo {author} {\bibfnamefont {Yangshuai}\
  \bibnamefont {Wang}},\ }\href {\doibase 10.48550/arXiv.2510.00721} {\enquote
  {\bibinfo {title} {Flexible {{Uncertainty Calibration}} for {{Machine-Learned
  Interatomic Potentials}}},}\ } (\bibinfo {year} {2025}),\ \Eprint
  {http://arxiv.org/abs/2510.00721} {arXiv:2510.00721 [physics]} \BibitemShut
  {NoStop}%
\bibitem [{\citenamefont {Zhu}\ \emph {et~al.}(2023)\citenamefont {Zhu},
  \citenamefont {Batzner}, \citenamefont {Musaelian},\ and\ \citenamefont
  {Kozinsky}}]{zhuFastUncertaintyEstimates2023}%
  \BibitemOpen
  \bibfield  {author} {\bibinfo {author} {\bibfnamefont {Albert}\ \bibnamefont
  {Zhu}}, \bibinfo {author} {\bibfnamefont {Simon}\ \bibnamefont {Batzner}},
  \bibinfo {author} {\bibfnamefont {Albert}\ \bibnamefont {Musaelian}}, \ and\
  \bibinfo {author} {\bibfnamefont {Boris}\ \bibnamefont {Kozinsky}},\
  }\bibfield  {title} {\enquote {\bibinfo {title} {Fast uncertainty estimates
  in deep learning interatomic potentials},}\ }\href {\doibase
  10.1063/5.0136574} {\bibfield  {journal} {\bibinfo  {journal} {The Journal of
  Chemical Physics}\ }\textbf {\bibinfo {volume} {158}},\ \bibinfo {pages}
  {164111} (\bibinfo {year} {2023})}\BibitemShut {NoStop}%
\bibitem [{\citenamefont {Garain}\ \emph {et~al.}(2025)\citenamefont {Garain},
  \citenamefont {Jr}, \citenamefont {Bispo},\ and\ \citenamefont
  {Barbatti}}]{garainUncertaintyCalibrationMolecular2025}%
  \BibitemOpen
  \bibfield  {author} {\bibinfo {author} {\bibfnamefont {Bidhan~Chandra}\
  \bibnamefont {Garain}}, \bibinfo {author} {\bibfnamefont {Max~Pinheiro}\
  \bibnamefont {Jr}}, \bibinfo {author} {\bibfnamefont {Matheus~O.}\
  \bibnamefont {Bispo}}, \ and\ \bibinfo {author} {\bibfnamefont {Mario}\
  \bibnamefont {Barbatti}},\ }\href {\doibase 10.26434/chemrxiv-2025-fx019-v2}
  {\enquote {\bibinfo {title} {Uncertainty {{Calibration}} in {{Molecular
  Machine Learning}}: {{Comparing Evidential}} and {{Ensemble Approaches}}},}\
  } (\bibinfo {year} {2025})\BibitemShut {NoStop}%
\bibitem [{\citenamefont {Kellner}\ and\ \citenamefont
  {Ceriotti}(2024)}]{kellnerUncertaintyQuantificationDirect2024a}%
  \BibitemOpen
  \bibfield  {author} {\bibinfo {author} {\bibfnamefont {Matthias}\
  \bibnamefont {Kellner}}\ and\ \bibinfo {author} {\bibfnamefont {Michele}\
  \bibnamefont {Ceriotti}},\ }\bibfield  {title} {\enquote {\bibinfo {title}
  {Uncertainty quantification by direct propagation of shallow ensembles},}\
  }\href {\doibase 10.1088/2632-2153/ad594a} {\bibfield  {journal} {\bibinfo
  {journal} {Machine Learning: Science and Technology}\ }\textbf {\bibinfo
  {volume} {5}},\ \bibinfo {pages} {035006} (\bibinfo {year}
  {2024})}\BibitemShut {NoStop}%
\bibitem [{\citenamefont {Mazitov}\ \emph {et~al.}(2025)\citenamefont
  {Mazitov}, \citenamefont {Bigi}, \citenamefont {Kellner}, \citenamefont
  {Pegolo}, \citenamefont {Tisi}, \citenamefont {Fraux}, \citenamefont
  {Pozdnyakov}, \citenamefont {Loche},\ and\ \citenamefont
  {Ceriotti}}]{mazitovPETMADLightweightUniversal2025a}%
  \BibitemOpen
  \bibfield  {author} {\bibinfo {author} {\bibfnamefont {Arslan}\ \bibnamefont
  {Mazitov}}, \bibinfo {author} {\bibfnamefont {Filippo}\ \bibnamefont {Bigi}},
  \bibinfo {author} {\bibfnamefont {Matthias}\ \bibnamefont {Kellner}},
  \bibinfo {author} {\bibfnamefont {Paolo}\ \bibnamefont {Pegolo}}, \bibinfo
  {author} {\bibfnamefont {Davide}\ \bibnamefont {Tisi}}, \bibinfo {author}
  {\bibfnamefont {Guillaume}\ \bibnamefont {Fraux}}, \bibinfo {author}
  {\bibfnamefont {Sergey}\ \bibnamefont {Pozdnyakov}}, \bibinfo {author}
  {\bibfnamefont {Philip}\ \bibnamefont {Loche}}, \ and\ \bibinfo {author}
  {\bibfnamefont {Michele}\ \bibnamefont {Ceriotti}},\ }\bibfield  {title}
  {\enquote {\bibinfo {title} {{{PET-MAD}} as a lightweight universal
  interatomic potential for advanced materials modeling},}\ }\href {\doibase
  10.1038/s41467-025-65662-7} {\bibfield  {journal} {\bibinfo  {journal}
  {Nature Communications}\ }\textbf {\bibinfo {volume} {16}},\ \bibinfo {pages}
  {10653} (\bibinfo {year} {2025})}\BibitemShut {NoStop}%
\bibitem [{\citenamefont {Kempen}\ \emph {et~al.}(2025)\citenamefont {Kempen},
  \citenamefont {Nielsen},\ and\ \citenamefont
  {Andersen}}]{kempenBreakingScalingRelations2025}%
  \BibitemOpen
  \bibfield  {author} {\bibinfo {author} {\bibfnamefont {Luuk H.~E.}\
  \bibnamefont {Kempen}}, \bibinfo {author} {\bibfnamefont {Marius~Juul}\
  \bibnamefont {Nielsen}}, \ and\ \bibinfo {author} {\bibfnamefont {Mie}\
  \bibnamefont {Andersen}},\ }\bibfield  {title} {\enquote {\bibinfo {title}
  {Breaking {{Scaling Relations}} with {{Inverse Catalysts}}: {{A Machine
  Learning Exploration}} of {{Trends}} in {{CO2-to-Formate Energy
  Barriers}}},}\ }\href {\doibase 10.1021/acscatal.5c05872} {\bibfield
  {journal} {\bibinfo  {journal} {ACS Catalysis}\ }\textbf {\bibinfo {volume}
  {15}},\ \bibinfo {pages} {17635--17644} (\bibinfo {year} {2025})}\BibitemShut
  {NoStop}%
\bibitem [{\citenamefont {Plé}\ \emph {et~al.}(2025)\citenamefont {Plé},
  \citenamefont {Adjoua}, \citenamefont {Benali}, \citenamefont {Posenitskiy},
  \citenamefont {Villot}, \citenamefont {Lagardère},\ and\ \citenamefont
  {Piquemal}}]{chemrxiv-2025-f1hgn-v4}%
  \BibitemOpen
  \bibfield  {author} {\bibinfo {author} {\bibfnamefont {Thomas}\ \bibnamefont
  {Plé}}, \bibinfo {author} {\bibfnamefont {Olivier}\ \bibnamefont {Adjoua}},
  \bibinfo {author} {\bibfnamefont {Anouar}\ \bibnamefont {Benali}}, \bibinfo
  {author} {\bibfnamefont {Evgeny}\ \bibnamefont {Posenitskiy}}, \bibinfo
  {author} {\bibfnamefont {Corentin}\ \bibnamefont {Villot}}, \bibinfo {author}
  {\bibfnamefont {Louis}\ \bibnamefont {Lagardère}}, \ and\ \bibinfo {author}
  {\bibfnamefont {Jean-Philip}\ \bibnamefont {Piquemal}},\ }\bibfield  {title}
  {\enquote {\bibinfo {title} {A foundation model for accurate atomistic
  simulations in drug design},}\ }\href {\doibase
  10.26434/chemrxiv-2025-f1hgn-v4} {\bibfield  {journal} {\bibinfo  {journal}
  {ChemRxiv}\ }\textbf {\bibinfo {volume} {2025}} (\bibinfo {year} {2025}),\
  10.26434/chemrxiv-2025-f1hgn-v4},\ \Eprint
  {http://arxiv.org/abs/https://chemrxiv.org/doi/pdf/10.26434/chemrxiv-2025-f1hgn-v4}
  {https://chemrxiv.org/doi/pdf/10.26434/chemrxiv-2025-f1hgn-v4} \BibitemShut
  {NoStop}%
\bibitem [{\citenamefont {Vinchurkar}\ \emph {et~al.}(2025)\citenamefont
  {Vinchurkar}, \citenamefont {Abdelmaqsoud},\ and\ \citenamefont
  {Kitchin}}]{vinchurkarUncertaintyQuantificationGraph2025}%
  \BibitemOpen
  \bibfield  {author} {\bibinfo {author} {\bibfnamefont {Tirtha}\ \bibnamefont
  {Vinchurkar}}, \bibinfo {author} {\bibfnamefont {Kareem}\ \bibnamefont
  {Abdelmaqsoud}}, \ and\ \bibinfo {author} {\bibfnamefont {John~R}\
  \bibnamefont {Kitchin}},\ }\bibfield  {title} {\enquote {\bibinfo {title}
  {Uncertainty quantification in graph neural networks with shallow
  ensembles},}\ }\href {\doibase 10.1088/2632-2153/ae0bf0} {\bibfield
  {journal} {\bibinfo  {journal} {Machine Learning: Science and Technology}\
  }\textbf {\bibinfo {volume} {6}},\ \bibinfo {pages} {045007} (\bibinfo {year}
  {2025})}\BibitemShut {NoStop}%
\bibitem [{\citenamefont {Chong}\ \emph {et~al.}(2023)\citenamefont {Chong},
  \citenamefont {Grasselli}, \citenamefont {Ben~Mahmoud}, \citenamefont
  {Morrow}, \citenamefont {Deringer},\ and\ \citenamefont
  {Ceriotti}}]{chon+23jctc}%
  \BibitemOpen
  \bibfield  {author} {\bibinfo {author} {\bibfnamefont {Sanggyu}\ \bibnamefont
  {Chong}}, \bibinfo {author} {\bibfnamefont {Federico}\ \bibnamefont
  {Grasselli}}, \bibinfo {author} {\bibfnamefont {Chiheb}\ \bibnamefont
  {Ben~Mahmoud}}, \bibinfo {author} {\bibfnamefont {Joe~D.}\ \bibnamefont
  {Morrow}}, \bibinfo {author} {\bibfnamefont {Volker~L.}\ \bibnamefont
  {Deringer}}, \ and\ \bibinfo {author} {\bibfnamefont {Michele}\ \bibnamefont
  {Ceriotti}},\ }\bibfield  {title} {\enquote {\bibinfo {title} {Robustness of
  {{Local Predictions}} in {{Atomistic Machine Learning Models}}},}\ }\href
  {\doibase 10.1021/acs.jctc.3c00704} {\bibfield  {journal} {\bibinfo
  {journal} {J. Chem. Theory Comput.}\ }\textbf {\bibinfo {volume} {19}},\
  \bibinfo {pages} {8020--8031} (\bibinfo {year} {2023})}\BibitemShut {NoStop}%
\bibitem [{\citenamefont {Bigi}\ \emph {et~al.}(2024)\citenamefont {Bigi},
  \citenamefont {Chong}, \citenamefont {Ceriotti},\ and\ \citenamefont
  {Grasselli}}]{bigiPredictionRigidityFormalism2024a}%
  \BibitemOpen
  \bibfield  {author} {\bibinfo {author} {\bibfnamefont {Filippo}\ \bibnamefont
  {Bigi}}, \bibinfo {author} {\bibfnamefont {Sanggyu}\ \bibnamefont {Chong}},
  \bibinfo {author} {\bibfnamefont {Michele}\ \bibnamefont {Ceriotti}}, \ and\
  \bibinfo {author} {\bibfnamefont {Federico}\ \bibnamefont {Grasselli}},\
  }\bibfield  {title} {\enquote {\bibinfo {title} {A prediction rigidity
  formalism for low-cost uncertainties in trained neural networks},}\ }\href
  {\doibase 10.1088/2632-2153/ad805f} {\bibfield  {journal} {\bibinfo
  {journal} {Machine Learning: Science and Technology}\ }\textbf {\bibinfo
  {volume} {5}},\ \bibinfo {pages} {045018} (\bibinfo {year}
  {2024})}\BibitemShut {NoStop}%
\bibitem [{\citenamefont {Chong}\ \emph {et~al.}(2025)\citenamefont {Chong},
  \citenamefont {Bigi}, \citenamefont {Grasselli}, \citenamefont {Loche},
  \citenamefont {Kellner},\ and\ \citenamefont {Ceriotti}}]{chon+25fd}%
  \BibitemOpen
  \bibfield  {author} {\bibinfo {author} {\bibfnamefont {Sanggyu}\ \bibnamefont
  {Chong}}, \bibinfo {author} {\bibfnamefont {Filippo}\ \bibnamefont {Bigi}},
  \bibinfo {author} {\bibfnamefont {Federico}\ \bibnamefont {Grasselli}},
  \bibinfo {author} {\bibfnamefont {Philip}\ \bibnamefont {Loche}}, \bibinfo
  {author} {\bibfnamefont {Matthias}\ \bibnamefont {Kellner}}, \ and\ \bibinfo
  {author} {\bibfnamefont {Michele}\ \bibnamefont {Ceriotti}},\ }\bibfield
  {title} {\enquote {\bibinfo {title} {Prediction rigidities for data-driven
  chemistry},}\ }\href {\doibase 10.1039/D4FD00101J} {\bibfield  {journal}
  {\bibinfo  {journal} {Faraday Discuss.}\ }\textbf {\bibinfo {volume} {256}},\
  \bibinfo {pages} {322--344} (\bibinfo {year} {2025})}\BibitemShut {NoStop}%
\bibitem [{\citenamefont {Batatia}\ \emph {et~al.}(2025)\citenamefont
  {Batatia}, \citenamefont {Benner}, \citenamefont {Chiang}, \citenamefont
  {Elena}, \citenamefont {Kovács}, \citenamefont {Riebesell}, \citenamefont
  {Advincula}, \citenamefont {Asta}, \citenamefont {Avaylon}, \citenamefont
  {Baldwin}, \citenamefont {Berger}, \citenamefont {Bernstein}, \citenamefont
  {Bhowmik}, \citenamefont {Bigi}, \citenamefont {Blau}, \citenamefont
  {Cărare}, \citenamefont {Ceriotti}, \citenamefont {Chong}, \citenamefont
  {Darby}, \citenamefont {De}, \citenamefont {Della~Pia}, \citenamefont
  {Deringer}, \citenamefont {Elijošius}, \citenamefont {El-Machachi},
  \citenamefont {Fako}, \citenamefont {Falcioni}, \citenamefont {Ferrari},
  \citenamefont {Gardner}, \citenamefont {Gawkowski}, \citenamefont
  {Genreith-Schriever}, \citenamefont {George}, \citenamefont {Goodall},
  \citenamefont {Grandel}, \citenamefont {Grey}, \citenamefont {Grigorev},
  \citenamefont {Han}, \citenamefont {Handley}, \citenamefont {Heenen},
  \citenamefont {Hermansson}, \citenamefont {Ho}, \citenamefont {Hofmann},
  \citenamefont {Holm}, \citenamefont {Jaafar}, \citenamefont {Jakob},
  \citenamefont {Jung}, \citenamefont {Kapil}, \citenamefont {Kaplan},
  \citenamefont {Karimitari}, \citenamefont {Kermode}, \citenamefont {Kourtis},
  \citenamefont {Kroupa}, \citenamefont {Kullgren}, \citenamefont {Kuner},
  \citenamefont {Kuryla}, \citenamefont {Liepuoniute}, \citenamefont {Lin},
  \citenamefont {Margraf}, \citenamefont {Magdău}, \citenamefont
  {Michaelides}, \citenamefont {Moore}, \citenamefont {Naik}, \citenamefont
  {Niblett}, \citenamefont {Norwood}, \citenamefont {O’Neill}, \citenamefont
  {Ortner}, \citenamefont {Persson}, \citenamefont {Reuter}, \citenamefont
  {Rosen}, \citenamefont {Rosset}, \citenamefont {Schaaf}, \citenamefont
  {Schran}, \citenamefont {Shi}, \citenamefont {Sivonxay}, \citenamefont
  {Stenczel}, \citenamefont {Sutton}, \citenamefont {Svahn}, \citenamefont
  {Swinburne}, \citenamefont {Tilly}, \citenamefont {van~der Oord},
  \citenamefont {Vargas}, \citenamefont {Varga-Umbrich}, \citenamefont {Vegge},
  \citenamefont {Vondrák}, \citenamefont {Wang}, \citenamefont {Witt},
  \citenamefont {Wolf}, \citenamefont {Zills},\ and\ \citenamefont
  {Csányi}}]{batatiaFoundationModelAtomistic2025}%
  \BibitemOpen
  \bibfield  {author} {\bibinfo {author} {\bibfnamefont {Ilyes}\ \bibnamefont
  {Batatia}}, \bibinfo {author} {\bibfnamefont {Philipp}\ \bibnamefont
  {Benner}}, \bibinfo {author} {\bibfnamefont {Yuan}\ \bibnamefont {Chiang}},
  \bibinfo {author} {\bibfnamefont {Alin~M.}\ \bibnamefont {Elena}}, \bibinfo
  {author} {\bibfnamefont {Dávid~P.}\ \bibnamefont {Kovács}}, \bibinfo
  {author} {\bibfnamefont {Janosh}\ \bibnamefont {Riebesell}}, \bibinfo
  {author} {\bibfnamefont {Xavier~R.}\ \bibnamefont {Advincula}}, \bibinfo
  {author} {\bibfnamefont {Mark}\ \bibnamefont {Asta}}, \bibinfo {author}
  {\bibfnamefont {Matthew}\ \bibnamefont {Avaylon}}, \bibinfo {author}
  {\bibfnamefont {William~J.}\ \bibnamefont {Baldwin}}, \bibinfo {author}
  {\bibfnamefont {Fabian}\ \bibnamefont {Berger}}, \bibinfo {author}
  {\bibfnamefont {Noam}\ \bibnamefont {Bernstein}}, \bibinfo {author}
  {\bibfnamefont {Arghya}\ \bibnamefont {Bhowmik}}, \bibinfo {author}
  {\bibfnamefont {Filippo}\ \bibnamefont {Bigi}}, \bibinfo {author}
  {\bibfnamefont {Samuel~M.}\ \bibnamefont {Blau}}, \bibinfo {author}
  {\bibfnamefont {Vlad}\ \bibnamefont {Cărare}}, \bibinfo {author}
  {\bibfnamefont {Michele}\ \bibnamefont {Ceriotti}}, \bibinfo {author}
  {\bibfnamefont {Sanggyu}\ \bibnamefont {Chong}}, \bibinfo {author}
  {\bibfnamefont {James~P.}\ \bibnamefont {Darby}}, \bibinfo {author}
  {\bibfnamefont {Sandip}\ \bibnamefont {De}}, \bibinfo {author} {\bibfnamefont
  {Flaviano}\ \bibnamefont {Della~Pia}}, \bibinfo {author} {\bibfnamefont
  {Volker~L.}\ \bibnamefont {Deringer}}, \bibinfo {author} {\bibfnamefont
  {Rokas}\ \bibnamefont {Elijošius}}, \bibinfo {author} {\bibfnamefont
  {Zakariya}\ \bibnamefont {El-Machachi}}, \bibinfo {author} {\bibfnamefont
  {Edvin}\ \bibnamefont {Fako}}, \bibinfo {author} {\bibfnamefont {Fabio}\
  \bibnamefont {Falcioni}}, \bibinfo {author} {\bibfnamefont {Andrea~C.}\
  \bibnamefont {Ferrari}}, \bibinfo {author} {\bibfnamefont {John L.~A.}\
  \bibnamefont {Gardner}}, \bibinfo {author} {\bibfnamefont {Mikołaj~J.}\
  \bibnamefont {Gawkowski}}, \bibinfo {author} {\bibfnamefont {Annalena}\
  \bibnamefont {Genreith-Schriever}}, \bibinfo {author} {\bibfnamefont
  {Janine}\ \bibnamefont {George}}, \bibinfo {author} {\bibfnamefont {Rhys
  E.~A.}\ \bibnamefont {Goodall}}, \bibinfo {author} {\bibfnamefont {Jonas}\
  \bibnamefont {Grandel}}, \bibinfo {author} {\bibfnamefont {Clare~P.}\
  \bibnamefont {Grey}}, \bibinfo {author} {\bibfnamefont {Petr}\ \bibnamefont
  {Grigorev}}, \bibinfo {author} {\bibfnamefont {Shuang}\ \bibnamefont {Han}},
  \bibinfo {author} {\bibfnamefont {Will}\ \bibnamefont {Handley}}, \bibinfo
  {author} {\bibfnamefont {Hendrik~H.}\ \bibnamefont {Heenen}}, \bibinfo
  {author} {\bibfnamefont {Kersti}\ \bibnamefont {Hermansson}}, \bibinfo
  {author} {\bibfnamefont {Cheuk~Hin}\ \bibnamefont {Ho}}, \bibinfo {author}
  {\bibfnamefont {Stephan}\ \bibnamefont {Hofmann}}, \bibinfo {author}
  {\bibfnamefont {Christian}\ \bibnamefont {Holm}}, \bibinfo {author}
  {\bibfnamefont {Jad}\ \bibnamefont {Jaafar}}, \bibinfo {author}
  {\bibfnamefont {Konstantin~S.}\ \bibnamefont {Jakob}}, \bibinfo {author}
  {\bibfnamefont {Hyunwook}\ \bibnamefont {Jung}}, \bibinfo {author}
  {\bibfnamefont {Venkat}\ \bibnamefont {Kapil}}, \bibinfo {author}
  {\bibfnamefont {Aaron~D.}\ \bibnamefont {Kaplan}}, \bibinfo {author}
  {\bibfnamefont {Nima}\ \bibnamefont {Karimitari}}, \bibinfo {author}
  {\bibfnamefont {James~R.}\ \bibnamefont {Kermode}}, \bibinfo {author}
  {\bibfnamefont {Panagiotis}\ \bibnamefont {Kourtis}}, \bibinfo {author}
  {\bibfnamefont {Namu}\ \bibnamefont {Kroupa}}, \bibinfo {author}
  {\bibfnamefont {Jolla}\ \bibnamefont {Kullgren}}, \bibinfo {author}
  {\bibfnamefont {Matthew~C.}\ \bibnamefont {Kuner}}, \bibinfo {author}
  {\bibfnamefont {Domantas}\ \bibnamefont {Kuryla}}, \bibinfo {author}
  {\bibfnamefont {Guoda}\ \bibnamefont {Liepuoniute}}, \bibinfo {author}
  {\bibfnamefont {Chen}\ \bibnamefont {Lin}}, \bibinfo {author} {\bibfnamefont
  {Johannes~T.}\ \bibnamefont {Margraf}}, \bibinfo {author} {\bibfnamefont
  {Ioan-Bogdan}\ \bibnamefont {Magdău}}, \bibinfo {author} {\bibfnamefont
  {Angelos}\ \bibnamefont {Michaelides}}, \bibinfo {author} {\bibfnamefont
  {J.~Harry}\ \bibnamefont {Moore}}, \bibinfo {author} {\bibfnamefont
  {Aakash~A.}\ \bibnamefont {Naik}}, \bibinfo {author} {\bibfnamefont
  {Samuel~P.}\ \bibnamefont {Niblett}}, \bibinfo {author} {\bibfnamefont
  {Sam~Walton}\ \bibnamefont {Norwood}}, \bibinfo {author} {\bibfnamefont
  {Niamh}\ \bibnamefont {O’Neill}}, \bibinfo {author} {\bibfnamefont
  {Christoph}\ \bibnamefont {Ortner}}, \bibinfo {author} {\bibfnamefont
  {Kristin~A.}\ \bibnamefont {Persson}}, \bibinfo {author} {\bibfnamefont
  {Karsten}\ \bibnamefont {Reuter}}, \bibinfo {author} {\bibfnamefont
  {Andrew~S.}\ \bibnamefont {Rosen}}, \bibinfo {author} {\bibfnamefont {Louise
  A.~M.}\ \bibnamefont {Rosset}}, \bibinfo {author} {\bibfnamefont {Lars~L.}\
  \bibnamefont {Schaaf}}, \bibinfo {author} {\bibfnamefont {Christoph}\
  \bibnamefont {Schran}}, \bibinfo {author} {\bibfnamefont {Benjamin~X.}\
  \bibnamefont {Shi}}, \bibinfo {author} {\bibfnamefont {Eric}\ \bibnamefont
  {Sivonxay}}, \bibinfo {author} {\bibfnamefont {Tamás~K.}\ \bibnamefont
  {Stenczel}}, \bibinfo {author} {\bibfnamefont {Christopher}\ \bibnamefont
  {Sutton}}, \bibinfo {author} {\bibfnamefont {Viktor}\ \bibnamefont {Svahn}},
  \bibinfo {author} {\bibfnamefont {Thomas~D.}\ \bibnamefont {Swinburne}},
  \bibinfo {author} {\bibfnamefont {Jules}\ \bibnamefont {Tilly}}, \bibinfo
  {author} {\bibfnamefont {Cas}\ \bibnamefont {van~der Oord}}, \bibinfo
  {author} {\bibfnamefont {Santiago}\ \bibnamefont {Vargas}}, \bibinfo {author}
  {\bibfnamefont {Eszter}\ \bibnamefont {Varga-Umbrich}}, \bibinfo {author}
  {\bibfnamefont {Tejs}\ \bibnamefont {Vegge}}, \bibinfo {author}
  {\bibfnamefont {Martin}\ \bibnamefont {Vondrák}}, \bibinfo {author}
  {\bibfnamefont {Yangshuai}\ \bibnamefont {Wang}}, \bibinfo {author}
  {\bibfnamefont {William~C.}\ \bibnamefont {Witt}}, \bibinfo {author}
  {\bibfnamefont {Thomas}\ \bibnamefont {Wolf}}, \bibinfo {author}
  {\bibfnamefont {Fabian}\ \bibnamefont {Zills}}, \ and\ \bibinfo {author}
  {\bibfnamefont {Gábor}\ \bibnamefont {Csányi}},\ }\bibfield  {title}
  {\enquote {\bibinfo {title} {A foundation model for atomistic materials
  chemistry},}\ }\href {\doibase 10.1063/5.0297006} {\bibfield  {journal}
  {\bibinfo  {journal} {The Journal of Chemical Physics}\ }\textbf {\bibinfo
  {volume} {163}},\ \bibinfo {pages} {184110} (\bibinfo {year}
  {2025})}\BibitemShut {NoStop}%
\bibitem [{\citenamefont {Behler}\ and\ \citenamefont
  {Parrinello}(2007)}]{behlerGeneralizedNeuralNetworkRepresentation2007a}%
  \BibitemOpen
  \bibfield  {author} {\bibinfo {author} {\bibfnamefont {Jörg}\ \bibnamefont
  {Behler}}\ and\ \bibinfo {author} {\bibfnamefont {Michele}\ \bibnamefont
  {Parrinello}},\ }\bibfield  {title} {\enquote {\bibinfo {title} {Generalized
  {{Neural-Network Representation}} of {{High-Dimensional Potential-Energy
  Surfaces}}},}\ }\href {\doibase 10.1103/PhysRevLett.98.146401} {\bibfield
  {journal} {\bibinfo  {journal} {Physical Review Letters}\ }\textbf {\bibinfo
  {volume} {98}},\ \bibinfo {pages} {146401} (\bibinfo {year}
  {2007})}\BibitemShut {NoStop}%
\bibitem [{\citenamefont {Schütt}\ \emph {et~al.}(2021)\citenamefont
  {Schütt}, \citenamefont {Unke},\ and\ \citenamefont
  {Gastegger}}]{schuttEquivariantMessagePassing2021b}%
  \BibitemOpen
  \bibfield  {author} {\bibinfo {author} {\bibfnamefont {Kristof}\ \bibnamefont
  {Schütt}}, \bibinfo {author} {\bibfnamefont {Oliver}\ \bibnamefont {Unke}},
  \ and\ \bibinfo {author} {\bibfnamefont {Michael}\ \bibnamefont
  {Gastegger}},\ }\bibfield  {title} {\enquote {\bibinfo {title} {Equivariant
  message passing for the prediction of tensorial properties and molecular
  spectra},}\ }in\ \href {https://proceedings.mlr.press/v139/schutt21a.html}
  {\emph {\bibinfo {booktitle} {Proceedings of the 38th International
  Conference on Machine Learning}}}\ (\bibinfo  {publisher} {{PMLR}},\ \bibinfo
  {year} {2021})\ pp.\ \bibinfo {pages} {9377--9388},\ \bibinfo {note} {{ISSN}:
  2640-3498}\BibitemShut {NoStop}%
\bibitem [{\citenamefont {Pozdnyakov}\ \emph {et~al.}(2020)\citenamefont
  {Pozdnyakov}, \citenamefont {Willatt}, \citenamefont {Bartók}, \citenamefont
  {Ortner}, \citenamefont {Csányi},\ and\ \citenamefont
  {Ceriotti}}]{pozdnyakovIncompletenessAtomicStructure2020b}%
  \BibitemOpen
  \bibfield  {author} {\bibinfo {author} {\bibfnamefont {Sergey~N.}\
  \bibnamefont {Pozdnyakov}}, \bibinfo {author} {\bibfnamefont {Michael~J.}\
  \bibnamefont {Willatt}}, \bibinfo {author} {\bibfnamefont {Albert~P.}\
  \bibnamefont {Bartók}}, \bibinfo {author} {\bibfnamefont {Christoph}\
  \bibnamefont {Ortner}}, \bibinfo {author} {\bibfnamefont {Gábor}\
  \bibnamefont {Csányi}}, \ and\ \bibinfo {author} {\bibfnamefont {Michele}\
  \bibnamefont {Ceriotti}},\ }\bibfield  {title} {\enquote {\bibinfo {title}
  {Incompleteness of atomic structure representations},}\ }\href {\doibase
  10.1103/PhysRevLett.125.166001} {\bibfield  {journal} {\bibinfo  {journal}
  {Physical Review Letters}\ }\textbf {\bibinfo {volume} {125}},\ \bibinfo
  {pages} {166001} (\bibinfo {year} {2020})},\ \bibinfo {note} {publisher:
  American Physical Society}\BibitemShut {NoStop}%
\bibitem [{\citenamefont {Zaverkin}\ and\ \citenamefont
  {K{\"a}stner}(2020)}]{zaverkinGaussianMomentsPhysically2020}%
  \BibitemOpen
  \bibfield  {author} {\bibinfo {author} {\bibfnamefont {Viktor}\ \bibnamefont
  {Zaverkin}}\ and\ \bibinfo {author} {\bibfnamefont {Johannes}\ \bibnamefont
  {K{\"a}stner}},\ }\bibfield  {title} {\enquote {\bibinfo {title} {Gaussian
  {{Moments}} as {{Physically Inspired Molecular Descriptors}} for {{Accurate}}
  and {{Scalable Machine Learning Potentials}}},}\ }\href {\doibase
  10.1021/acs.jctc.0c00347} {\bibfield  {journal} {\bibinfo  {journal} {Journal
  of Chemical Theory and Computation}\ }\textbf {\bibinfo {volume} {16}},\
  \bibinfo {pages} {5410--5421} (\bibinfo {year} {2020})}\BibitemShut {NoStop}%
\bibitem [{\citenamefont {Zaverkin}\ \emph {et~al.}(2021)\citenamefont
  {Zaverkin}, \citenamefont {Holzmüller}, \citenamefont {Steinwart},\ and\
  \citenamefont {Kästner}}]{zaverkinFastSampleEfficientInteratomic2021}%
  \BibitemOpen
  \bibfield  {author} {\bibinfo {author} {\bibfnamefont {Viktor}\ \bibnamefont
  {Zaverkin}}, \bibinfo {author} {\bibfnamefont {David}\ \bibnamefont
  {Holzmüller}}, \bibinfo {author} {\bibfnamefont {Ingo}\ \bibnamefont
  {Steinwart}}, \ and\ \bibinfo {author} {\bibfnamefont {Johannes}\
  \bibnamefont {Kästner}},\ }\bibfield  {title} {\enquote {\bibinfo {title}
  {Fast and sample-efficient interatomic neural network potentials for
  molecules and materials based on gaussian moments},}\ }\href {\doibase
  10.1021/acs.jctc.1c00527} {\bibfield  {journal} {\bibinfo  {journal} {Journal
  of Chemical Theory and Computation}\ }\textbf {\bibinfo {volume} {17}},\
  \bibinfo {pages} {6658--6670} (\bibinfo {year} {2021})},\ \Eprint
  {http://arxiv.org/abs/2109.09569} {2109.09569} \BibitemShut {NoStop}%
\bibitem [{\citenamefont {Sch{\"a}fer}\ \emph {et~al.}(2025)\citenamefont
  {Sch{\"a}fer}, \citenamefont {Segreto}, \citenamefont {Zills}, \citenamefont
  {Holm},\ and\ \citenamefont
  {K{\"a}stner}}]{schaferApaxFlexiblePerformant2025}%
  \BibitemOpen
  \bibfield  {author} {\bibinfo {author} {\bibfnamefont {Moritz~R.}\
  \bibnamefont {Sch{\"a}fer}}, \bibinfo {author} {\bibfnamefont {Nico}\
  \bibnamefont {Segreto}}, \bibinfo {author} {\bibfnamefont {Fabian}\
  \bibnamefont {Zills}}, \bibinfo {author} {\bibfnamefont {Christian}\
  \bibnamefont {Holm}}, \ and\ \bibinfo {author} {\bibfnamefont {Johannes}\
  \bibnamefont {K{\"a}stner}},\ }\bibfield  {title} {\enquote {\bibinfo {title}
  {Apax: {{A Flexible}} and {{Performant Framework}} for the {{Development}} of
  {{Machine-Learned Interatomic Potentials}}},}\ }\href {\doibase
  10.1021/acs.jcim.5c01221} {\bibfield  {journal} {\bibinfo  {journal} {Journal
  of Chemical Information and Modeling}\ }\textbf {\bibinfo {volume} {65}},\
  \bibinfo {pages} {8066--8078} (\bibinfo {year} {2025})}\BibitemShut {NoStop}%
\bibitem [{\citenamefont {Frank}\ \emph {et~al.}(2022)\citenamefont {Frank},
  \citenamefont {Unke},\ and\ \citenamefont
  {M{\"u}ller}}]{frankSo3kratesEquivariantAttention2022}%
  \BibitemOpen
  \bibfield  {author} {\bibinfo {author} {\bibfnamefont {Thorben}\ \bibnamefont
  {Frank}}, \bibinfo {author} {\bibfnamefont {Oliver}\ \bibnamefont {Unke}}, \
  and\ \bibinfo {author} {\bibfnamefont {Klaus-Robert}\ \bibnamefont
  {M{\"u}ller}},\ }\bibfield  {title} {\enquote {\bibinfo {title} {So3krates:
  {{Equivariant}} attention for interactions on arbitrary length-scales in
  molecular systems},}\ }\href@noop {} {\bibfield  {journal} {\bibinfo
  {journal} {Advances in Neural Information Processing Systems}\ }\textbf
  {\bibinfo {volume} {35}},\ \bibinfo {pages} {29400--29413} (\bibinfo {year}
  {2022})}\BibitemShut {NoStop}%
\bibitem [{\citenamefont {Batzner}\ \emph {et~al.}(2022)\citenamefont
  {Batzner}, \citenamefont {Musaelian}, \citenamefont {Sun}, \citenamefont
  {Geiger}, \citenamefont {Mailoa}, \citenamefont {Kornbluth}, \citenamefont
  {Molinari}, \citenamefont {Smidt},\ and\ \citenamefont
  {Kozinsky}}]{batznerE3equivariantGraphNeural2022}%
  \BibitemOpen
  \bibfield  {author} {\bibinfo {author} {\bibfnamefont {Simon}\ \bibnamefont
  {Batzner}}, \bibinfo {author} {\bibfnamefont {Albert}\ \bibnamefont
  {Musaelian}}, \bibinfo {author} {\bibfnamefont {Lixin}\ \bibnamefont {Sun}},
  \bibinfo {author} {\bibfnamefont {Mario}\ \bibnamefont {Geiger}}, \bibinfo
  {author} {\bibfnamefont {Jonathan~P.}\ \bibnamefont {Mailoa}}, \bibinfo
  {author} {\bibfnamefont {Mordechai}\ \bibnamefont {Kornbluth}}, \bibinfo
  {author} {\bibfnamefont {Nicola}\ \bibnamefont {Molinari}}, \bibinfo {author}
  {\bibfnamefont {Tess~E.}\ \bibnamefont {Smidt}}, \ and\ \bibinfo {author}
  {\bibfnamefont {Boris}\ \bibnamefont {Kozinsky}},\ }\bibfield  {title}
  {\enquote {\bibinfo {title} {E(3)-equivariant graph neural networks for
  data-efficient and accurate interatomic potentials},}\ }\href {\doibase
  10.1038/s41467-022-29939-5} {\bibfield  {journal} {\bibinfo  {journal}
  {Nature Communications}\ }\textbf {\bibinfo {volume} {13}},\ \bibinfo {pages}
  {2453} (\bibinfo {year} {2022})}\BibitemShut {NoStop}%
\bibitem [{\citenamefont {Zills}\ \emph
  {et~al.}(2024{\natexlab{b}})\citenamefont {Zills}, \citenamefont {Schäfer},
  \citenamefont {Segreto}, \citenamefont {Kästner}, \citenamefont {Holm},\
  and\ \citenamefont {Tovey}}]{zillsCollaborationMachineLearnedPotentials2024}%
  \BibitemOpen
  \bibfield  {author} {\bibinfo {author} {\bibfnamefont {Fabian}\ \bibnamefont
  {Zills}}, \bibinfo {author} {\bibfnamefont {Moritz~René}\ \bibnamefont
  {Schäfer}}, \bibinfo {author} {\bibfnamefont {Nico}\ \bibnamefont
  {Segreto}}, \bibinfo {author} {\bibfnamefont {Johannes}\ \bibnamefont
  {Kästner}}, \bibinfo {author} {\bibfnamefont {Christian}\ \bibnamefont
  {Holm}}, \ and\ \bibinfo {author} {\bibfnamefont {Samuel}\ \bibnamefont
  {Tovey}},\ }\bibfield  {title} {\enquote {\bibinfo {title} {Collaboration on
  {{Machine-Learned Potentials}} with {{IPSuite}}: {{A Modular Framework}} for
  {{Learning-on-the-Fly}}},}\ }\href {\doibase 10.1021/acs.jpcb.3c07187}
  {\bibfield  {journal} {\bibinfo  {journal} {The Journal of Physical Chemistry
  B}\ }\textbf {\bibinfo {volume} {128}},\ \bibinfo {pages} {3662--3676}
  (\bibinfo {year} {2024}{\natexlab{b}})}\BibitemShut {NoStop}%
\bibitem [{\citenamefont {Welling}\ and\ \citenamefont
  {Teh}(2011)}]{welling2011bayesian}%
  \BibitemOpen
  \bibfield  {author} {\bibinfo {author} {\bibfnamefont {Max}\ \bibnamefont
  {Welling}}\ and\ \bibinfo {author} {\bibfnamefont {Yee~W}\ \bibnamefont
  {Teh}},\ }\bibfield  {title} {\enquote {\bibinfo {title} {Bayesian learning
  via stochastic gradient langevin dynamics},}\ }in\ \href@noop {} {\emph
  {\bibinfo {booktitle} {Proceedings of the 28th international conference on
  machine learning (ICML-11)}}}\ (\bibinfo {year} {2011})\ pp.\ \bibinfo
  {pages} {681--688}\BibitemShut {NoStop}%
\bibitem [{\citenamefont {Thaler}\ \emph {et~al.}(2023)\citenamefont {Thaler},
  \citenamefont {Doehner},\ and\ \citenamefont
  {Zavadlav}}]{thalerScalableBayesianUncertainty2023a}%
  \BibitemOpen
  \bibfield  {author} {\bibinfo {author} {\bibfnamefont {Stephan}\ \bibnamefont
  {Thaler}}, \bibinfo {author} {\bibfnamefont {Gregor}\ \bibnamefont
  {Doehner}}, \ and\ \bibinfo {author} {\bibfnamefont {Julija}\ \bibnamefont
  {Zavadlav}},\ }\bibfield  {title} {\enquote {\bibinfo {title} {Scalable
  {{Bayesian Uncertainty Quantification}} for {{Neural Network Potentials}}:
  {{Promise}} and {{Pitfalls}}},}\ }\href {\doibase 10.1021/acs.jctc.2c01267}
  {\bibfield  {journal} {\bibinfo  {journal} {Journal of Chemical Theory and
  Computation}\ }\textbf {\bibinfo {volume} {19}},\ \bibinfo {pages}
  {4520--4532} (\bibinfo {year} {2023})}\BibitemShut {NoStop}%
\bibitem [{\citenamefont {Fort}\ and\ \citenamefont
  {Jastrzebski}(2019)}]{fortLargeScaleStructure2019}%
  \BibitemOpen
  \bibfield  {author} {\bibinfo {author} {\bibfnamefont {Stanislav}\
  \bibnamefont {Fort}}\ and\ \bibinfo {author} {\bibfnamefont {Stanislaw}\
  \bibnamefont {Jastrzebski}},\ }\bibfield  {title} {\enquote {\bibinfo {title}
  {Large scale structure of neural network loss landscapes},}\ }in\ \href@noop
  {} {\emph {\bibinfo {booktitle} {Proceedings of the 33rd {{International
  Conference}} on {{Neural Information Processing Systems}}}}},\ \bibinfo
  {series and number} {\bibinfo {number} {602}}\ (\bibinfo  {publisher} {Curran
  Associates Inc.},\ \bibinfo {year} {2019})\ pp.\ \bibinfo {pages}
  {6709--6717}\BibitemShut {NoStop}%
\bibitem [{\citenamefont {Fort}\ \emph {et~al.}(2020)\citenamefont {Fort},
  \citenamefont {Hu},\ and\ \citenamefont
  {Lakshminarayanan}}]{fortDeepEnsemblesLoss2020}%
  \BibitemOpen
  \bibfield  {author} {\bibinfo {author} {\bibfnamefont {Stanislav}\
  \bibnamefont {Fort}}, \bibinfo {author} {\bibfnamefont {Huiyi}\ \bibnamefont
  {Hu}}, \ and\ \bibinfo {author} {\bibfnamefont {Balaji}\ \bibnamefont
  {Lakshminarayanan}},\ }\href {\doibase 10.48550/arXiv.1912.02757} {\enquote
  {\bibinfo {title} {Deep {{Ensembles}}: {{A Loss Landscape Perspective}}},}\ }
  (\bibinfo {year} {2020}),\ \Eprint {http://arxiv.org/abs/1912.02757}
  {1912.02757} \BibitemShut {NoStop}%
\bibitem [{\citenamefont {Zhang}\ \emph {et~al.}(2019)\citenamefont {Zhang},
  \citenamefont {Lin}, \citenamefont {Wang}, \citenamefont {Car},\ and\
  \citenamefont {E}}]{zhangActiveLearningUniformly2019}%
  \BibitemOpen
  \bibfield  {author} {\bibinfo {author} {\bibfnamefont {Linfeng}\ \bibnamefont
  {Zhang}}, \bibinfo {author} {\bibfnamefont {De-Ye}\ \bibnamefont {Lin}},
  \bibinfo {author} {\bibfnamefont {Han}\ \bibnamefont {Wang}}, \bibinfo
  {author} {\bibfnamefont {Roberto}\ \bibnamefont {Car}}, \ and\ \bibinfo
  {author} {\bibfnamefont {Weinan}\ \bibnamefont {E}},\ }\bibfield  {title}
  {\enquote {\bibinfo {title} {Active learning of uniformly accurate
  interatomic potentials for materials simulation},}\ }\href {\doibase
  10.1103/PhysRevMaterials.3.023804} {\bibfield  {journal} {\bibinfo  {journal}
  {Physical Review Materials}\ }\textbf {\bibinfo {volume} {3}},\ \bibinfo
  {pages} {023804} (\bibinfo {year} {2019})}\BibitemShut {NoStop}%
\bibitem [{\citenamefont {Podryabinkin}\ and\ \citenamefont
  {Shapeev}(2017)}]{podryabinkinActiveLearningLinearly2017a}%
  \BibitemOpen
  \bibfield  {author} {\bibinfo {author} {\bibfnamefont {Evgeny~V.}\
  \bibnamefont {Podryabinkin}}\ and\ \bibinfo {author} {\bibfnamefont
  {Alexander~V.}\ \bibnamefont {Shapeev}},\ }\bibfield  {title} {\enquote
  {\bibinfo {title} {Active learning of linearly parametrized interatomic
  potentials},}\ }\href {\doibase 10.1016/j.commatsci.2017.08.031} {\bibfield
  {journal} {\bibinfo  {journal} {Computational Materials Science}\ }\textbf
  {\bibinfo {volume} {140}},\ \bibinfo {pages} {171--180} (\bibinfo {year}
  {2017})}\BibitemShut {NoStop}%
\bibitem [{\citenamefont {Bui}\ \emph {et~al.}(2025)\citenamefont {Bui},
  \citenamefont {Maifeld-Carucci},\ and\ \citenamefont
  {Liu}}]{buiCalibratedUncertaintySampling2025}%
  \BibitemOpen
  \bibfield  {author} {\bibinfo {author} {\bibfnamefont {Ha~Manh}\ \bibnamefont
  {Bui}}, \bibinfo {author} {\bibfnamefont {Iliana}\ \bibnamefont
  {Maifeld-Carucci}}, \ and\ \bibinfo {author} {\bibfnamefont {Anqi}\
  \bibnamefont {Liu}},\ }\href {\doibase 10.48550/arXiv.2510.03162} {\enquote
  {\bibinfo {title} {Calibrated {{Uncertainty Sampling}} for {{Active
  Learning}}},}\ } (\bibinfo {year} {2025}),\ \Eprint
  {http://arxiv.org/abs/2510.03162} {2510.03162} \BibitemShut {NoStop}%
\bibitem [{\citenamefont {Heid}\ \emph {et~al.}(2024)\citenamefont {Heid},
  \citenamefont {Schörghuber}, \citenamefont {Wanzenböck},\ and\
  \citenamefont {Madsen}}]{heidSpatiallyResolvedUncertainties2024a}%
  \BibitemOpen
  \bibfield  {author} {\bibinfo {author} {\bibfnamefont {Esther}\ \bibnamefont
  {Heid}}, \bibinfo {author} {\bibfnamefont {Johannes}\ \bibnamefont
  {Schörghuber}}, \bibinfo {author} {\bibfnamefont {Ralf}\ \bibnamefont
  {Wanzenböck}}, \ and\ \bibinfo {author} {\bibfnamefont {Georg K.~H.}\
  \bibnamefont {Madsen}},\ }\bibfield  {title} {\enquote {\bibinfo {title}
  {Spatially {{Resolved Uncertainties}} for {{Machine Learning Potentials}}},}\
  }\href {\doibase 10.1021/acs.jcim.4c00904} {\bibfield  {journal} {\bibinfo
  {journal} {Journal of Chemical Information and Modeling}\ }\textbf {\bibinfo
  {volume} {64}},\ \bibinfo {pages} {6377--6387} (\bibinfo {year}
  {2024})}\BibitemShut {NoStop}%
\bibitem [{\citenamefont {Willow}\ \emph {et~al.}(2025)\citenamefont {Willow},
  \citenamefont {Park}, \citenamefont {Sim}, \citenamefont {Moon},
  \citenamefont {Min}, \citenamefont {Yang}, \citenamefont {Kim}, \citenamefont
  {Lee},\ and\ \citenamefont {Myung}}]{willow2025bayesian}%
  \BibitemOpen
  \bibfield  {author} {\bibinfo {author} {\bibfnamefont {Soohaeng~Yoo}\
  \bibnamefont {Willow}}, \bibinfo {author} {\bibfnamefont {Tae~Hyeon}\
  \bibnamefont {Park}}, \bibinfo {author} {\bibfnamefont {Gi~Beom}\
  \bibnamefont {Sim}}, \bibinfo {author} {\bibfnamefont {Sung~Wook}\
  \bibnamefont {Moon}}, \bibinfo {author} {\bibfnamefont {Seung~Kyu}\
  \bibnamefont {Min}}, \bibinfo {author} {\bibfnamefont {D~ChangMo}\
  \bibnamefont {Yang}}, \bibinfo {author} {\bibfnamefont {Hyun~Woo}\
  \bibnamefont {Kim}}, \bibinfo {author} {\bibfnamefont {Juho}\ \bibnamefont
  {Lee}}, \ and\ \bibinfo {author} {\bibfnamefont {Chang~Woo}\ \bibnamefont
  {Myung}},\ }\bibfield  {title} {\enquote {\bibinfo {title} {Bayesian e
  (3)-equivariant interatomic potential with iterative restratification of
  many-body message passing},}\ }\href@noop {} {\bibfield  {journal} {\bibinfo
  {journal} {arXiv preprint arXiv:2510.03046}\ } (\bibinfo {year}
  {2025})}\BibitemShut {NoStop}%
\bibitem [{\citenamefont {Guo}\ \emph {et~al.}(2017)\citenamefont {Guo},
  \citenamefont {Pleiss}, \citenamefont {Sun},\ and\ \citenamefont
  {Weinberger}}]{guoCalibrationModernNeural2017}%
  \BibitemOpen
  \bibfield  {author} {\bibinfo {author} {\bibfnamefont {Chuan}\ \bibnamefont
  {Guo}}, \bibinfo {author} {\bibfnamefont {Geoff}\ \bibnamefont {Pleiss}},
  \bibinfo {author} {\bibfnamefont {Yu}~\bibnamefont {Sun}}, \ and\ \bibinfo
  {author} {\bibfnamefont {Kilian~Q.}\ \bibnamefont {Weinberger}},\ }\bibfield
  {title} {\enquote {\bibinfo {title} {On calibration of modern neural
  networks},}\ }in\ \href {https://dl.acm.org/doi/10.5555/3305381.3305518}
  {\emph {\bibinfo {booktitle} {Proceedings of the 34th {{International
  Conference}} on {{Machine Learning}} - {{Volume}} 70}}},\ \bibinfo {series
  and number} {{{ICML}}'17}\ (\bibinfo  {publisher} {JMLR.org},\ \bibinfo
  {year} {2017})\ pp.\ \bibinfo {pages} {1321--1330}\BibitemShut {NoStop}%
\bibitem [{\citenamefont {Musil}\ \emph {et~al.}(2019)\citenamefont {Musil},
  \citenamefont {Willatt}, \citenamefont {Langovoy},\ and\ \citenamefont
  {Ceriotti}}]{musi+19jctc}%
  \BibitemOpen
  \bibfield  {author} {\bibinfo {author} {\bibfnamefont {F{\'e}lix}\
  \bibnamefont {Musil}}, \bibinfo {author} {\bibfnamefont {Michael~J.}\
  \bibnamefont {Willatt}}, \bibinfo {author} {\bibfnamefont {Mikhail~A.}\
  \bibnamefont {Langovoy}}, \ and\ \bibinfo {author} {\bibfnamefont {Michele}\
  \bibnamefont {Ceriotti}},\ }\bibfield  {title} {\enquote {\bibinfo {title}
  {Fast and {{Accurate Uncertainty Estimation}} in {{Chemical Machine
  Learning}}},}\ }\href {\doibase 10.1021/acs.jctc.8b00959} {\bibfield
  {journal} {\bibinfo  {journal} {Journal of Chemical Theory and Computation}\
  }\textbf {\bibinfo {volume} {15}},\ \bibinfo {pages} {906--915} (\bibinfo
  {year} {2019})}\BibitemShut {NoStop}%
\bibitem [{\citenamefont {Beck}\ \emph {et~al.}(2025)\citenamefont {Beck},
  \citenamefont {Simko}, \citenamefont {Schaaf}, \citenamefont {Marsalek},\
  and\ \citenamefont {Schran}}]{beckMultiheadCommitteesEnable2025}%
  \BibitemOpen
  \bibfield  {author} {\bibinfo {author} {\bibfnamefont {Hubert}\ \bibnamefont
  {Beck}}, \bibinfo {author} {\bibfnamefont {Pavol}\ \bibnamefont {Simko}},
  \bibinfo {author} {\bibfnamefont {Lars~L.}\ \bibnamefont {Schaaf}}, \bibinfo
  {author} {\bibfnamefont {Ondrej}\ \bibnamefont {Marsalek}}, \ and\ \bibinfo
  {author} {\bibfnamefont {Christoph}\ \bibnamefont {Schran}},\ }\bibfield
  {title} {\enquote {\bibinfo {title} {Multi-head committees enable direct
  uncertainty prediction for atomistic foundation models},}\ }\href {\doibase
  10.1063/5.0302097} {\bibfield  {journal} {\bibinfo  {journal} {The Journal of
  Chemical Physics}\ }\textbf {\bibinfo {volume} {163}},\ \bibinfo {pages}
  {234103} (\bibinfo {year} {2025})}\BibitemShut {NoStop}%
\bibitem [{\citenamefont {Rasmussen}\ \emph {et~al.}(2023)\citenamefont
  {Rasmussen}, \citenamefont {Duan}, \citenamefont {Kulik},\ and\ \citenamefont
  {Jensen}}]{rasmussenUncertainUncertaintiesComparison2023}%
  \BibitemOpen
  \bibfield  {author} {\bibinfo {author} {\bibfnamefont {Maria~H.}\
  \bibnamefont {Rasmussen}}, \bibinfo {author} {\bibfnamefont {Chenru}\
  \bibnamefont {Duan}}, \bibinfo {author} {\bibfnamefont {Heather~J.}\
  \bibnamefont {Kulik}}, \ and\ \bibinfo {author} {\bibfnamefont {Jan~H.}\
  \bibnamefont {Jensen}},\ }\bibfield  {title} {\enquote {\bibinfo {title}
  {Uncertain of uncertainties? {{A}} comparison of uncertainty quantification
  metrics for chemical data sets},}\ }\href {\doibase
  10.1186/s13321-023-00790-0} {\bibfield  {journal} {\bibinfo  {journal}
  {Journal of Cheminformatics}\ }\textbf {\bibinfo {volume} {15}},\ \bibinfo
  {pages} {121} (\bibinfo {year} {2023})}\BibitemShut {NoStop}%
\bibitem [{\citenamefont {Brandenburg}\ \emph {et~al.}(2018)\citenamefont
  {Brandenburg}, \citenamefont {Bannwarth}, \citenamefont {Hansen},\ and\
  \citenamefont {Grimme}}]{brandenburgB973cRevisedLowcost2018}%
  \BibitemOpen
  \bibfield  {author} {\bibinfo {author} {\bibfnamefont {Jan~Gerit}\
  \bibnamefont {Brandenburg}}, \bibinfo {author} {\bibfnamefont {Christoph}\
  \bibnamefont {Bannwarth}}, \bibinfo {author} {\bibfnamefont {Andreas}\
  \bibnamefont {Hansen}}, \ and\ \bibinfo {author} {\bibfnamefont {Stefan}\
  \bibnamefont {Grimme}},\ }\bibfield  {title} {\enquote {\bibinfo {title}
  {B97-3c: {{A}} revised low-cost variant of the {{B97-D}} density functional
  method},}\ }\href {\doibase 10.1063/1.5012601} {\bibfield  {journal}
  {\bibinfo  {journal} {J. Chem. Phys.}\ }\textbf {\bibinfo {volume} {148}},\
  \bibinfo {pages} {064104} (\bibinfo {year} {2018})}\BibitemShut {NoStop}%
\bibitem [{\citenamefont
  {Becke}(1997)}]{beckeDensityfunctionalThermochemistrySystematic1997}%
  \BibitemOpen
  \bibfield  {author} {\bibinfo {author} {\bibfnamefont {Axel~D.}\ \bibnamefont
  {Becke}},\ }\bibfield  {title} {\enquote {\bibinfo {title}
  {Density-functional thermochemistry. {{V}}. {{Systematic}} optimization of
  exchange-correlation functionals},}\ }\href {\doibase 10.1063/1.475007}
  {\bibfield  {journal} {\bibinfo  {journal} {J. Chem. Phys.}\ }\textbf
  {\bibinfo {volume} {107}},\ \bibinfo {pages} {8554--8560} (\bibinfo {year}
  {1997})}\BibitemShut {NoStop}%
\bibitem [{\citenamefont {Grimme}\ \emph {et~al.}(2010)\citenamefont {Grimme},
  \citenamefont {Antony}, \citenamefont {Ehrlich},\ and\ \citenamefont
  {Krieg}}]{grimmeConsistentAccurateInitio2010}%
  \BibitemOpen
  \bibfield  {author} {\bibinfo {author} {\bibfnamefont {Stefan}\ \bibnamefont
  {Grimme}}, \bibinfo {author} {\bibfnamefont {Jens}\ \bibnamefont {Antony}},
  \bibinfo {author} {\bibfnamefont {Stephan}\ \bibnamefont {Ehrlich}}, \ and\
  \bibinfo {author} {\bibfnamefont {Helge}\ \bibnamefont {Krieg}},\ }\bibfield
  {title} {\enquote {\bibinfo {title} {A consistent and accurate ab initio
  parametrization of density functional dispersion correction ({{DFT-D}}) for
  the 94 elements {{H-Pu}}},}\ }\href {\doibase 10.1063/1.3382344} {\bibfield
  {journal} {\bibinfo  {journal} {J. Chem. Phys.}\ }\textbf {\bibinfo {volume}
  {132}},\ \bibinfo {pages} {154104} (\bibinfo {year} {2010})}\BibitemShut
  {NoStop}%
\bibitem [{\citenamefont {Gigli}\ \emph {et~al.}(2022)\citenamefont {Gigli},
  \citenamefont {Veit}, \citenamefont {Kotiuga}, \citenamefont {Pizzi},
  \citenamefont {Marzari},\ and\ \citenamefont {Ceriotti}}]{gigl+22npjcm}%
  \BibitemOpen
  \bibfield  {author} {\bibinfo {author} {\bibfnamefont {Lorenzo}\ \bibnamefont
  {Gigli}}, \bibinfo {author} {\bibfnamefont {Max}\ \bibnamefont {Veit}},
  \bibinfo {author} {\bibfnamefont {Michele}\ \bibnamefont {Kotiuga}}, \bibinfo
  {author} {\bibfnamefont {Giovanni}\ \bibnamefont {Pizzi}}, \bibinfo {author}
  {\bibfnamefont {Nicola}\ \bibnamefont {Marzari}}, \ and\ \bibinfo {author}
  {\bibfnamefont {Michele}\ \bibnamefont {Ceriotti}},\ }\bibfield  {title}
  {\enquote {\bibinfo {title} {Thermodynamics and dielectric response of
  {{BaTiO3}} by data-driven modeling},}\ }\href {\doibase
  10.1038/s41524-022-00845-0} {\bibfield  {journal} {\bibinfo  {journal} {npj
  Comput Mater}\ }\textbf {\bibinfo {volume} {8}},\ \bibinfo {pages} {209}
  (\bibinfo {year} {2022})}\BibitemShut {NoStop}%
\bibitem [{\citenamefont {Cheng}\ \emph {et~al.}(2019)\citenamefont {Cheng},
  \citenamefont {Engel}, \citenamefont {Behler}, \citenamefont {Dellago},\ and\
  \citenamefont {Ceriotti}}]{chengInitioThermodynamicsLiquid2019}%
  \BibitemOpen
  \bibfield  {author} {\bibinfo {author} {\bibfnamefont {Bingqing}\
  \bibnamefont {Cheng}}, \bibinfo {author} {\bibfnamefont {Edgar~A.}\
  \bibnamefont {Engel}}, \bibinfo {author} {\bibfnamefont {Jörg}\ \bibnamefont
  {Behler}}, \bibinfo {author} {\bibfnamefont {Christoph}\ \bibnamefont
  {Dellago}}, \ and\ \bibinfo {author} {\bibfnamefont {Michele}\ \bibnamefont
  {Ceriotti}},\ }\bibfield  {title} {\enquote {\bibinfo {title} {Ab initio
  thermodynamics of liquid and solid water},}\ }\href {\doibase
  10.1073/pnas.1815117116} {\bibfield  {journal} {\bibinfo  {journal}
  {Proceedings of the National Academy of Sciences}\ }\textbf {\bibinfo
  {volume} {116}},\ \bibinfo {pages} {1110--1115} (\bibinfo {year}
  {2019})}\BibitemShut {NoStop}%
\bibitem [{\citenamefont {Chmiela}\ \emph {et~al.}(2023)\citenamefont
  {Chmiela}, \citenamefont {Vassilev-Galindo}, \citenamefont {Unke},
  \citenamefont {Kabylda}, \citenamefont {Sauceda}, \citenamefont
  {Tkatchenko},\ and\ \citenamefont
  {Müller}}]{chmielaAccurateGlobalMachine2023a}%
  \BibitemOpen
  \bibfield  {author} {\bibinfo {author} {\bibfnamefont {Stefan}\ \bibnamefont
  {Chmiela}}, \bibinfo {author} {\bibfnamefont {Valentin}\ \bibnamefont
  {Vassilev-Galindo}}, \bibinfo {author} {\bibfnamefont {Oliver~T.}\
  \bibnamefont {Unke}}, \bibinfo {author} {\bibfnamefont {Adil}\ \bibnamefont
  {Kabylda}}, \bibinfo {author} {\bibfnamefont {Huziel~E.}\ \bibnamefont
  {Sauceda}}, \bibinfo {author} {\bibfnamefont {Alexandre}\ \bibnamefont
  {Tkatchenko}}, \ and\ \bibinfo {author} {\bibfnamefont {Klaus-Robert}\
  \bibnamefont {Müller}},\ }\bibfield  {title} {\enquote {\bibinfo {title}
  {Accurate global machine learning force fields for molecules with hundreds of
  atoms},}\ }\href {\doibase 10.1126/sciadv.adf0873} {\bibfield  {journal}
  {\bibinfo  {journal} {Science Advances}\ }\textbf {\bibinfo {volume} {9}},\
  \bibinfo {pages} {eadf0873} (\bibinfo {year} {2023})}\BibitemShut {NoStop}%
\bibitem [{\citenamefont {Deringer}\ and\ \citenamefont
  {Cs{\'a}nyi}(2017)}]{deringerMachineLearningBased2017a}%
  \BibitemOpen
  \bibfield  {author} {\bibinfo {author} {\bibfnamefont {Volker~L.}\
  \bibnamefont {Deringer}}\ and\ \bibinfo {author} {\bibfnamefont {G{\'a}bor}\
  \bibnamefont {Cs{\'a}nyi}},\ }\bibfield  {title} {\enquote {\bibinfo {title}
  {Machine learning based interatomic potential for amorphous carbon},}\ }\href
  {\doibase 10.1103/PhysRevB.95.094203} {\bibfield  {journal} {\bibinfo
  {journal} {Physical Review B}\ }\textbf {\bibinfo {volume} {95}},\ \bibinfo
  {pages} {094203} (\bibinfo {year} {2017})}\BibitemShut {NoStop}%
\bibitem [{\citenamefont {Schran}\ \emph {et~al.}(2020)\citenamefont {Schran},
  \citenamefont {Brezina},\ and\ \citenamefont
  {Marsalek}}]{schranCommitteeNeuralNetwork2020}%
  \BibitemOpen
  \bibfield  {author} {\bibinfo {author} {\bibfnamefont {Christoph}\
  \bibnamefont {Schran}}, \bibinfo {author} {\bibfnamefont {Krystof}\
  \bibnamefont {Brezina}}, \ and\ \bibinfo {author} {\bibfnamefont {Ondrej}\
  \bibnamefont {Marsalek}},\ }\bibfield  {title} {\enquote {\bibinfo {title}
  {Committee neural network potentials control generalization errors and enable
  active learning},}\ }\href {\doibase 10.1063/5.0016004} {\bibfield  {journal}
  {\bibinfo  {journal} {The Journal of Chemical Physics}\ }\textbf {\bibinfo
  {volume} {153}},\ \bibinfo {pages} {104105} (\bibinfo {year}
  {2020})}\BibitemShut {NoStop}%
\end{thebibliography}
\end{document}